\documentclass[12pt,preprint]{aastex}

\shorttitle{DENSITY STATISTICS IN MHD TURBULENCE}
\shortauthors{KOWAL, LAZARIAN \& BERESNYAK}

\begin{document}

\title{Density Fluctuations in MHD turbulence: spectra, intermittency and topology}
\author{G. Kowal, A. Lazarian  and A. Beresnyak}
\affil{Department of Astronomy, University of Wisconsin, Madison, WI}

\begin{abstract}

We perform three-dimensional (3D) compressible MHD simulations over many dynamical times for an extended range of sonic and Alfv\'{e}n Mach numbers and analyze the statistics of 3D density and 2D column density, which include probability distribution functions, spectra, skewness, kurtosis, She-L\'{e}v\^{e}que exponents, and genus. In order to establish the relation between the statistics of the observables, i.e. column densities, and the underlying 3D statistics of density, we analyze the effects of cloud boundaries. We define the parameter space for 3D measures to be recovered from column densities. In addition, we show that for subsonic turbulence the spectra of density fluctuations are consistent with $k^{-7/3}$ in the case of a strong magnetic field and $k^{-5/3}$ in the case of a weak magnetic field. For supersonic turbulence we confirm the earlier findings of the shallow spectra of density and Kolmogorov spectra of the logarithm of density. We find that the intermittencies of the density and velocity are very different.

\end{abstract}

\keywords{ISM: structure --- MHD --- turbulence}

\section{Introduction}

Magnetohydrodynamic (MHD) turbulence is a key element for understanding star formation \citep[see][and references therein]{mckee02,maclow04}. In a new emerging paradigm, density perturbations produced by compressible MHD turbulence create molecular clouds and clumps within the clouds. This makes studies of density statistics very timely. In addition, recent advances in the understanding of the properties of compressible magnetic turbulence \citep[see review by][and references therein]{cho05} call for observational testing. While velocity statistics are currently emerging as an essential tool for turbulence studies \citep[see review by][and references therein]{lazarian04}, density statistics are more readily available from observations.

Variations of density have been extensively studied  for a long time \cite[see][]{kaplan70,dickman85}. Electron density fluctuations have become widely used to gain insight into small-scale properties of interstellar turbulence \cite[see][and reference therein]{armstrong95}. Fluctuations of column density, at the same time, became a useful source of information about turbulence at larger scales \citep{falgarone05,padoan03}. Those studies of column densities are not limited to studies of spectra. Higher order structure functions have been employed recently \cite[see][]{falgarone05,padoan03}. In fact, it was claimed in \cite{padoan04}, that studies of higher order correlations could determine the Mach number of the interstellar flows. This motivates further studies of this type of statistics.

In addition, the topology of density in the interstellar medium (ISM) can provide an important insight into ISM physics. For instance, the topology of the ISM is expected to be different for models of the ISM in which most of the gas is in the hot phase with islands of other phases embedded within \cite[see][]{mckee77} and in models where a substantial part of the volume is filled with cold and warm gas \cite[see][]{cox74}. It was pointed out in \cite{lazarian02}, that the topology of density can also be studied using genus statistics \cite[see][]{gott86}. However, no detailed study has been performed as far as we know.

The issues above call for a comprehensive study of density statistics, which we undertake in this paper. We use an extended set of MHD numerical simulations, obtained for different sonic ${\cal M}_s$ and Alfv\'{e}nic ${\cal M}_A$ Mach numbers, to study spectra, higher order statistics, and the genus of the density. We follow turbulence for many dynamical time steps (e.g. 25 for 128$^3$ simulations) to provide good sampling. In all cases we provide results of synthetic observations to compare the statistics of density with the statistics of the column density that are available through observations.

In this article we investigate the scaling properties of the structure functions of density and the logarithm of density for compressible MHD turbulence with different sonic and Alfv\'{e}nic Mach numbers. In \S\ref{sec:models} we describe the numerical models of compressible MHD turbulence. In \S\ref{sec:pdfs} we discuss one-point probability distribution functions (PDFs). In \S\ref{sec:spectra} we study spectra of density and the logarithm of density and the anisotropy of density structures, and we show their dependence on Mach numbers. In \S\ref{sec:self-similarity} we perform tests of the property of self-similarity in the case of density structures. In \S\ref{sec:intermittency} we present the intermittency of density and the logarithm of density and compare the intermittency of density and column density. In \S\ref{sec:moments} we analyze the statistical moments, skewness and kurtosis, as possible measures of the ${\cal M}_s$ of turbulence. In \S\ref{sec:considerations} we study the effects of cloud boundaries, resolution and time averaging. In \S\ref{sec:topology} we analyze density structure using genus statistics. We compare the genus lines to those of projected density. In \S\ref{sec:discussion} we discuss our results and their relation to the previous studies. We draw our conclusions in \S\ref{sec:summary}.

\section{Numerical simulations}
\label{sec:models}

We use a third-order-accurate hybrid essentially non oscillatory (ENO) scheme \citep[see][CL02]{cho02a} to solve the ideal MHD equations in a periodic box,
\begin{eqnarray}
 \frac{\partial \rho}{\partial t} + \nabla \cdot (\rho {\bf v}) = 0, \\
 \frac{\partial \rho {\bf v}}{\partial t} + \nabla \cdot \left[ \rho {\bf v} {\bf v} + \left( p + \frac{B^2}{8 \pi} \right) {\bf I} - \frac{1}{4 \pi}{\bf B}{\bf B} \right] = {\bf f},  \\
 \frac{\partial {\bf B}}{\partial t} - \nabla \times ({\bf v} \times{\bf B}) = 0,
\end{eqnarray}
with zero-divergence condition $\nabla \cdot {\bf B} = 0$, and an isothermal equation of state $p = c_s^2 \rho$, where $\rho$ is density, ${\bf v}$ is velocity, ${\bf B}$ is magnetic field, $p$ is the gas pressure, and $c_s$ is the isothermal speed of sound. On the right-hand side, the source term $\bf{f}$ is a random large-scale driving force. The rms velocity $\delta V$ is maintained to be approximately unity, so that ${\bf v}$ can be viewed as the velocity measured in units of the rms velocity of the system and ${\bf B}/\left( 4 \pi \rho \right)^{1/2}$ as the Alfv\'{e}n velocity in the same units. The time $t$ is in units of the large eddy turnover time ($\sim L/\delta V$) and the length in units of $L$, the scale of the energy injection. The magnetic field consists of the uniform background field and a fluctuating field: ${\bf B}= {\bf B}_\mathrm{ext} + {\bf b}$. Initially ${\bf b}=0$.

We drive turbulence solenoidally in Fourier space to minimize the influence of the forcing on the generation of density structures. Density structures in turbulence can be associated with the slow and fast modes \cite[see][BLC05]{lithwick01,cho03,beresnyak05}. We use units in which $V_A=B_\mathrm{ext}/\left( 4 \pi \rho \right)^{1/2}=1$ and $\rho_0=1$. The values of $B_\mathrm{ext}$ have been chosen to be similar to those observed in the ISM turbulence. The average rms velocity $\delta V$ in a statistically stationary state is around $1$.

For our calculations, similar to our earlier studies (BLC05), we assume that $B_\mathrm{ext}/ \left( 4 \pi \rho \right)^{1/2} \sim \delta B/ \left( 4 \pi \rho \right)^{1/2} \sim \delta V$. In this case, the sound speed is the controlling parameter, and basically two regimes can exist: supersonic and subsonic. Note that within our model, supersonic means low $\beta$, i.e. the magnetic pressure dominates, and subsonic means high $\beta$, i.e. the gas pressure dominates.

We present 3D numerical experiments of compressible (MHD) turbulence for a broad range of Mach numbers ($0.2\le{\cal M}_s\le7.1$ and ${\cal M}_A\sim 0.7$ or $\sim7.3$; see Table \ref{tab:models}). The model name contains two letters: ``P'' and ``B'' followed by a number. The letters B and P mean the external magnetic field and the initial gas pressure, respectively, and the numbers designate the value of the corresponding quantity. For example, a name ``B.1P.01'' points to an experiment with $B_\mathrm{ext}=0.1$ and $P=0.01$. We understand the Mach number to be defined as the mean value of the ratio of the absolute value of the local velocity $V$ to the local value of the characteristic speed $c_s$ or $V_A$ (for the sonic and Alfv\'{e}nic Mach number, respectively). We analyzed models with three different resolutions: 128$^3$, 256$^3$, and 512$^3$ on a Cartesian grid with periodic boundary conditions. We drove the turbulence solenoidally at wave scale $k$ equal to about 2.5 (2.5 times smaller than the size of the box). This scale defines the injection scale in our models. We do not set the viscosity and diffusion explicitly in our models. The scale at which the dissipation starts to act is defined by the numerical diffusivity of the scheme. The ENO-type schemes are considered to be relatively low diffusion ones \cite[see][e.g.]{liu98,levy99}. The numerical diffusion depends not only on the adopted numerical scheme but also on the ``smoothness'' of the solution, so it changes locally in the system. In addition, it is also a time-varying quantity. All these problems make its estimation very difficult and incomparable between different applications. However, the dissipation scales can be estimated approximately from the velocity spectra. In the case of our models we estimated the dissipation scale $k_{\nu}$ at 15, 22, and 30 for the low, medium, and high resolution, respectively.

We divided our models into two groups corresponding to sub-Alfv\'{e}nic ($B_\mathrm{ext}=1.0$) and super-Alfv\'{e}nic ($B_\mathrm{ext}=0.1$) turbulence. For each group we computed several models with different values of pressure (see Table \ref{tab:models}).

\section{Probability Distribution Function}
\label{sec:pdfs}

Observations usually provide 2D maps of column density in the direction of the observer along the line of sight (LOS). Thus, it is very important in computational astrophysics to study the relation between the actual 3D properties of the simulated turbulence and the synthetic observations prepared on the basis of these simulations.

First, we compare the probability distribution functions (PDFs) of density and column density. PDFs give us information about the fraction of the total volume occupied by a given value of a measured quantity. For the case of compressible turbulence in which evolution is described by the Navier-Stokes equation and the isothermal equation of state, the PDF of density obeys a lognormal form \citep{passot98}. If we select a point in space and assume that the density at this point results from subsequent events perturbing the previous density, then the final density is a product $\rho_0\Pi_i(1+\delta_i)$, where $\rho_0$ is the initial density at the selected point and $\delta_i$ is a small compression/rarefaction factor. By the power of the Central Limit Theorem \citep{kallenberg97}, the logarithm of the resulting density, $\log (\rho/\rho_0)=\sum_i \log (1+\delta_i)$, should obey a Gaussian distribution.

The process of perturbation of the column density $\Sigma$ is similar to the perturbations of density described above, but here a perturbation occurs only in the fraction of the LOS smaller than unity; thus, each individual event induces a perturbation of column density with a smaller relative amplitude. This implies a smaller dispersion of the column density distribution \cite[see][]{ostriker03}.

\cite{vazquez01} proposed the parameter $\eta$ for determining the form of the column density PDF in molecular clouds, which is the ratio of the cloud size to the decorrelation length of the density field. They defined the decorrelation length as the lag at which the density autocorrelation function (ACF) has decayed to its 10\% level. If the density perturbation events are uncorrelated for $\eta \gtrsim 1$, large values of the ratio $\eta$ imply that the Central Limit Theorem can be applied to those events.

The decorrelation length estimated from the ACF of the density for our medium-resolution models ranges from about 20 cell sizes of the computational mesh for supersonic models to about 50 cell sizes for subsonic models, which corresponds to $5 \lesssim \eta \lesssim 13$ if we take the size of the computational box as the column length (in the case of medium resolution, it is equal to 256 cell sizes). Found values of $\eta$ signify that in our models at least partial convergence to a Gaussian PDF should occur.

\begin{figure}  
 \epsscale{1.0}
 \plottwo{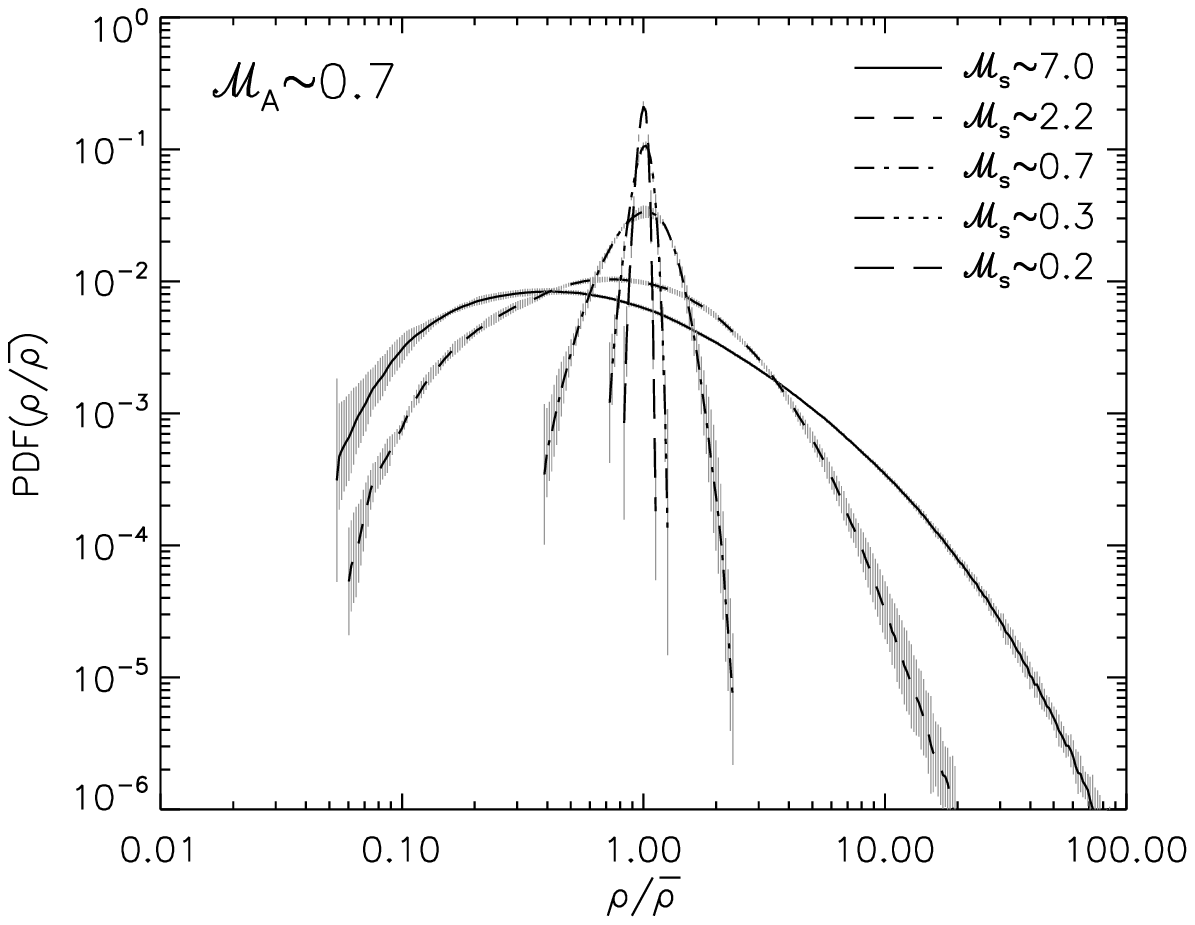}{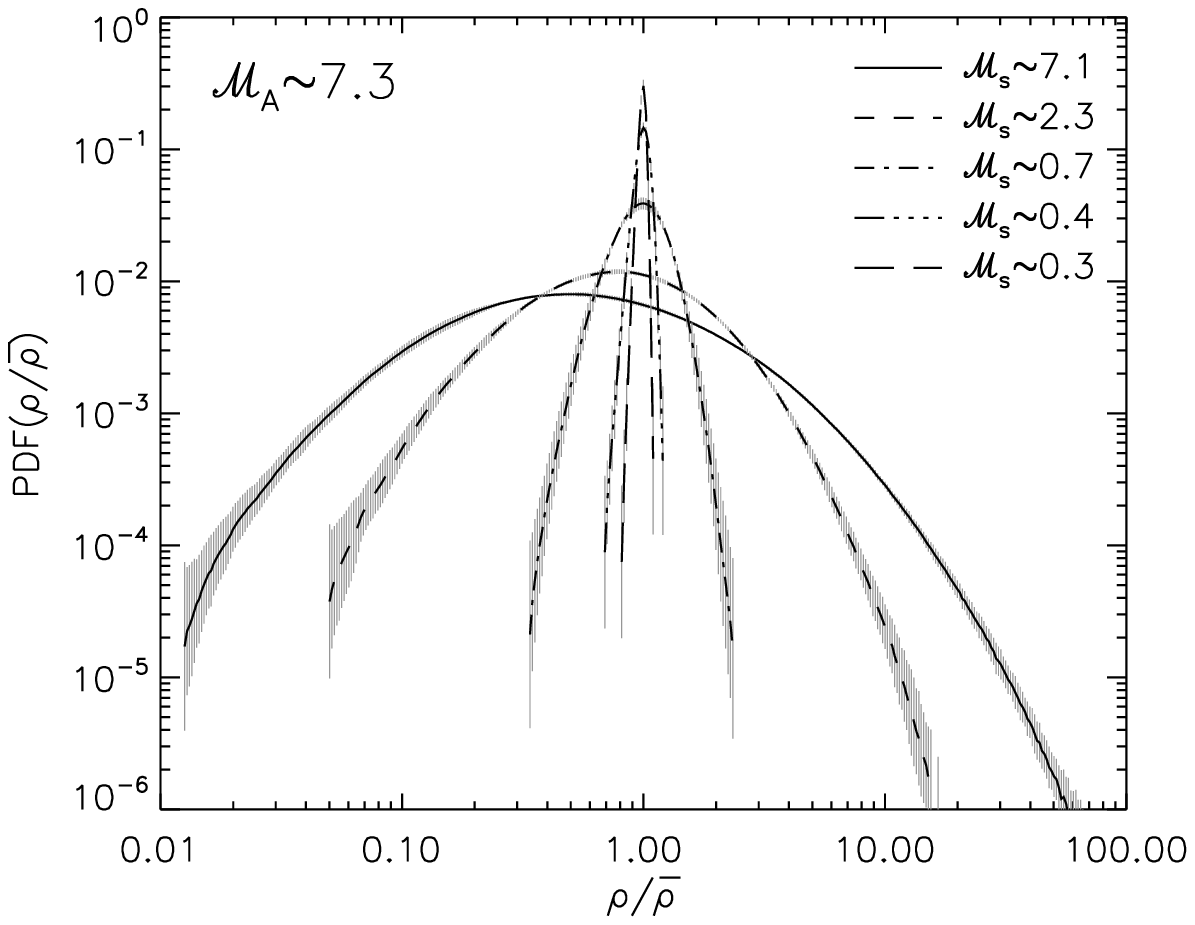}
 \plottwo{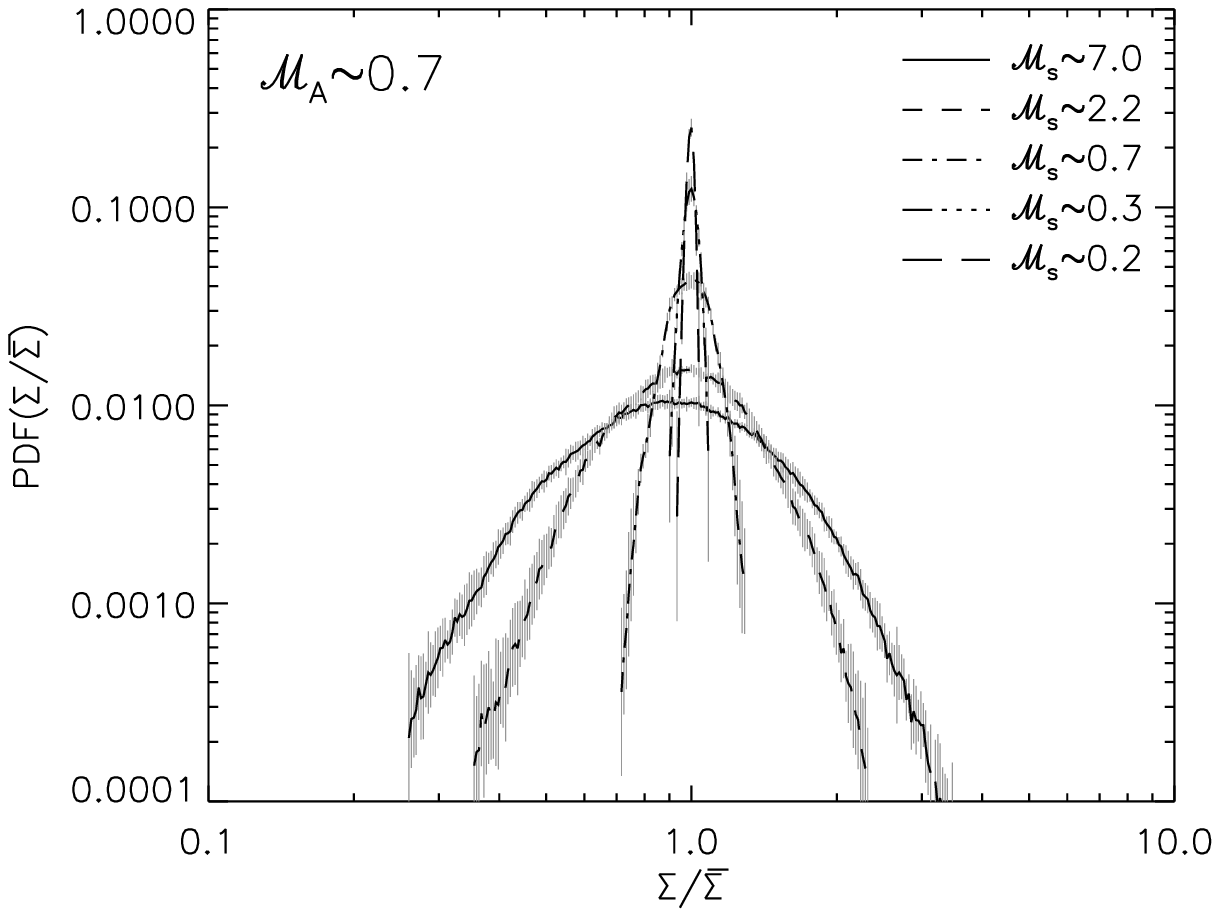}{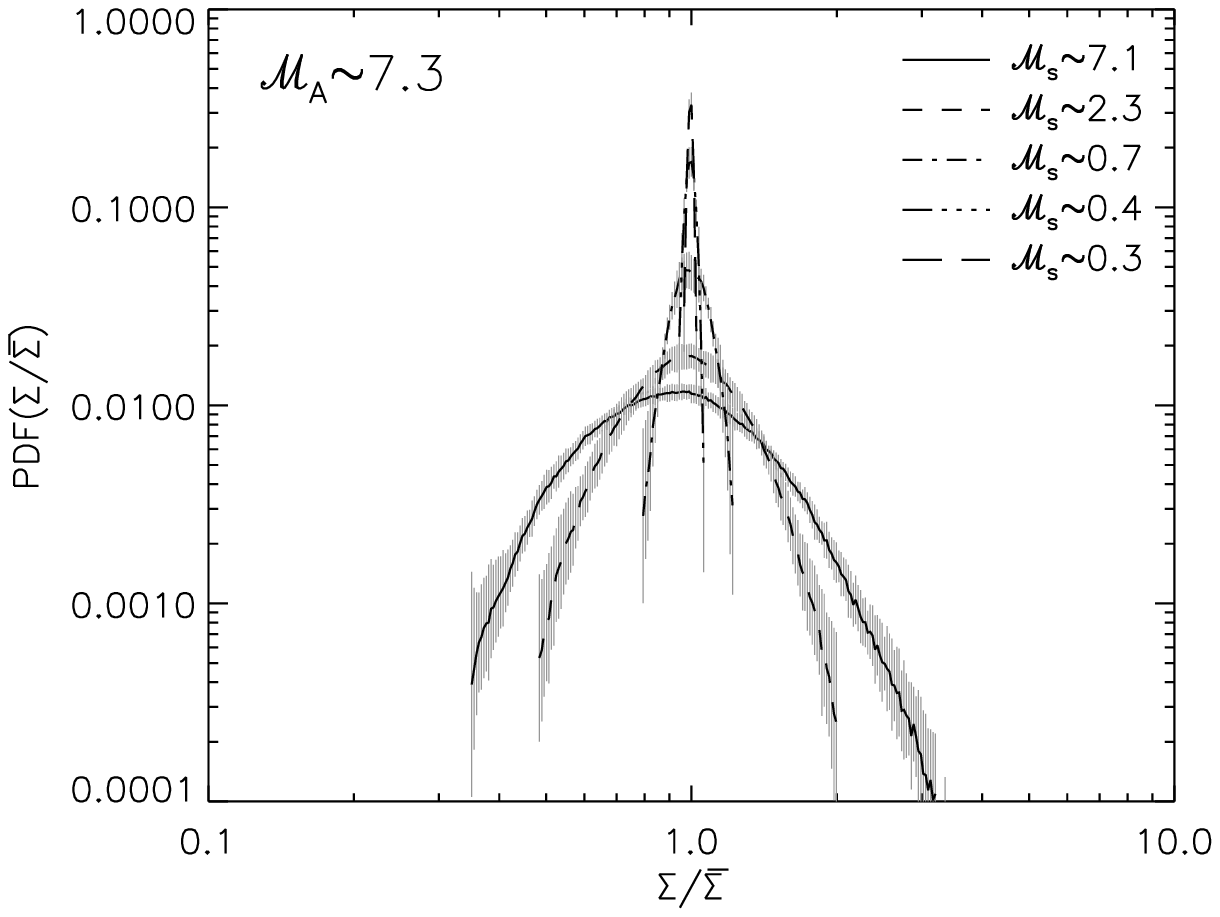}
 \caption{Normalized probability distribution functions (PDFs) of density $\rho/\bar{\rho}$ ({\em top row}) and column density $\Sigma/\bar{\Sigma}$ integrated along the line of sight parallel to $B_\mathrm{ext}$ ({\em bottom row}) for models with different values of ${\cal M}_s$ and for ${\cal M}_A\sim0.7$ ({\em left column}) and ${\cal M}_A\sim7.3$ ({\em right column}). Grey error bars signify the standard deviation of the PDFs, showing the strength of the departure at a single time snapshot from its mean profile averaged over time from $t\ge5$ to the last available snapshot. In this plot we include all models listed in Table \ref{tab:models} computed with medium resolution (256$^3$). \label{fig:dens_pdfs}}
\end{figure}

In the top row of Figure \ref{fig:dens_pdfs} we show PDFs of density normalized by its mean value for all models with medium resolution (256$^3$). The plot on the left shows results obtained from sub-Alfv\'{e}nic experiments (${\cal M}_A\sim0.7$), and the plot on the right results from super-Alfv\'{e}nic models (${\cal M}_A\sim7.3$). The PDF is additionally divided by the total number of mesh cells, which allows us to compare models with different resolutions. In the bottom row of Figure \ref{fig:dens_pdfs} we also see PDFs for column densities for the same models. The widths of PDF lines for $\log (\Sigma/\bar{\Sigma})$ are smaller then the widths of the corresponding lines for $\log (\rho/\bar{\rho})$. This confirms that a contribution from density perturbations to the perturbations of column density comes only from a fraction of the distance along the LOS giving smaller variations of the resulting column density.

Our plots confirm the strong dependence of PDFs of density on the sonic Mach number, an already known and well-studied property of density fluctuations in compressible turbulence \citep[see][and references therein]{vazquez01,ostriker03}. For most of the models PDFs are lognormal functions. For super-Alfv\'{e}nic turbulence PDFs are very symmetric about a vertical line crossing their maxima. However, for models with a strong external magnetic field (sub-Alfv\'{e}nic turbulence, ${\cal M}_A\sim0.7$) and very low pressure (supersonic turbulence, ${\cal M}_s\gtrsim1.0$), the shape of the density PDFs is significantly deformed, and its lower value arm ends in higher densities than in the case with a weak magnetic field (compare models for ${\cal M}_s\sim7$ drawn with solid lines in the left and right plots of Figure \ref{fig:dens_pdfs}). This supports a hypothesis that the gauge symmetry for $\log \rho$ that exists in Navier-Stokes equations is broken in MHD equations because of the magnetic tension term (BLC05), which physically manifests itself by preventing the formation of highly underdense regions. In the higher density part of the distribution we do not see a similar effect, because the highly dense structures are created mainly due to shocks, and in this case, the magnetic tension is too weak to prevent the condensation.

We show the PDFs of column density normalized to the mean value of $\Sigma$ and divided by the number of pixels in the 2D map of density projected along the direction of $B_\mathrm{ext}$. We see that the maximum and dispersion of PDFs still strongly depend on the sonic Mach number. Potentially this means that direct measurements of one-point statistics taken from the maps of molecular clouds can give us information about the speed of sound and the regime of turbulence taking place in these objects. See, however, discussion of the boundary effect in \S\ref{sec:boundary}.

In Figure \ref{fig:dens_pdfs} we included the degree of variation of PDFs in time as grey error bars. We note that the departure of PDFs from their mean profiles is very small in the middle part around the mean value. The strongest time changes are observed in the low- and high-density tails, but the PDFs for different models are still separable.

\section{Spectra and anisotropy of density and the logarithm of density}
\label{sec:spectra}

The spectrum of density fluctuations is an important property of a compressible flow. In some cases, the spectrum of density can be derived analytically. In nearly incompressible motions with a relatively strong magnetic field, the spectrum of density scales similarly to the pressure as $\sim k^{-7/3}$ if we consider the polytropic equation of state $p=a\rho^{\gamma}$ \citep{biskamp03}. In weakly magnetized MHD turbulence, however, the density spectrum is predicted to scale as $\sim k^{-5/3}$ \citep{montgomery87}. In supersonic flows, these relations might not be valid anymore because of shocks accumulating matter into the local and highly dense structures. Due to the high contrast of density, the linear relation $\delta p = c_s^2 \delta \rho$ is no longer valid, and the spectrum of density cannot be related to pressure so straightforwardly. In addition, the strong asymmetry of density fluctuations suggests the need to analyze the logarithm of density instead of density itself.

\begin{figure}  
 \epsscale{.95}
 \plottwo{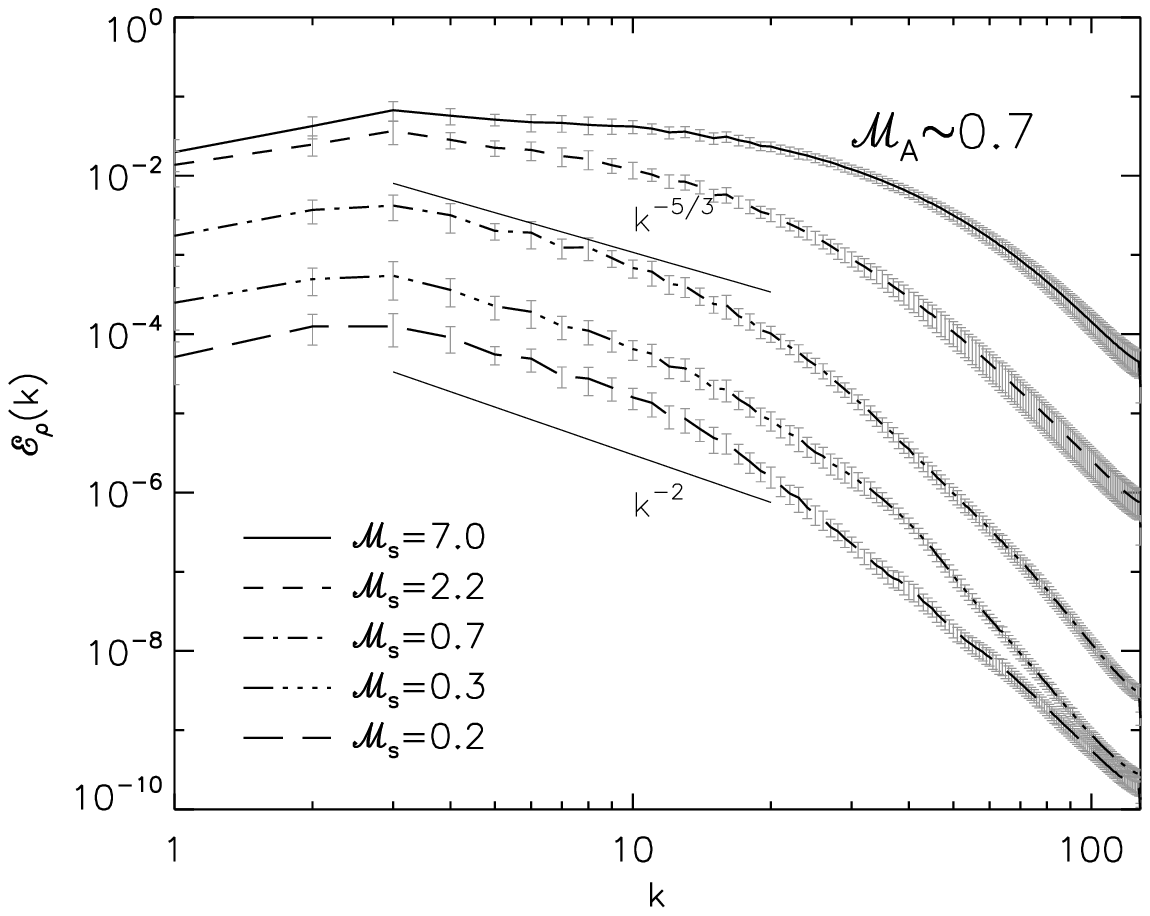}{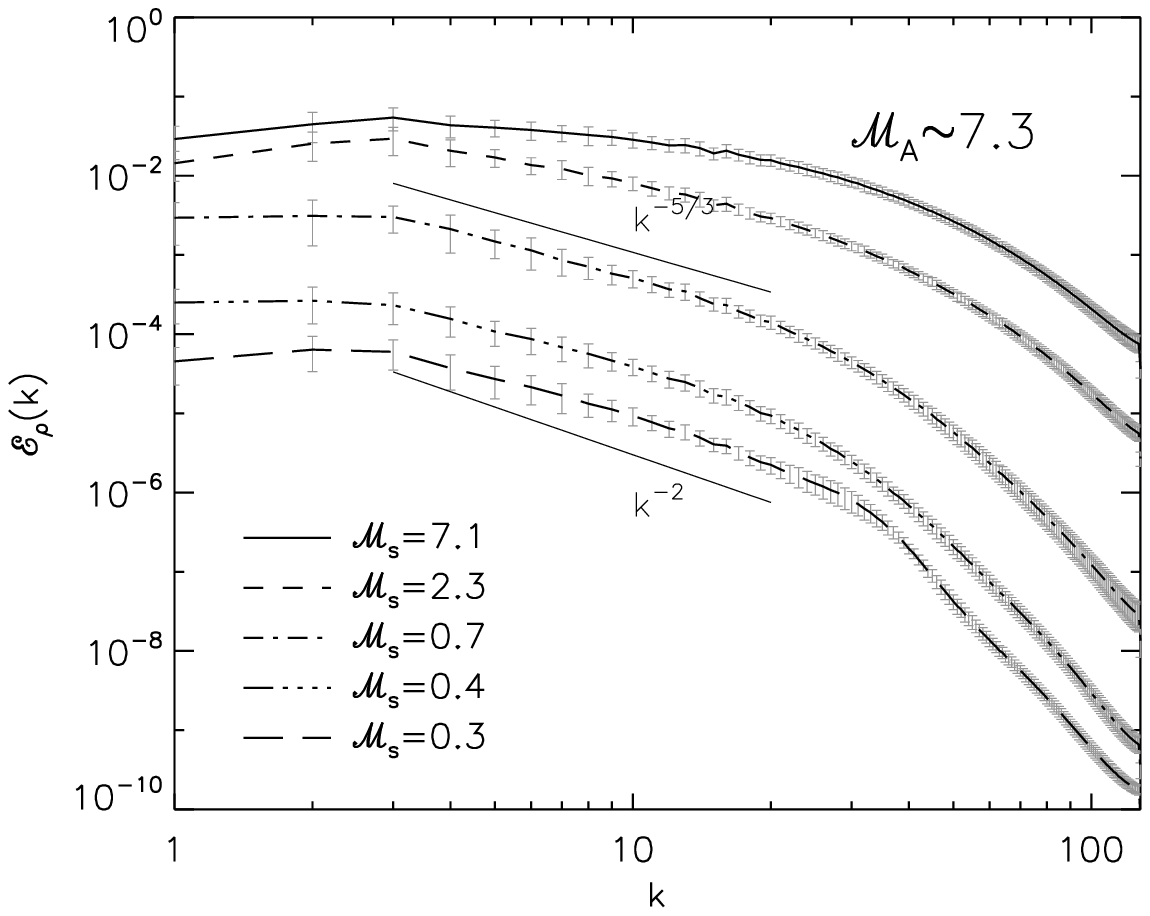}
 \plottwo{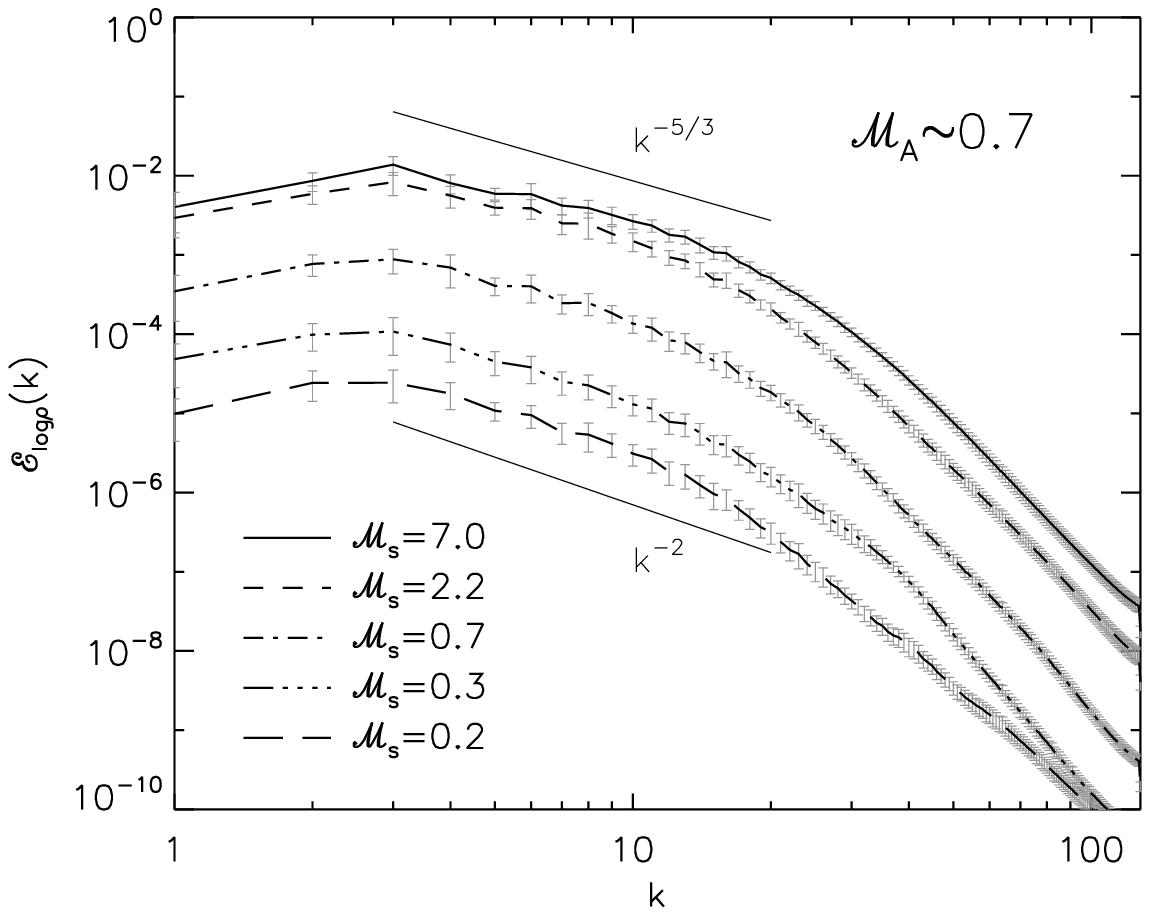}{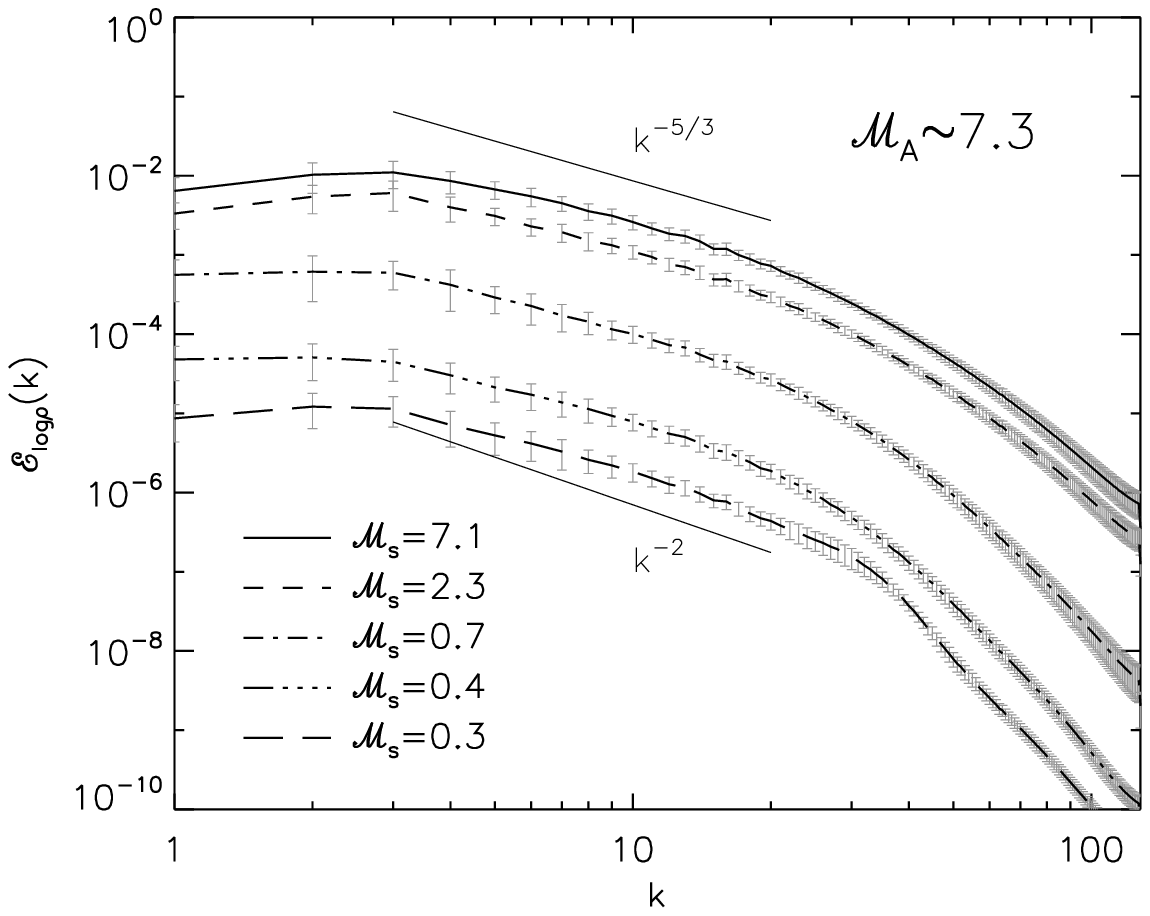}
 \plottwo{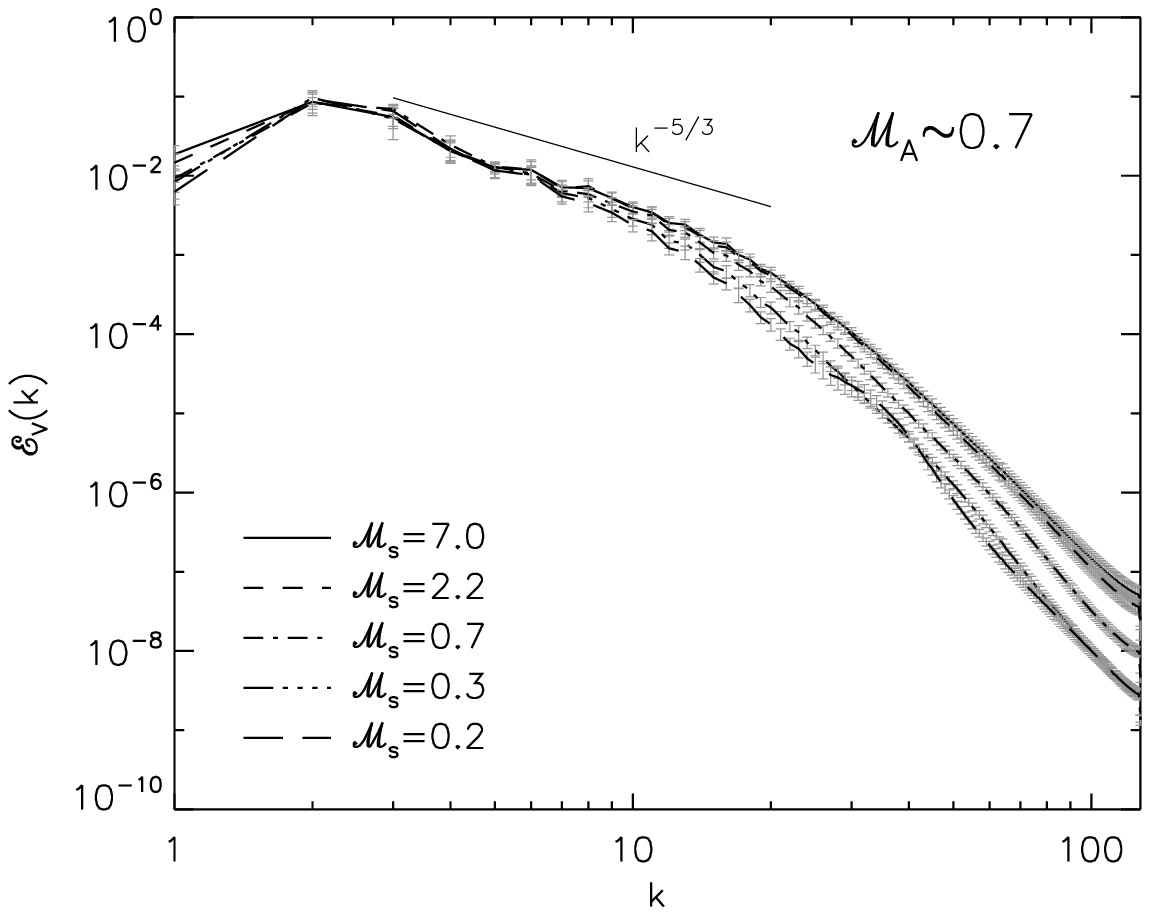}{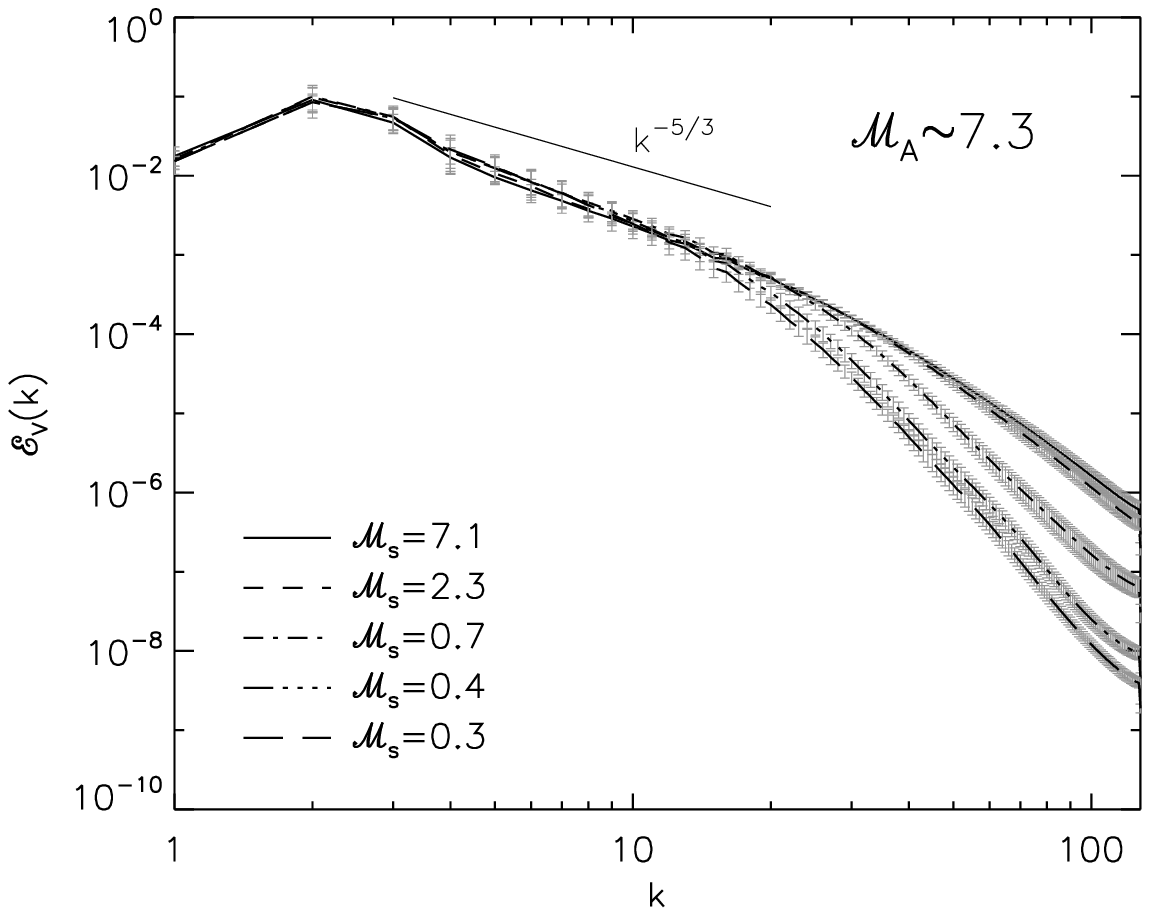}
 \caption{Spectra of density ({\em top row}), the logarithm of density ({\em middle row}) and velocity ({\em bottom row}) for experiments with different values of ${\cal M}_s$ and with ${\cal M}_A\sim0.7$ ({\em left column}) and ${\cal M}_A\sim7.3$ ({\em right column}) for models with medium resolution ($256^3$). Grey error bars signify the variance of spectra with time. The solid lines with slopes $-5/3$ and $-2$ cover the inertial range used to estimate the spectral indices of $\rho$ and $\log \rho$ spectra. \label{fig:spectra}}
\end{figure}

In Figure \ref{fig:spectra} we present the spectra of fluctuations of density and the logarithm of density ({\em top and middle rows, respectively}) for models with different ${\cal M}_s$. For comparison we also show the spectra of velocity fluctuations (Fig. \ref{fig:spectra}, {\em bottom row}). The two columns of plots in Figure \ref{fig:spectra} correspond to experiments with ${\cal M}_A\sim0.7$ ({\em left column}) and ${\cal M}_A\sim7.3$ ({\em right column}). As expected, we note a strong growth of the amplitude of density fluctuations with the sonic Mach number at all scales. This behavior is observed both in sub-Alfv\'{e}nic as well as in super-Alfv\'{e}nic turbulence (see Fig. \ref{fig:spectra}, {\em top left and top right, respectively}). Although the spectra significantly deviate from the power law, in Table \ref{tab:slopes} we calculate the spectral index of density $\alpha_\rho$ and the logarithm of density $\alpha_{\log \rho}$ within the inertial ranges estimated from the power spectrum of velocity. The width of the inertial range is shown by the range of solid lines with slopes $-5/3$ and $-2$ in all spectra plots. It is estimated to be within $k \in (3, 20)$. In Table \ref{tab:slopes} we also show the errors of estimation which combine the error of the fitting of the spectral index at each time snapshot and the standard deviation of variance of $\alpha_{\rho, \log \rho}$ in time. The slopes of the density spectra do not change significantly with ${\cal M}_s$ for subsonic experiments and correspond to analytical estimations (about $-2.2$, which is slightly less than $-7/3$, for turbulence with ${\cal M}_A\sim0.7$ and about $-1.7$, which is slightly more than $-5/3$, for weakly magnetized turbulence with ${\cal M}_A\sim7.3$). Such an agreement confirms the validity of the theoretical approximations. Nevertheless, these theoretical considerations do not cover the entire regime of compressible turbulence. While the fluid motions become supersonic, they strongly influence the density structure, making the small-scale structures more pronounced, which implies flattening of the spectra of density fluctuations (see values for ${\cal M}_s>1.0$ in Table \ref{tab:slopes}). Although, the spectral indices are clearly different for sub- and super-Alfv\'{e}nic turbulence, their errors are relatively large (up to $\pm0.3$). These errors have been calculated by summing the maximum value of the uncertainty of fits for individual spectra at each time snapshot and the standard deviation of time variation of spectral indices. The uncertainty of fit contributes the most to the total error bar (about 60\%-70\%). Nevertheless, there is still a small probability that both spectra, for sub- and super-Alfv\'{e}nic models, could overlap, resulting in less of a difference between these cases.

Note that in the case of a large contrast, a more proper quantity to analyze is the logarithm of density; we also show spectra of the logarithm of density fluctuations (see Fig. \ref{fig:spectra}, {\em middle row}). As we can see, the amplitude of fluctuations of $\log \rho$, like the case of $\rho$, grows with ${\cal M}_s$ at all scales. However, the logarithmic operation significantly ``filters'' the extreme values of the density,\footnote{As it is discussed in BLC05, the use of the logarithm of density is well justified from the point of view of the Navier-Stokes equations. Use of the logarithm of density provides better results than filtering with a threshold \cite[see][]{lazarian05}.} and they do not distort the spectra anymore. The slopes of spectra of $\log \rho$ are much less sensitive to the sonic Mach number, which is confirmed by small changes of the slope with the value of ${\cal M}_s$ (see respective columns in Table \ref{tab:slopes}). In addition, comparing sub- and super-Alfv\'{e}nic models in Table \ref{tab:slopes}, we see that, consistent with the theoretical considerations, the density gets a more shallow spectrum as the external magnetic field decreases (or ${\cal M}_A$ increases). For subsonic turbulence, the energy of the density fluctuations (and the energy of fluctuations of the logarithm of density) scales as $\sim k^{-7/3}$ in the case of strong $B_\mathrm{ext}$ and as $\sim k^{-5/3}$ in the case of weak $B_\mathrm{ext}$.

For comparison we also show the power spectra of velocity (Figure \ref{fig:spectra}, {\em bottom row}). While the spectra of density and the logarithm of density change both their amplitude and slope with the sonic Mach number, for velocity fluctuations we see almost the same-shaped spectra within the inertial range, insensitive to the strength of sonic motions. The only visible differences are present in the dissipation range, where the motions are strongly influenced by small-scale dissipation. This dissipation depends on the pressure of the ambient medium, and thus on ${\cal M}_s$, as this is the case for shocks.

\begin{figure}  
 \plottwo{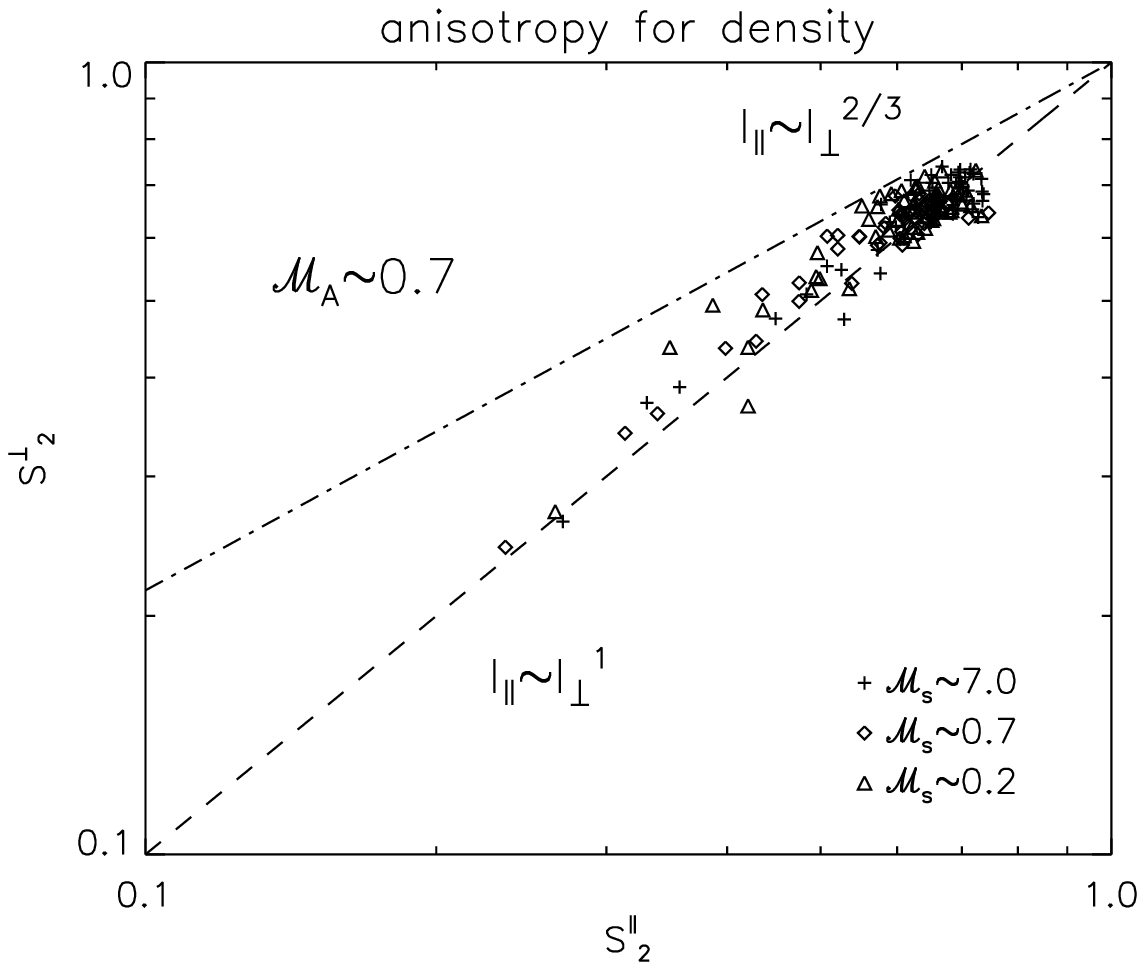}{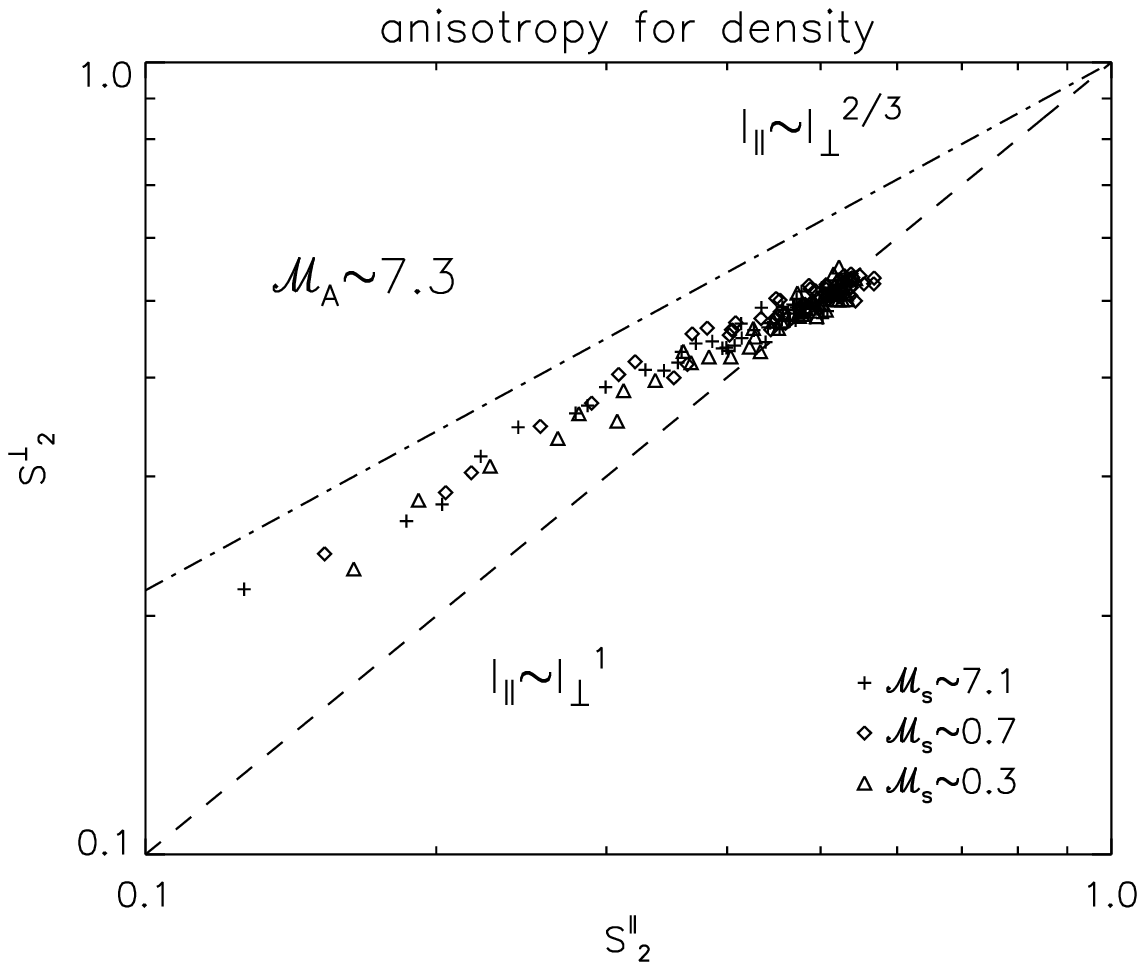}
 \plottwo{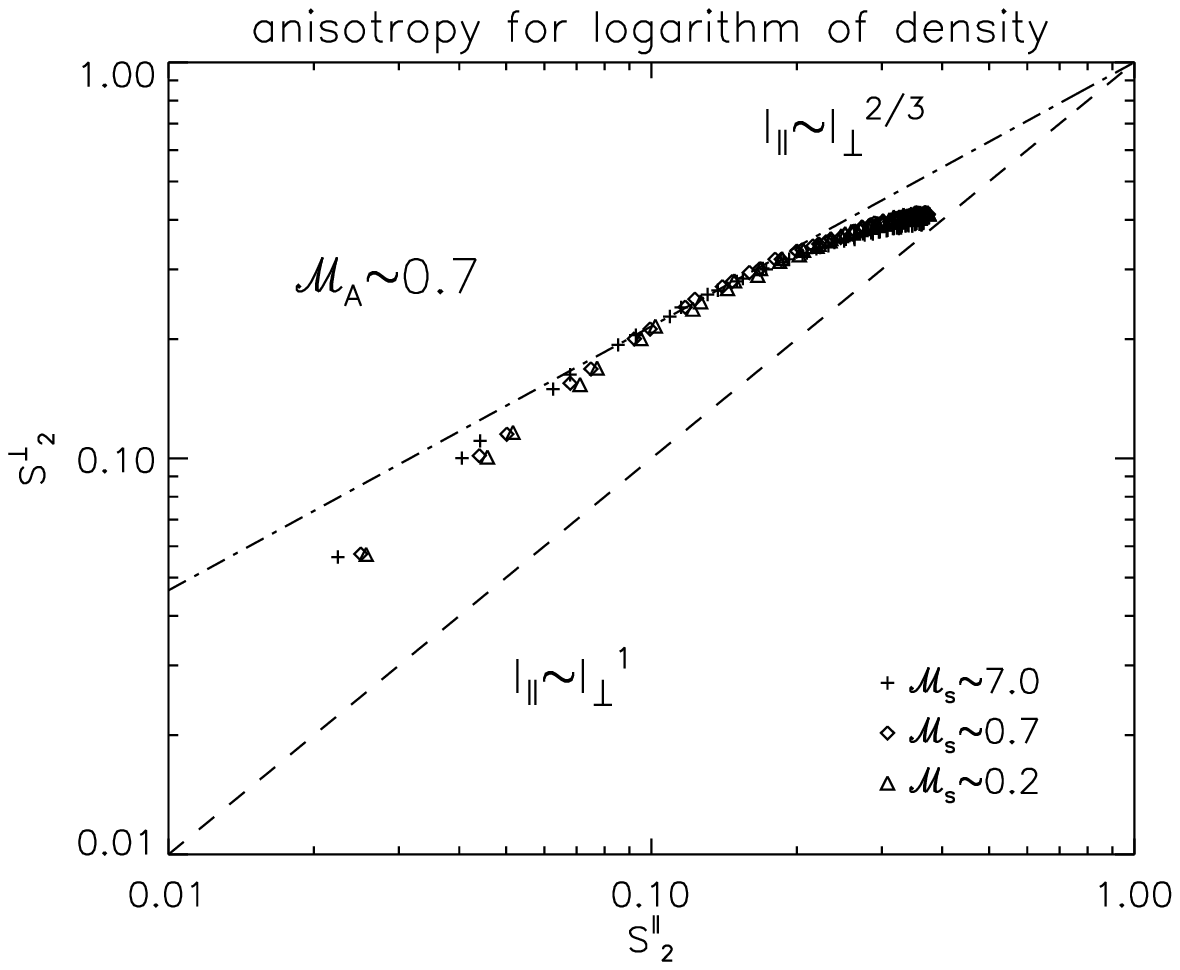}{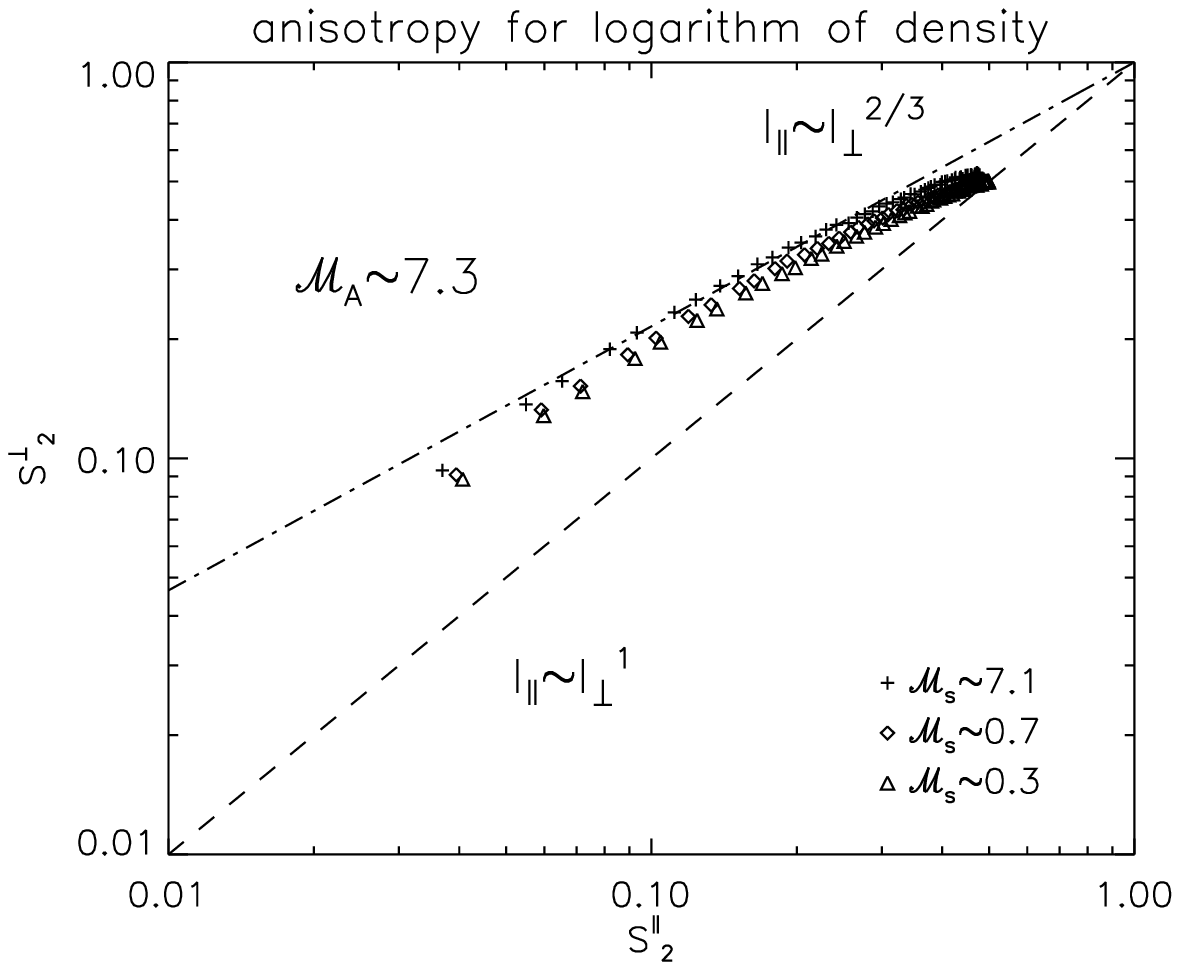}
 \plottwo{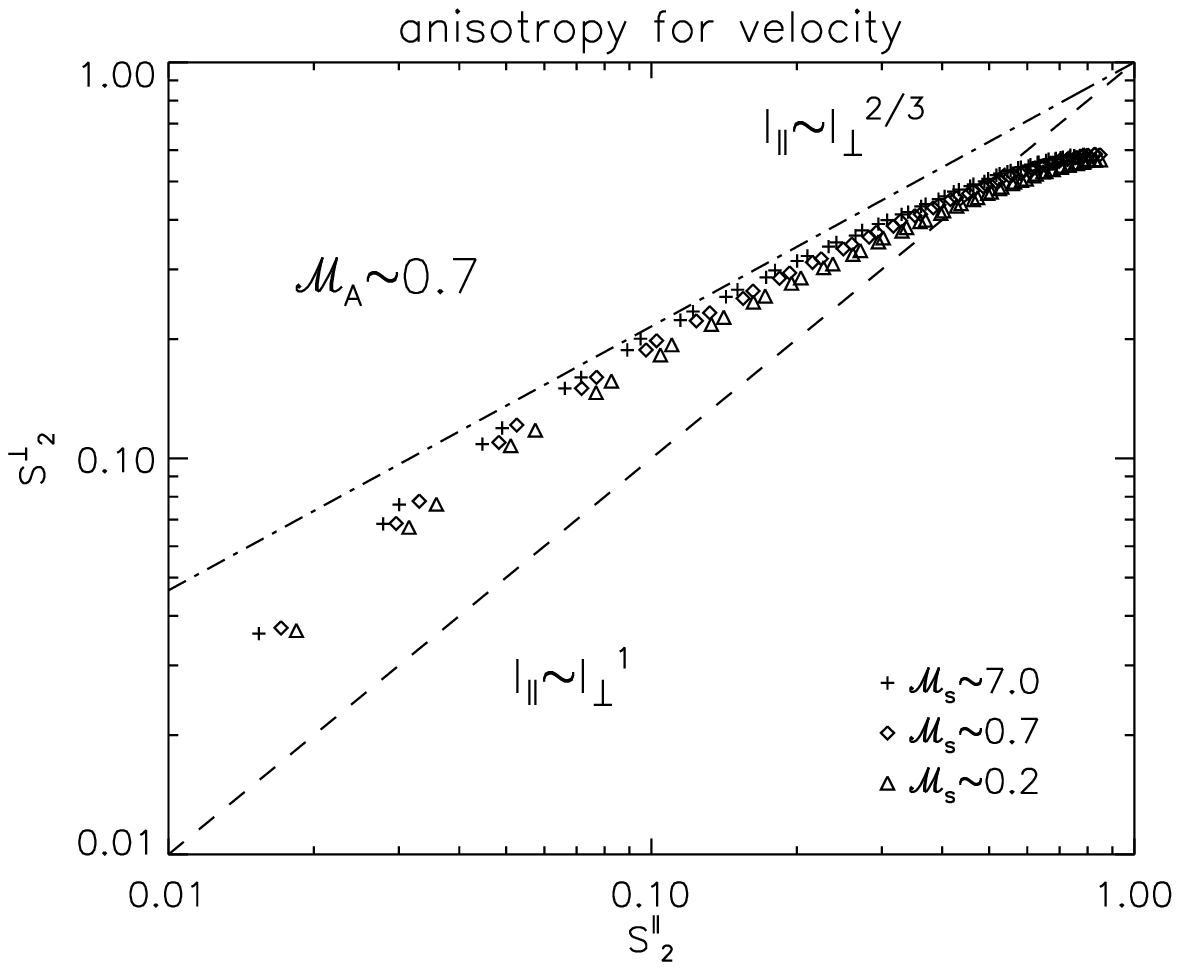}{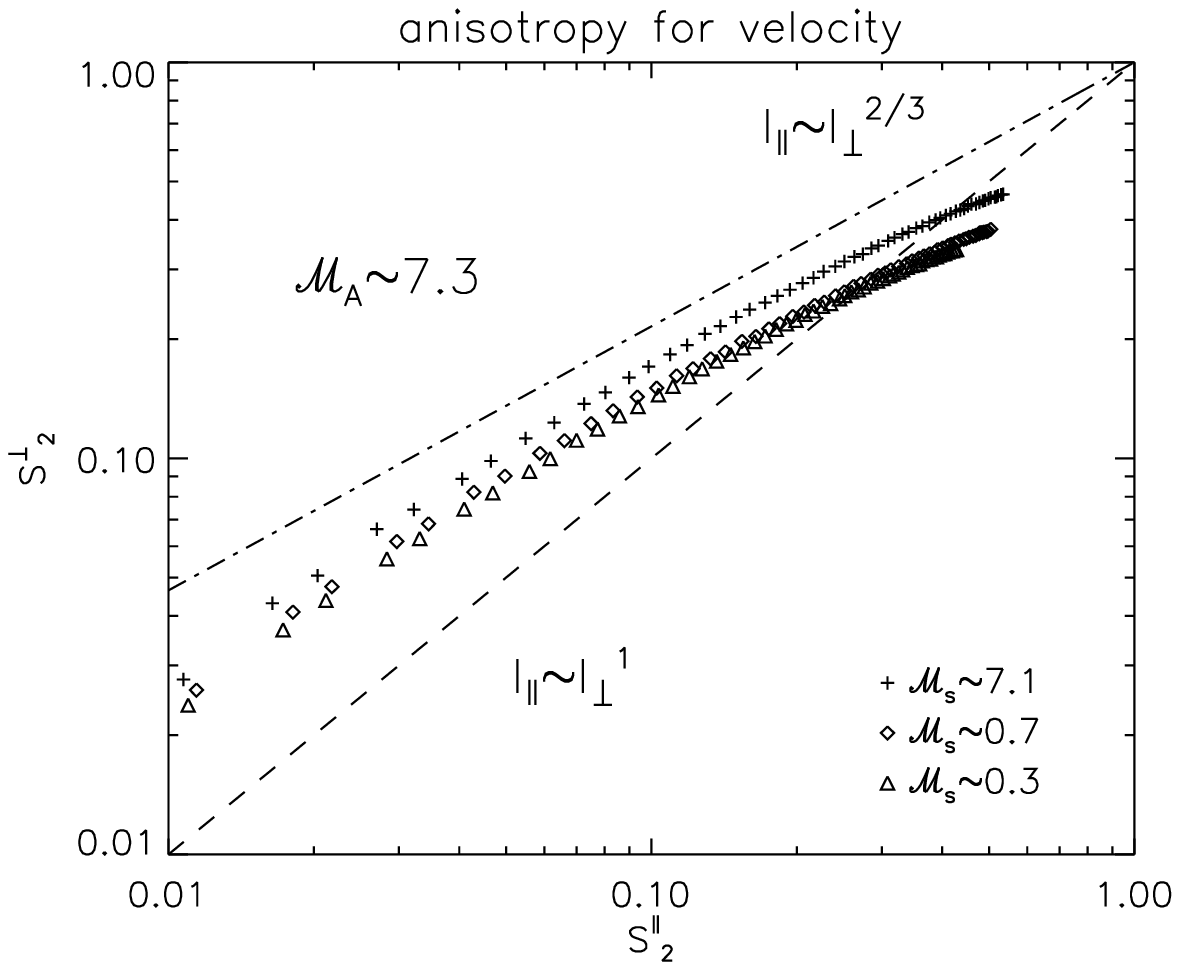}
 \caption{Anisotropy for the second-order structure function for density ({\em top row}), the logarithm of density ({\em middle row}), and the velocity field ({\em bottom row}) for experiments with different values of ${\cal M}_s$ and with ${\cal M}_A\sim0.7$ ({\em left column}) and ${\cal M}_A\sim7.3$ ({\em right column}) for models with resolution 256$^3$. \label{fig:anisotropy}}
\end{figure}

Another question is the anisotropy of density and the logarithm of density structures. In Figure \ref{fig:anisotropy} we show lines that mark the corresponding separation lengths for the second-order structure functions parallel and perpendicular to the local mean magnetic field.\footnote{The local mean magnetic field was computed using the procedure of smoothing by a 3D Gaussian profile with the width equal to the separation length. Because the volume of smoothing grows with the separation length $l$, the direction of the local mean magnetic field might change with $l$ at an arbitrary point. This is an extension of the procedures employed in \cite{cho00} and \cite{cho02b}.} In the case of sub-Alfv\'{e}nic turbulence, the degree of anisotropy for density is very difficult to estimate due to the high disperion of points. However, rough estimates suggest more isotropic density structures, because the points extend along the line $l_\parallel \sim l_\perp^1$. For models with ${\cal M}_A \sim 7.3$ the points in Figure~\ref{fig:anisotropy} have lower dispersion, and the anisotropy is more like the type from \citet[hereafter GS95]{goldreich95}, i.e. $l_\parallel \sim l_\perp^{2/3}$. In both Alfv\'{e}nic regimes, the anisotropy of density does not change significantly with ${\cal M}_s$. Plots for the logarithm of density show more smooth relations between parallel and perpendicular structure functions. The dispersion of points is very small. Moreover, we note the change of anisotropy with the scale. Lower values of structure functions correspond to lower values of the separation length (small-scale structures), so we might note that the logarithm of density structures are more isotropic than the GS95 model at small scales, but the anisotropy grows a bit larger than the GS95 prediction at larger scales. This difference is somewhat larger in the case of models with stronger external magnetic field (compare plots in the left and right columns of Figure \ref{fig:anisotropy}), which may signify their dependence on the strength of $B_\mathrm{ext}$. The anisotropy of $\log \rho$ structures is marginally dependent on the sonic Mach number, similar to the density structures. All these observations allow us to confirm the previous studies (see BLC05) that suggest that the anisotropy depends not only on the scale but on ${\cal M}_A$. In comparison to the discussed anisotropy of density and the logarithm of density, we also show the anisotropy of velocity (see Fig. \ref{fig:anisotropy}, {\em bottom row}). The anisotropy of velocity is independent of ${\cal M}_s$ too, but it slightly changes with ${\cal M}_A$ (compare the bottom left and right plots in Figure \ref{fig:anisotropy}), especially at small scales. For both the sub-Alfv\'{e}nic and super-Alfv\'{e}nic turbulence we see the correspondence of the anisotropy to GS95 model. From these observations we see that the anisotropy of the logarithm of density structures is very similar to the anisotropy of velocity.

\section{Extended self-similarity for density and the logarithm of density}
\label{sec:self-similarity}

The statistical distribution of velocity eddies of size $l$ is described by the moments of the velocity increments called structure functions
\begin{equation}
 S^{(p)}(l)=\langle | \delta v ({\bf l}) | ^p \rangle,
 \label{eqn:str_func}
\end{equation}
where $\delta v$ is a longitudinal increment
\begin{equation}
 \delta v_\parallel(l) = [{\bf v}({\bf x} + {\bf l}) - {\bf v}({\bf x})]\cdot\hat{l},
 \label{eqn:vel_incr}
\end{equation}
and $\langle \ldots \rangle$ denotes an ensemble averaging. Note that the structure function is defined here with the absolute value of the argument. This ensures the positivity of the structure functions also for the odd orders.

An interesting and not yet fully understood property of these functions is their scaling behavior. A structure function of an arbitrary order scales with the separation length $l$ within the intertial range as $S^{(p)}(l)\sim l^{\zeta(p)}$. From such a scaling relation, one can deduce a self-similarity between structure functions of two arbitrary orders. Taking two different orders $p$ and $p'$, we need only a few transformations to derive a relation \cite[see][]{biskamp03},
\begin{equation}
 S^{(p)}(l) \sim [S^{(p')}(l)]^{\zeta(p)/\zeta(p')}.
 \label{eqn:sf_relation}
\end{equation}
Although simple, this relation implies two important properties of the function $S^{(p)}(l)$. First, it shows how self-similarity connects the structure functions, and second, it lets us introduce a normalization using a structure function of an arbitrary order $p'$. In Kolmogorov phenomenology, the relation $S^{(p)}(l) \sim l^{\zeta(p)}$ has been derived analytically, implying $\zeta(p)=p/3$; thus, the third moment has been taken as the normalization. Let us note again that the above considerations concern only the part of the structure function within the inertial range.

\cite{benzi93} reported that for hydrodynamic turbulence, functions $S^{(p)}(S^{(3)})$ exhibit a much broader power-law range compared to $S^{(p)}(l)$. While for the inertial range a similarity in the scaling of two functions stems from the Kolmogorov scaling $S^{(3)}(l) \sim l$, the power-law scaling of $S^{(p)}(S^{(3)})$ protrudes well beyond the inertial, into the dissipation range.\footnote{In practical terms this means that instead of obtaining $S^{(p)}$ as a function of $l$, one gets $S^{(p)}$ as a function of $S^{(3)}$, which is nonlinear in a way that corrects for the distortions of $S^{(p)}$.} This observation shows that the dissipation ``spoils'' different orders of the structure function in the same manner. Therefore, there is no particular need to use the third moment, but one can use any other moment $S^{(p')}(l) \sim l^{\zeta(p')}$ and obtain a good power law of the form analogous to Eq. (\ref{eqn:sf_relation}) \cite[see][]{biskamp03}. Though not understood, the above property of $S^{(p)}$ allows insight into the scaling of realistically high Re flows with limited Re number simulations or laboratory experiments. In this way, the work of \cite{benzi93} allowed much progress in studies of the properties of self-similar flows.

All the above considerations originally concerned the properties of velocity and magnetic field structure. Relations between the fluctuations of density and the fluctuations of the other quantities like pressure, velocity and magnetic fields in the nonlinear regime of MHD turbulence are nontrivial to define; therefore, the scaling function and its self-similarity are still not well understood. Although, there is no clear analytical reason for a self-similarity to be valid for density structures, we check this property by calculating analogous structure functions
\begin{equation}
 S^{(p)}(l)=\langle | f({\bf r} + {\bf l}) - f({\bf r}) |^{p} \rangle ,
 \label{eqn:sf_scalar}
\end{equation}
where $f({\bf r})$ might be an arbitrary scalar function of position, in particular, density or the logarithm of density. Note that we do not assume any property of scaling so far.

We calculate structure functions for density and the logarithm of density according to Eq. (\ref{eqn:sf_scalar}). We point out that in general, for a normalization function one can take an arbitrary order of the structure function, not necessarily that resulting from the Kolmogorov law $S^{(3)}(l)=-\frac{4}{5}\epsilon l$ \cite[where the structure function is defined without the absolute value before averaging; see][]{biskamp03}. The latter is only valid in the case of incompressible hydrodynamic turbulence for the velocity field within the inertial range where the viscous term is negligible. The difference in definition of the structure function requires additional comment. The averaging in both definitions is taken over the ensemble. It means that if we get the negative value of the increment of the quantity $\delta f = f({\bf r} + {\bf l}) - f({\bf r})$, we can switch positions and take the increment $\delta f = f({\bf r}) - f({\bf r} + {\bf l})$ into the averaging. This operation is permitted, because the Kolmogorov law is valid only for the absolute values of $l$.

\begin{figure}  
 \plottwo{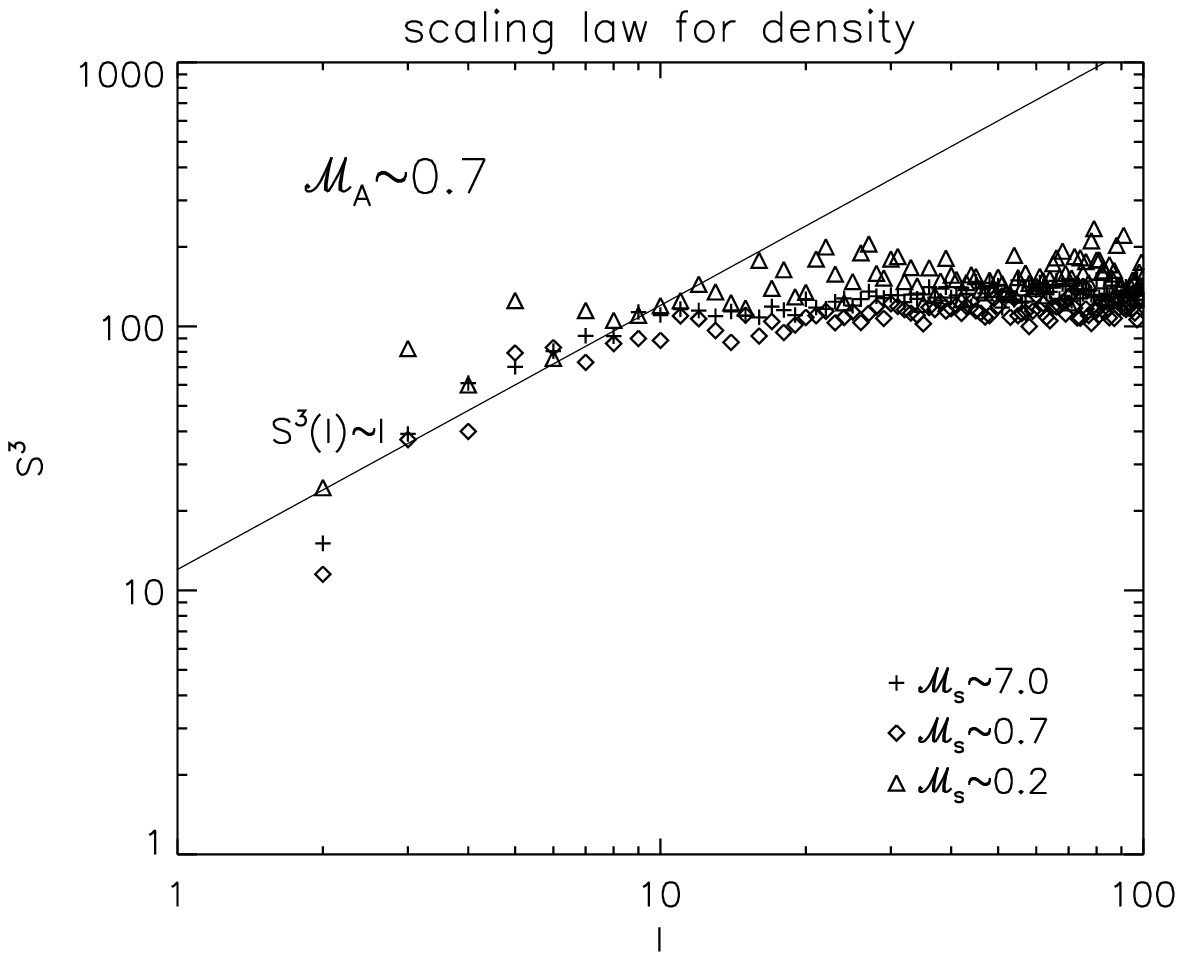}{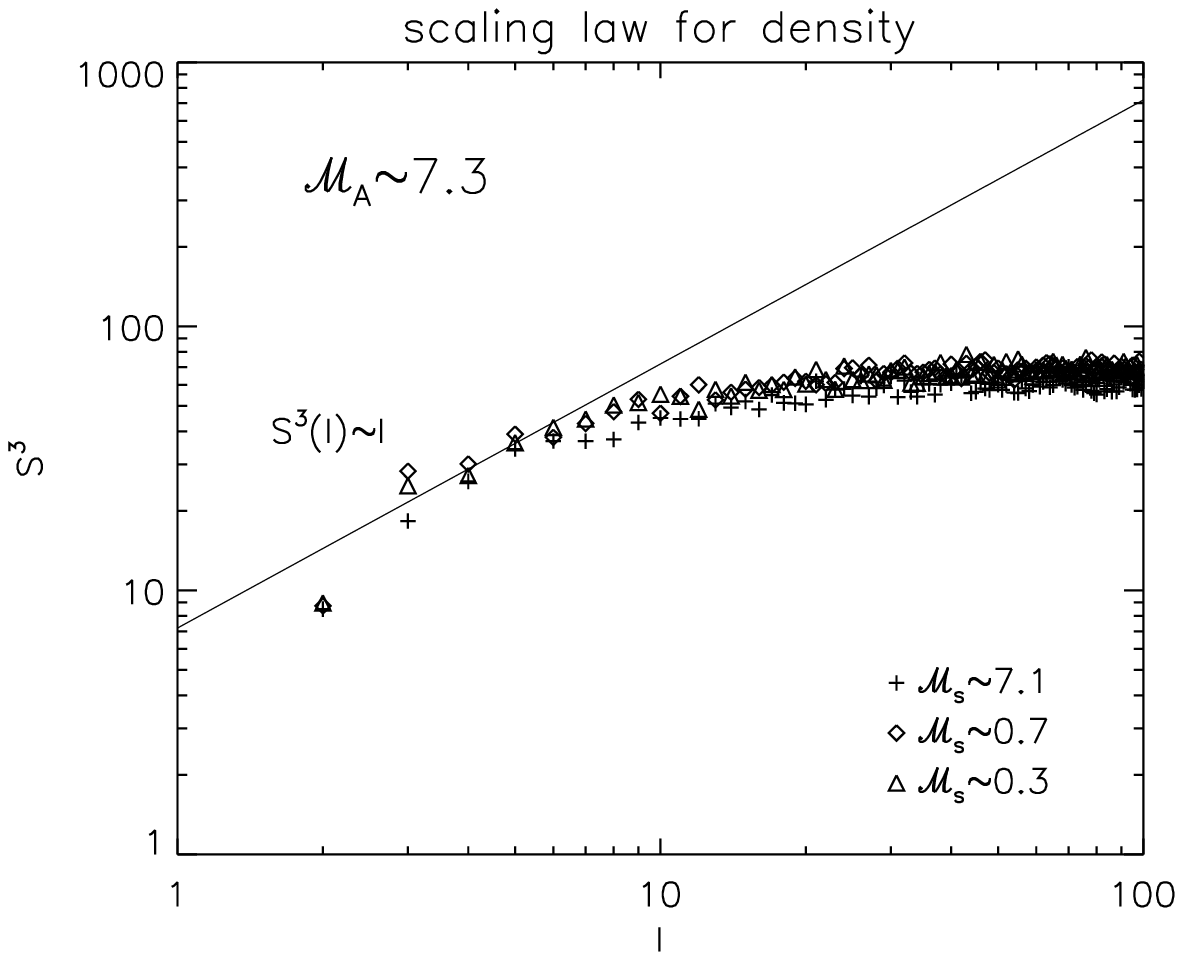}
 \plottwo{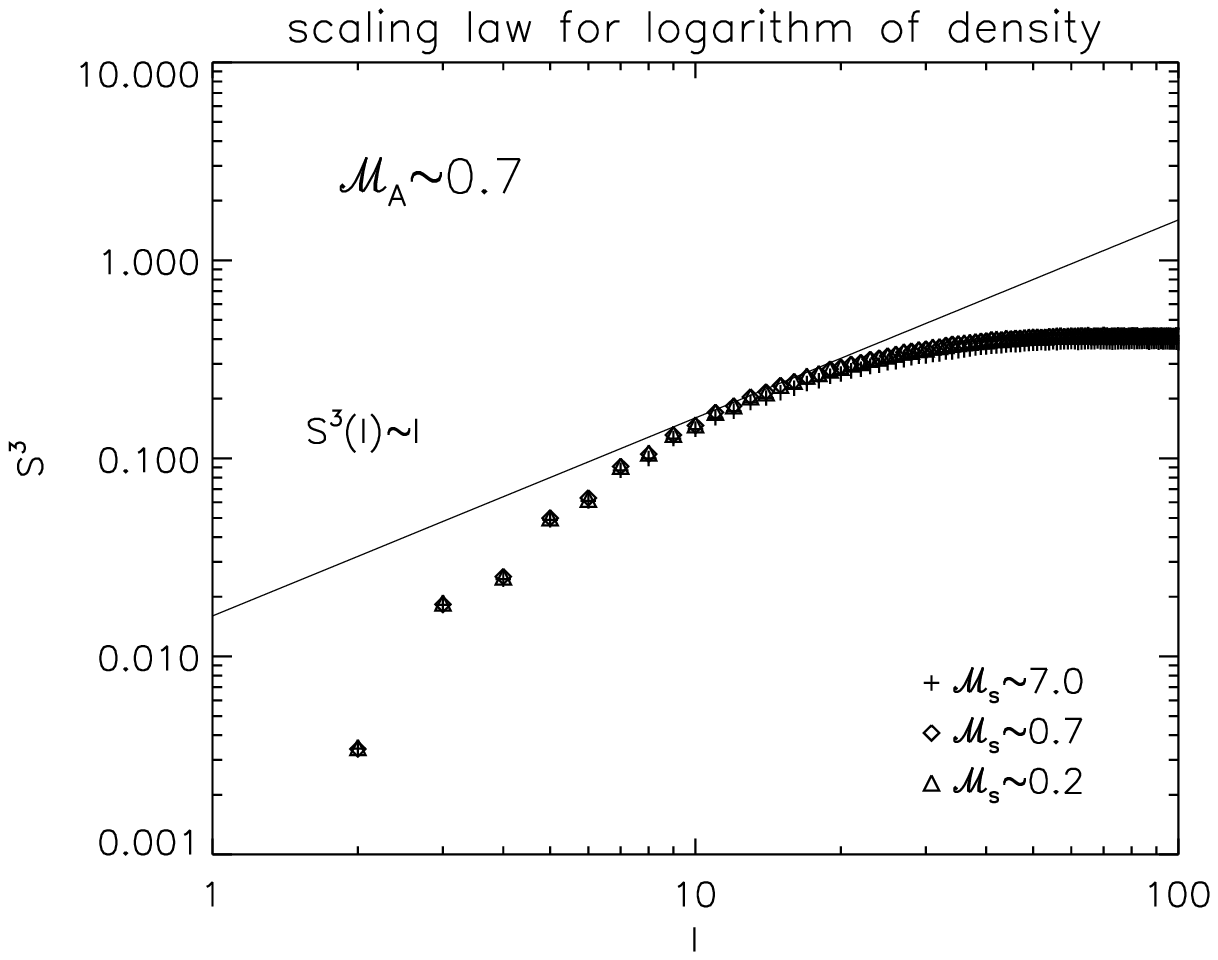}{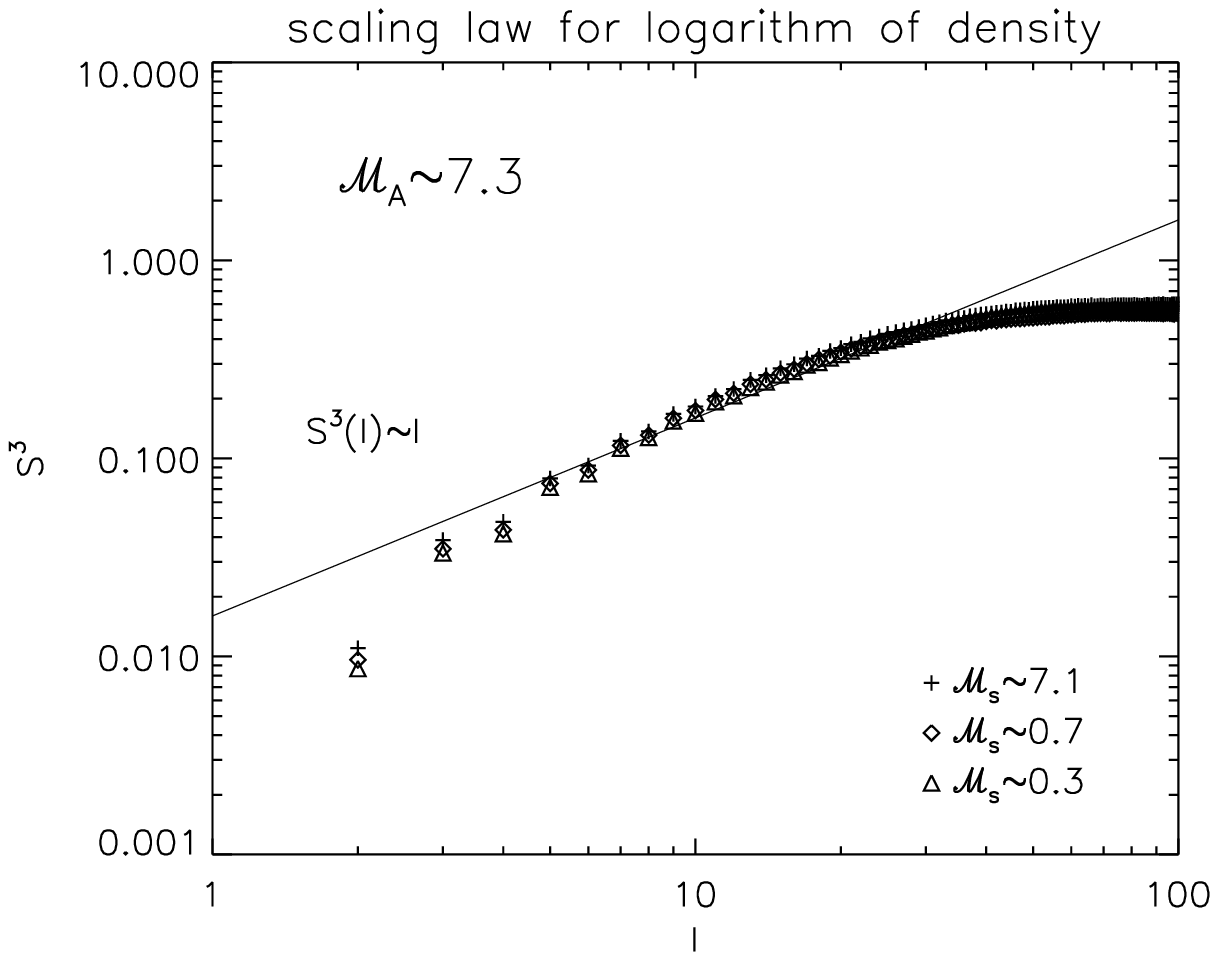}
 \plottwo{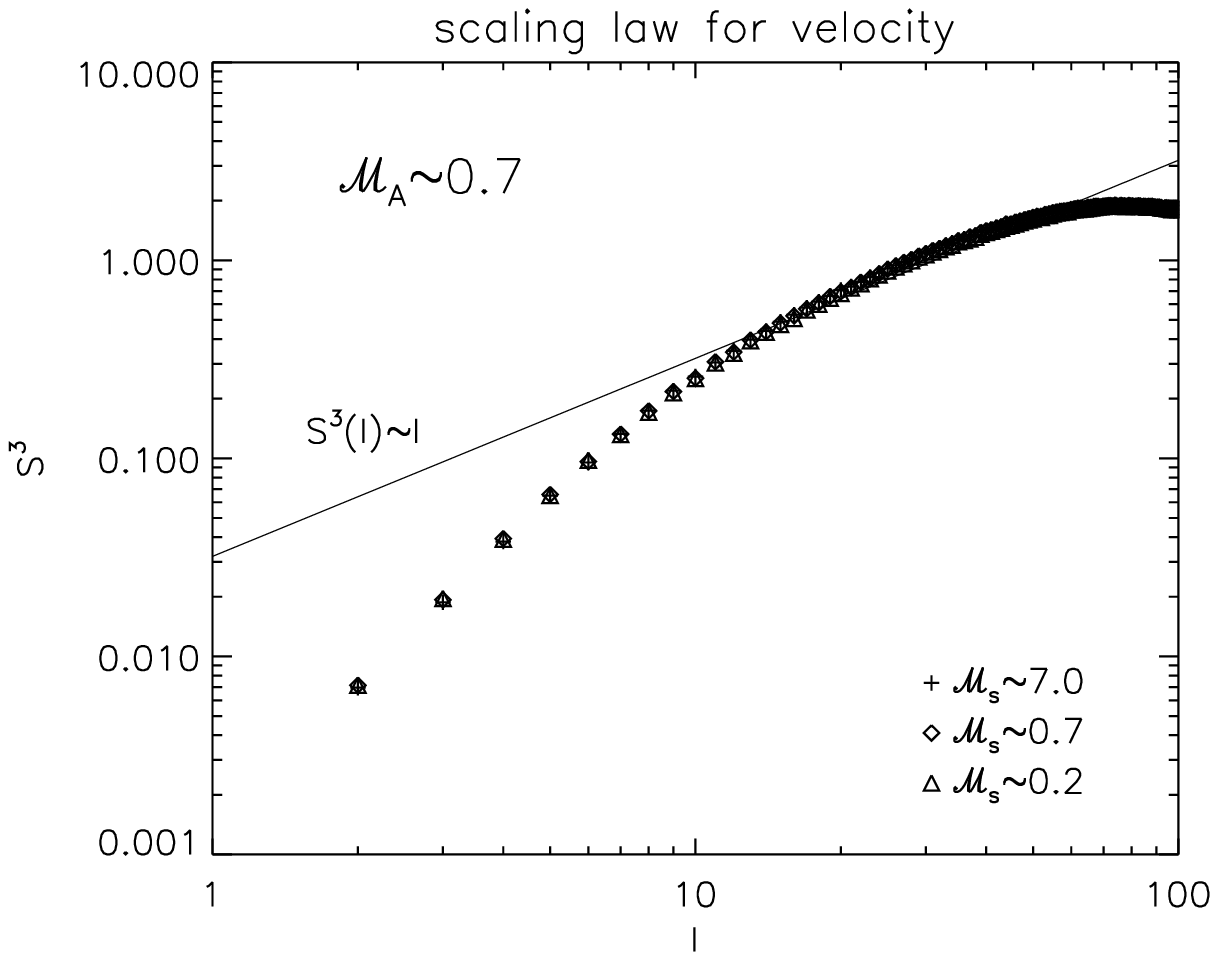}{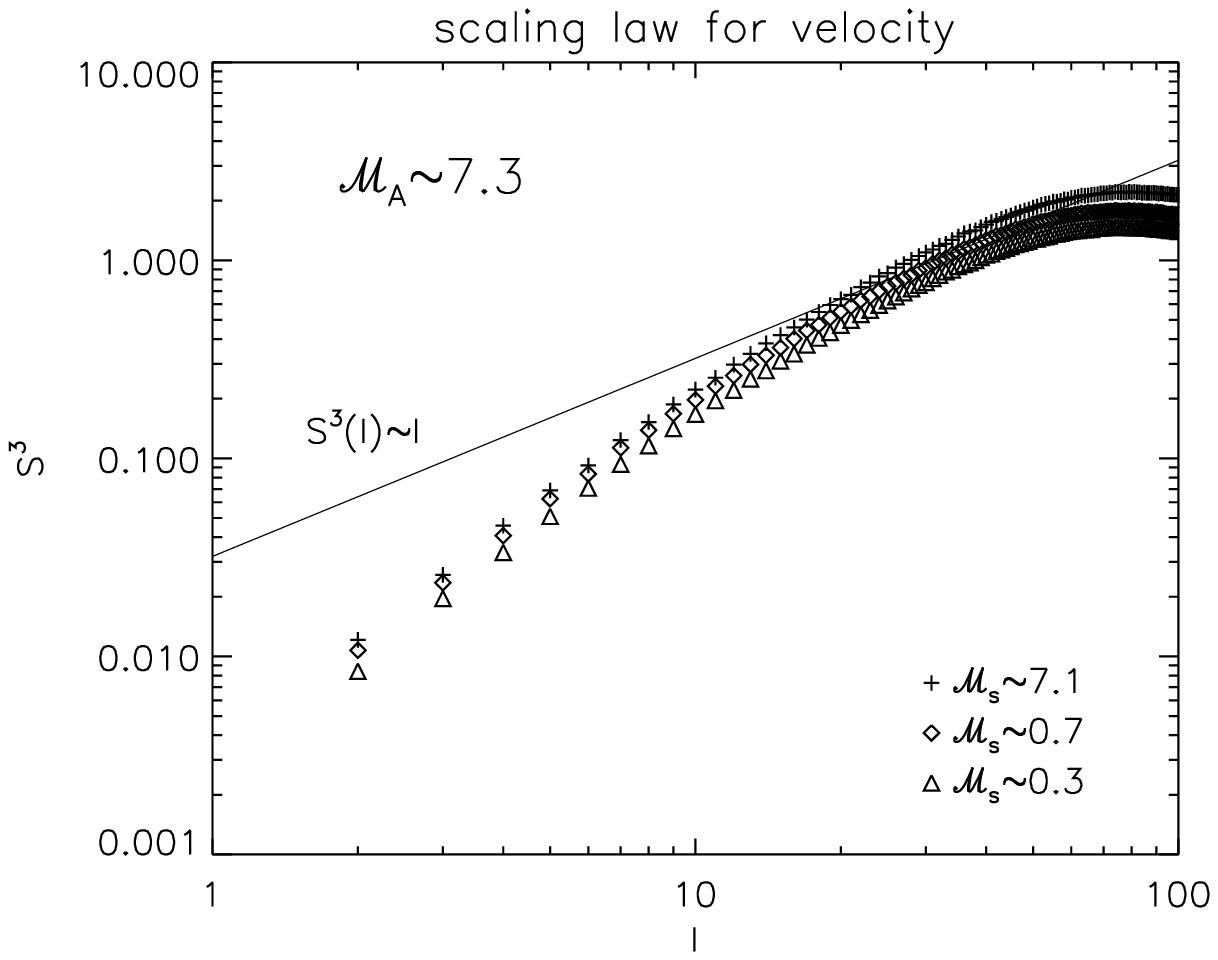}
 \caption{Scaling law for density, the logarithm of density, and velocity ({\em top, middle, and bottom rows, respectively}) for medium-resolution experiments with ${\cal M}_A\sim0.7$ ({\em left column}) and ${\cal M}_A\sim7.3$ ({\em right column}). \label{fig:scaling_law}}
\end{figure}

To test if a similar scaling law exists also for density and the logarithm of density in the magnetized compressible turbulence, we present Figure~\ref{fig:scaling_law}, which shows third-order structure functions for density, the logarithm of density, and velocity as functions of the separation length. The solid lines show the hypothetical Kolmogorov scaling law $S^{(3)}(l) = a l$, where $a$ is a constant. In the case of density, the models with a strong magnetic field show relatively large dispersion of points. In addition, the line flattens very quickly ($l \gtrsim 10$). In this situation, the region where the structure function scales linearly with the separation length is difficult to find. The situation is somewhat better in the case of the logarithm of density. Although the relations are far from linear, we can estimate the region of linear scaling ($8 \lesssim l \lesssim 20$ for sub-Alfv\'{e}nic models and $5 \lesssim l \lesssim 30$ for super-Alfv\'{e}nic models). In Figure \ref{fig:scaling_law} ({\em bottom row}) we show the third-order structure functions for velocity. We can clearly see the best linear scaling withing the range $15 \lesssim l \lesssim 60$ for subAlfv\'{e}nic models and $20 \lesssim l \lesssim 60$ for superAlfv\'{e}nic models. The linear scaling regions for the logarithm of density and velocity do not correspond. The dependence of third-order structure functions on the sonic Mach number is marginal. The only dependence can be observed in the case of velocity in models with a weak magnetic field. However, the sonic Mach number does not change the slope of the structure function.

\begin{figure}  
 \epsscale{0.9}
 \plottwo{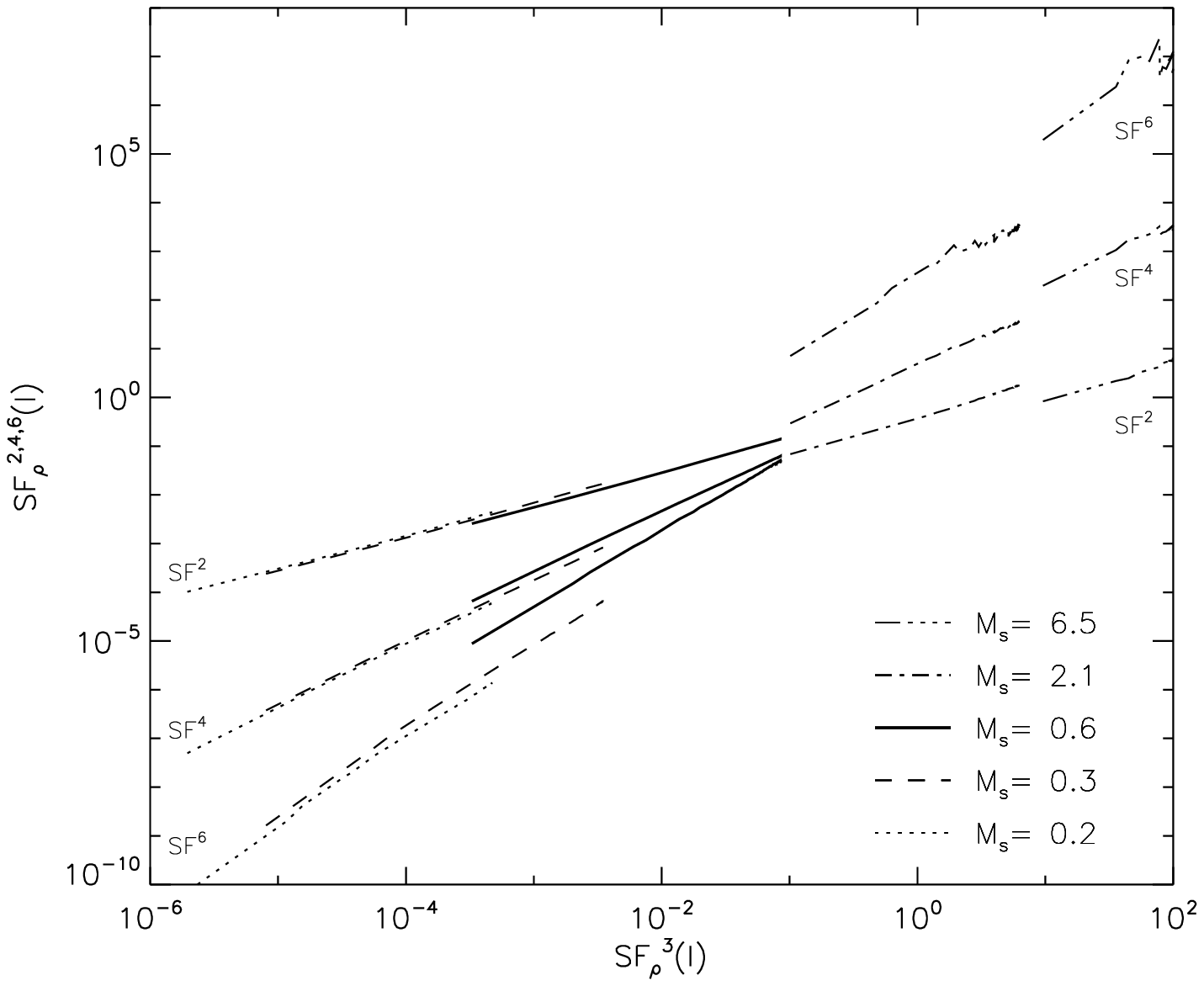}{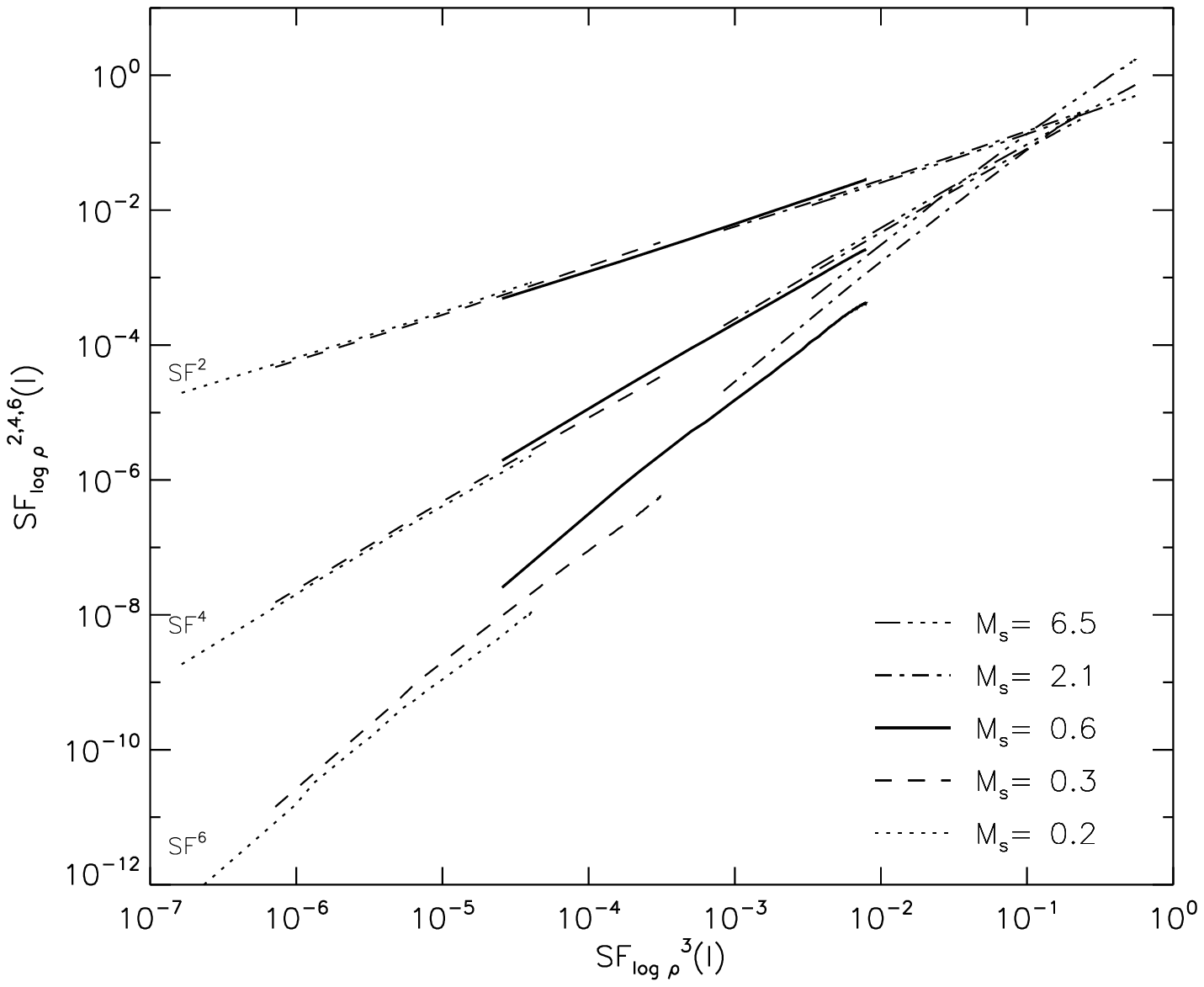}
 \plottwo{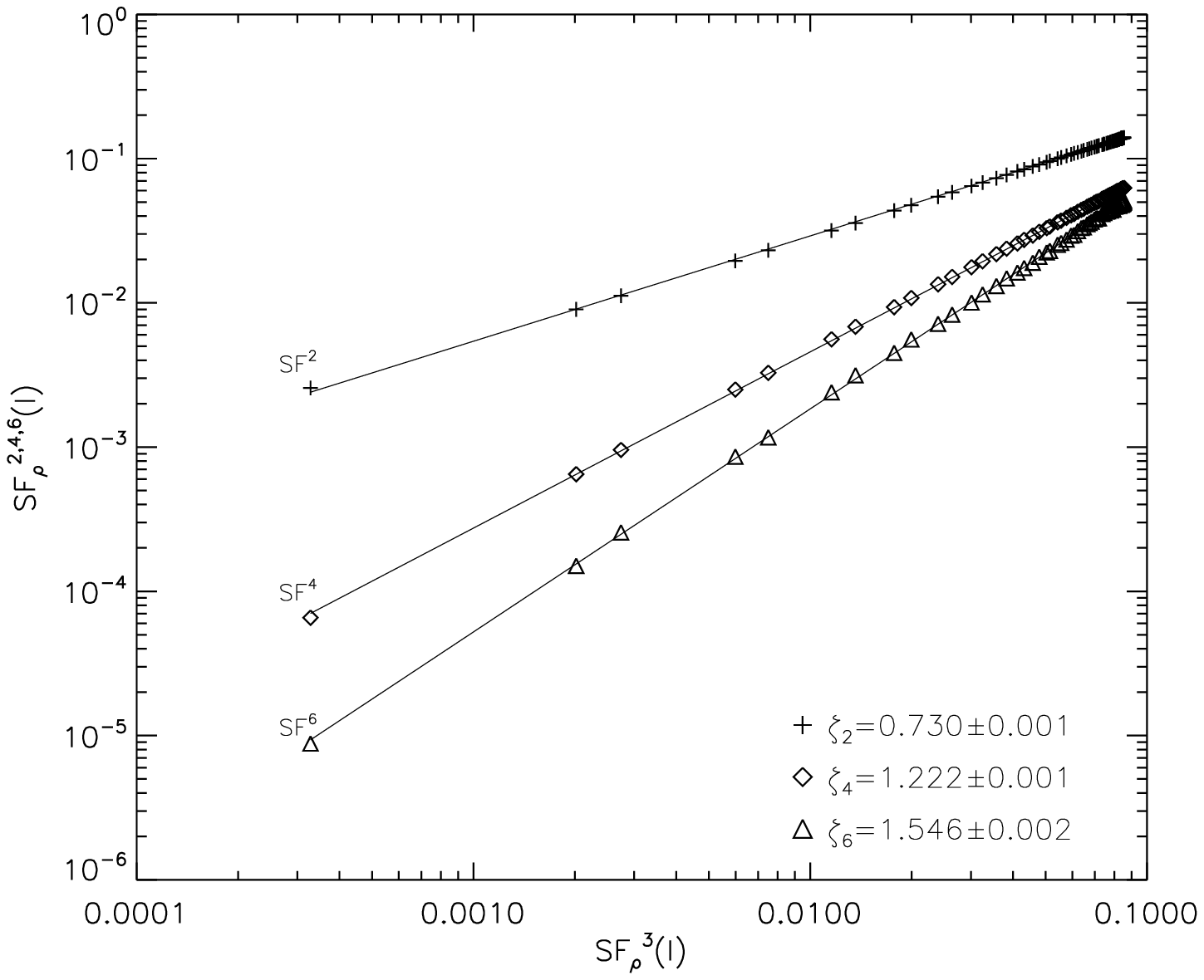}{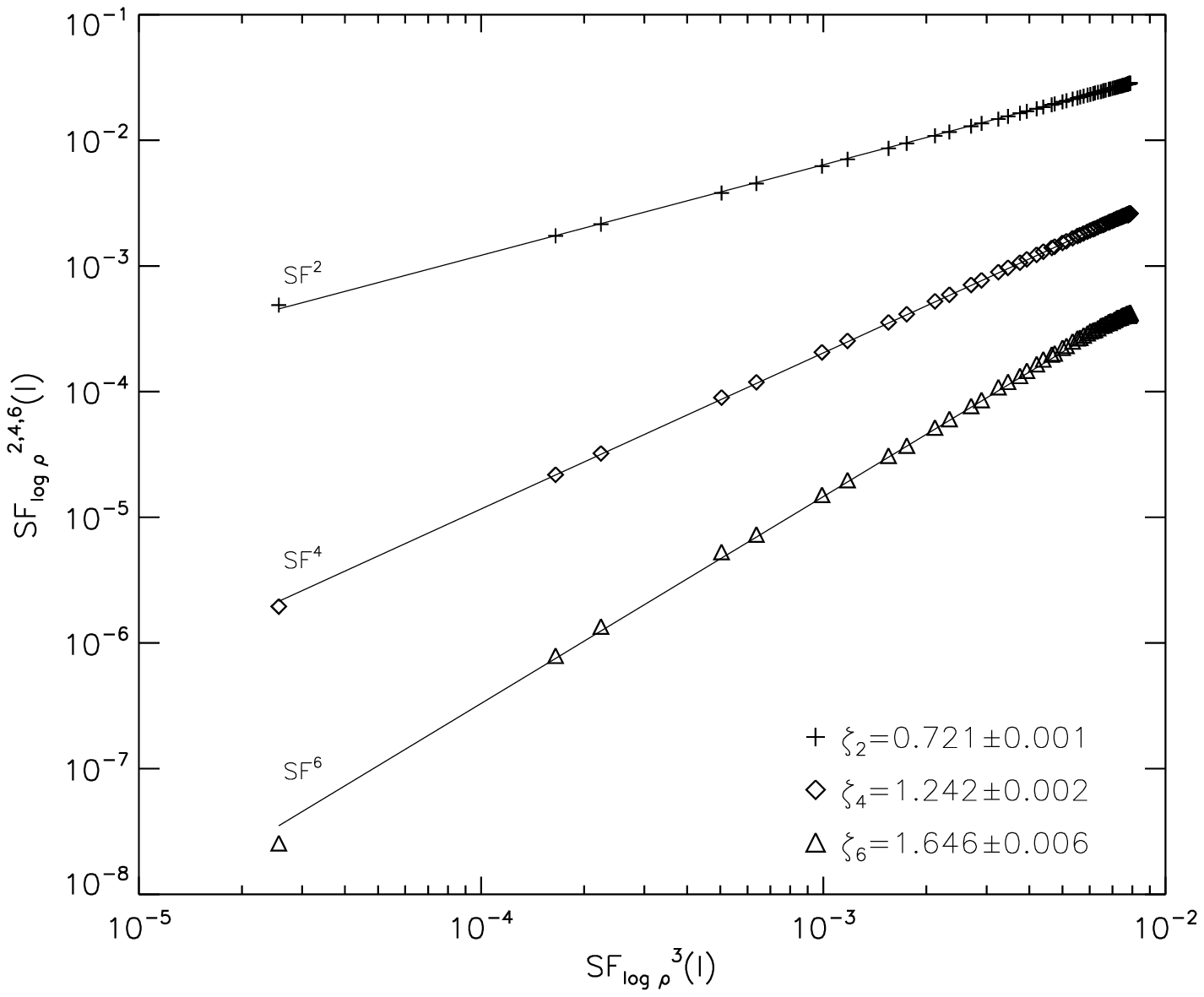}
 \caption{{\em Top row}: Tests for some sort of self-similarity for density ({\em left}) and the logarithm of density ({\em right}). Plots show structure functions of three different orders p=2, 4, and 6 as a function of the third-order structure function for models with ${\cal M}_A\sim0.7$. Structure functions of models with different ${\cal M}_s$ form lines in different ranges of the plot. The linear shape of the lines suggests the existence of self-similarity. For high values of ${\cal M}_s$, the dispersion of points is higher, but the points are still aligned along a line. In the case of $\log \rho$ ({\em top right}) the structure functions of lower orders preserve the value of the slope for a wide range of sonic Mach numbers. Strong density fluctuations disturb the slope of higher order structure functions. {\em bottom row}: Structure functions $SF^p(l)$ of density ({\em left}) and the logarithm of density ({\em right}) for three different orders p=2, 4, and 6 as a function of the third-order structure function for models with ${\cal M}_{A}\sim0.7$ and ${\cal M}_{s}\sim0.7$. Points represent the computed values of structure functions, and lines are a linear fit to the points. The fitting procedure gives the value of the slope with very small uncertainty (less than 1.5\% for density and less than 0.4\% for the logarithm of density).\label{fig:dens_logd_ess}}
\end{figure}

In Figure \ref{fig:dens_logd_ess} we show a test of an extended self-similarity. We plot the structure functions of second, fourth, and sixth order as a function of the structure function of third order. Plots contain functions for sub-Alfv\'{e}nic models only with different sonic Mach numbers for density and the logarithm of density (top left and right plots in Figure \ref{fig:dens_logd_ess}, respectively). The alignment of points along straight lines in Figure \ref{fig:dens_logd_ess} clearly suggests the existence of a self-similarity both for density and the logarithm of density. In the case of density, this nice picture is slightly disturbed for supersonic models, but only for $S^{(p)}$ with order $p>3$. Structures with large density contrasts significantly distort structure functions of higher orders. Nevertheless, the linear relation is still well seen. Applying the logarithmic operation on density before computing the structure function significantly improves a linear alignment of points and introduces some ordering in the placement of lines (see Fig. \ref{fig:dens_logd_ess}, {\em top right}). We see that lines corresponding to the structure functions of the same order are placed on the same line. This is especially well seen for functions of second order. This draws a hypothesis that the ``perfect'' low-order structure functions of $\log \rho$ have the same slope independent of the sonic Mach number and the structure functions are only shifted along the line of a fixed inclination according to the strength of the sonic motions. For higher orders the strong asymmetries in the distribution of $\log \rho$ dominate, giving the reason for the flattening of the relation between structure functions of different orders.

The existence of a self-similarity of structure functions of density and the logarithm of density is very useful. The question of to what degree these relations are linear induces us to perform a fitting, which is shown in Figure \ref{fig:dens_logd_ess} ({\em bottom row}). We have chosen one particular model with ${\cal M}_s\sim0.7$ and ${\cal M}_A\sim0.7$. For this model we perform linear fitting in the log-log domain of second, fourth and sixth order structure functions of density (Fig.~\ref{fig:dens_logd_ess} {\em bottom left}) and the logarithm of density (Fig.~\ref{fig:dens_logd_ess} {\em bottom right}) as a function of the third order structure function of the corresponding quantity. In this case, points are regularly spaced and fitting on the log-log axes gives very accurate values of slopes even for high-order structure functions (fractional errors in the order both for density structure functions and the logarithm of density for the lines are presented in Fig. \ref{fig:dens_logd_ess}). Such a high degree of linearity suggests the existence of highly self-similar structures independent of scale, also self-similar in some way, because at small scales the dissipation starts to play an important role, which we clearly see in spectra of density, velocity, or magnetic fields. The effect of dissipation is not seen for relations between structure functions of different orders, which suggests that the dissipation acts in a consistent way at different scales. Moreover, the dissipation must undergo a similar scaling relation, otherwise we would get different slopes of the SFs within the inertial and dissipative ranges.

\begin{figure}  
 \plottwo{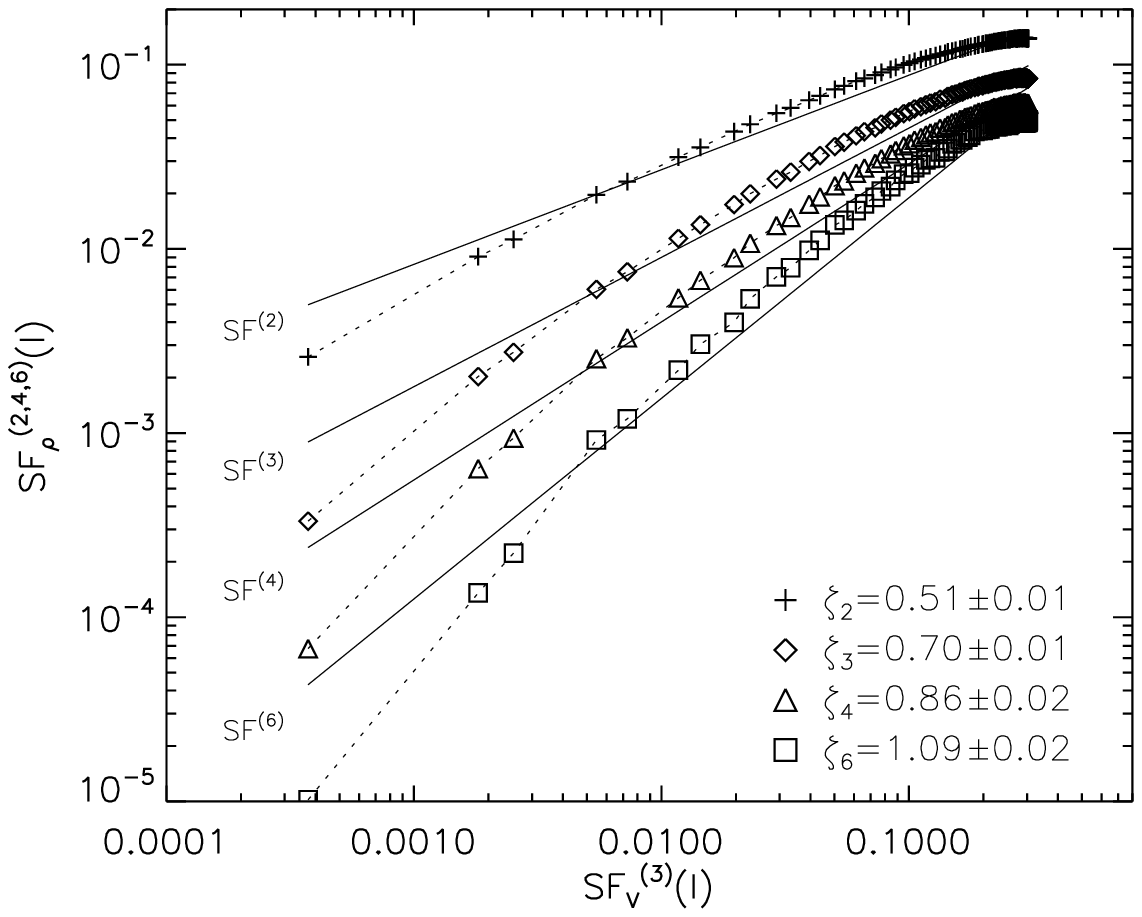}{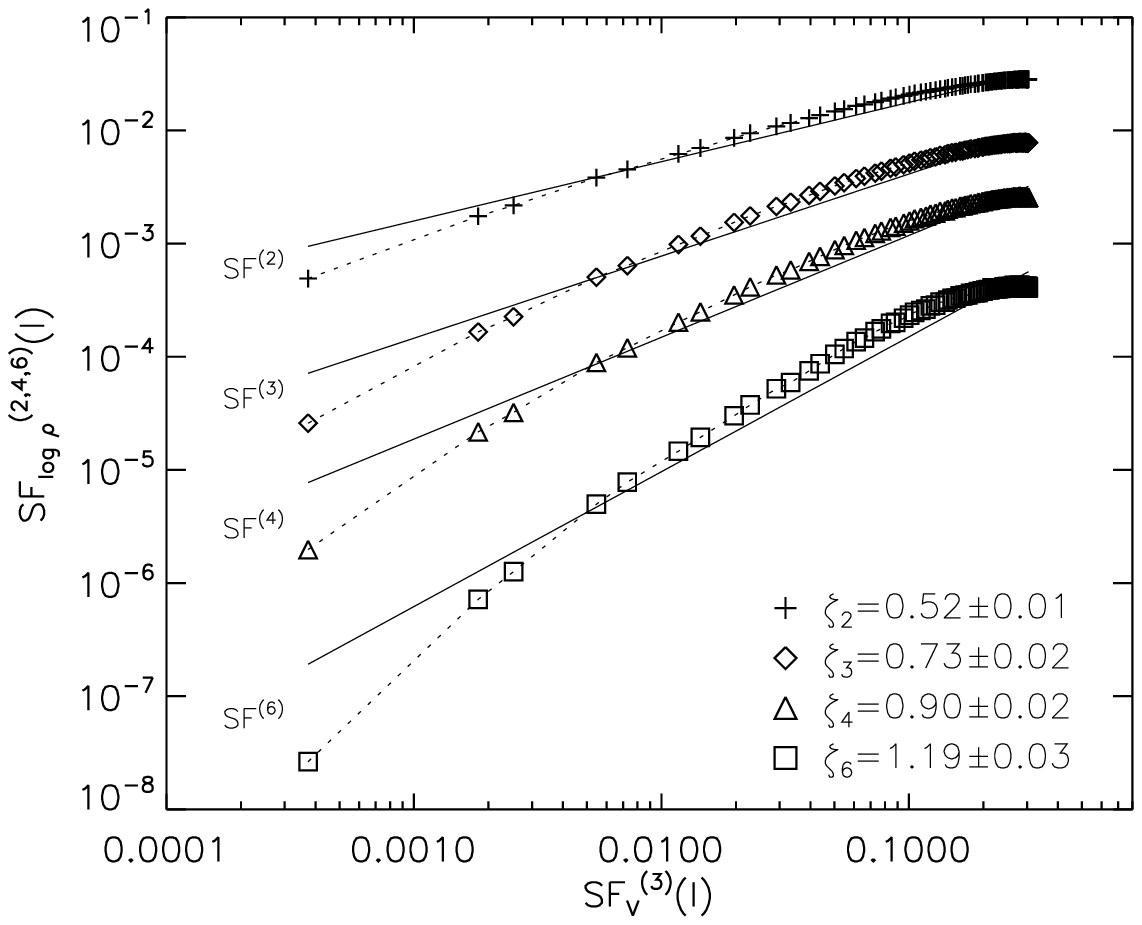}
 \caption{Structure functions of density ({\em left}) and the logarithm of density ({\em right}) as a function of the third-order structure function of velocity. The relation is not linear, so there is no simple similarity between density and velocity structure functions. An important consequence of this nonlinearity is an arbitrary normalization, which is not because of the Kolmogorov law relating third-order structure functions with the separation length $l$. \label{fig:sf3velo}}
\end{figure}

Going further, we could ask if the extendend self-similarity is valid for structure functions of different quantities. In Figure \ref{fig:sf3velo} we draw plots of the structure functions of the second, third, fourth and sixth order of density ({\em left}) and the logarithm of density ({\em right}) as a function of the structure function of third order of the velocity. The existence of any similarity should be manifest by the linear relation here also, but as we see, this relation is far from linear. There are at least two reasons for this. At first we may suppose that scaling relations of the structure functions of a particular order, which are used as a normalization (in this paper we use the third moment), look different for different quantities, but it might also be the influence of dissipation acting differently on different quantities. Whatever the reason is, the problem of relations, if any exist, between the structures of different quantities involved in turbulence requires further investigation.

\section{Intermittency of density and the logarithm of density}
\label{sec:intermittency}

An important insight into the turbulence can be achieved by studies of turbulence self-similarity. This property, which is also called scale invariance, implies that fluid turbulence can be reproduced by the magnification of some part of it. Take the famous model of incompressible Kolmogorov turbulence as an example. In this model, the energy is injected at a large scale $L$ and forms eddies that transfer the energy to smaller and smaller scales. At the scales where the corresponding Reynolds number Re$_l\sim lv_l/\nu$ is much larger than unity, the dissipation over the eddy turnover time $t_l\sim l/v_l$ is negligible. As a result, the energy cascades to smaller and smaller scales without much dissipation, i.e. $v_l^2/t_l\sim \mathrm{const}$, which gives the well-known Kolmogorov power-law scaling for the eddies of scale $l$, namely, $v_l\sim l^{1/3}$. The cascade terminates at the dissipation scale, which provides an inertial range from $L$ to $L \mathrm{Re}^{-3/4}$, where Re is the Reynolds number corresponding to the flow at the injection scale $L$. At the dissipation scales, self-similarity is known to fail with turbulence forming non-Gaussian dissipation structures as exemplified, e.g. in \cite{biskamp03}. Interestingly enough, present-day research shows that self-similarity is not exactly true even along the inertial range. Instead, fluctuations tend to get increasingly sparse in time and space at smaller scales. This property is called intermittency. Note that power-law scaling does not guarantee scale invariance or the absence of intermittency.

Intermittency is an essential property of astrophysical fluids. As intermittency violates the self-similarity of motions, it becomes impossible to naively extrapolate the properties of fluids obtained computationally with relatively low resolution to the actual astrophysical situations. In terms of astrophysics, intermittency affects turbulent heating, momentum transfer, interaction with cosmic rays, radio waves, and many more essential processes. Physical interpretation of intermittency started after the work by Kolmogorov, but the first successful model was presented by \cite{she94}. The scaling relations suggested by \cite{she94} related $\zeta(p)$ to the scaling of the velocity $v_l \sim l^{1/g}$, the energy cascade rate $t^{-1}_l \sim l^{-x}$, and the co-dimension of the dissipative structures $C$,
\begin{equation}
 \zeta(p) = \frac{p}{g} (1-x) + C \left( 1 - (1 - x/C)^{p/g} \right).
\end{equation}
Parameter $C$ is connected to the dimension of the dissipative structures $D$ through relation $C = 3 - D$ \citep{mueller00}. In hydrodynamical turbulence, according to Kolmogorov scaling, we have $g = 3$ and $x = \frac{2}{3}$. Vortex filaments, which are 1D structures, correspond to $C=2$ ($D=1$). For MHD turbulence we also observe current sheets, which are 2D dissipative structures, corresponding to $C=1$ ($D=2$). For these two types of dissipative structures we obtain two different scaling relations (substituting $g = 3$ and $x = \frac{2}{3}$),
\begin{equation}
 \zeta(p) = \frac{p}{9} + 2 \left[ 1 - (2/3)^{p/3} \right] \ \mathrm{for} \ C=2
 \label{eqn:sl}
\end{equation}
and
\begin{equation}
 \zeta(p) = \frac{p}{9} + 1 - (1/3)^{p/3} \ \mathrm{for} \ C=1 .
 \label{eqn:mb}
\end{equation}
Relation (\ref{eqn:sl}) is often called the She \& L\'{e}v\^{e}que scaling \citep{she94} and equation (\ref{eqn:mb}) the M\"uller-Biskamp scaling \citep{mueller00}. There are theoretical arguments against the model \citep[see][]{novikov94}, but so far the She \& L\'{e}v\^{e}que scaling is the best for reproducing the intermittency of incompressible hydrodynamic turbulence.

So far there is no similar theory for density scaling in compressible MHD turbulence. The presence of shocks disturbs the spectra and structure functions of density. However, as we discussed in \S\ref{sec:spectra}, the logarithm of density exhibits both spectrum and Goldreich-Sidhar scale-dependent anisotropy (see also BLC05).

\begin{figure}  
 \epsscale{0.9}
 \plottwo{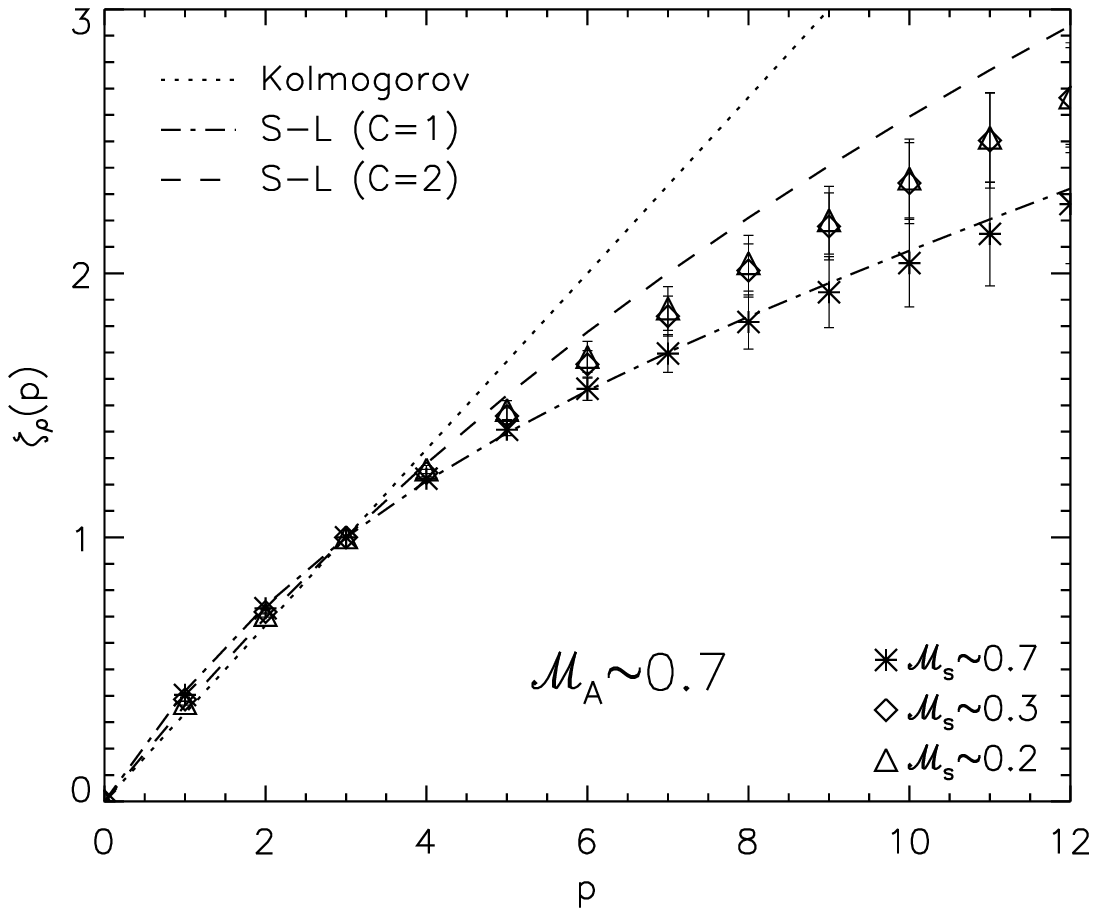}{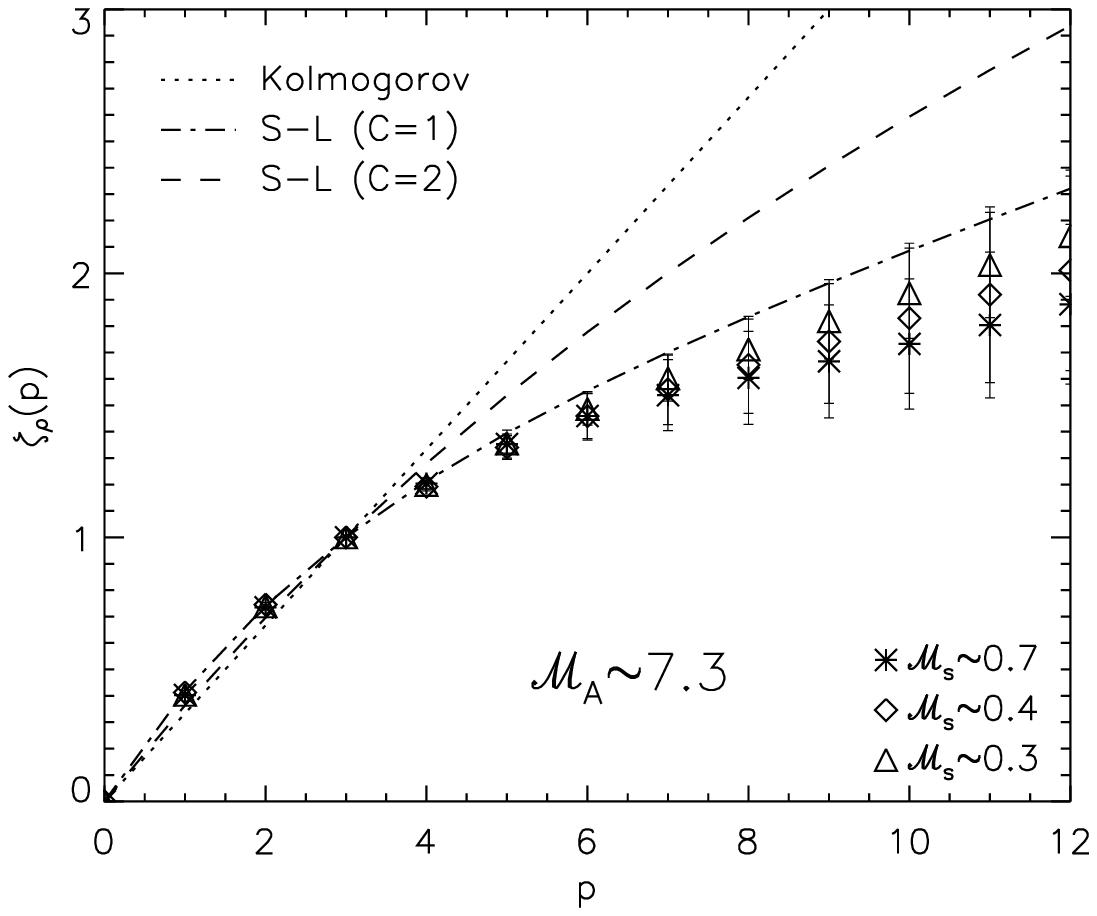}
 \plottwo{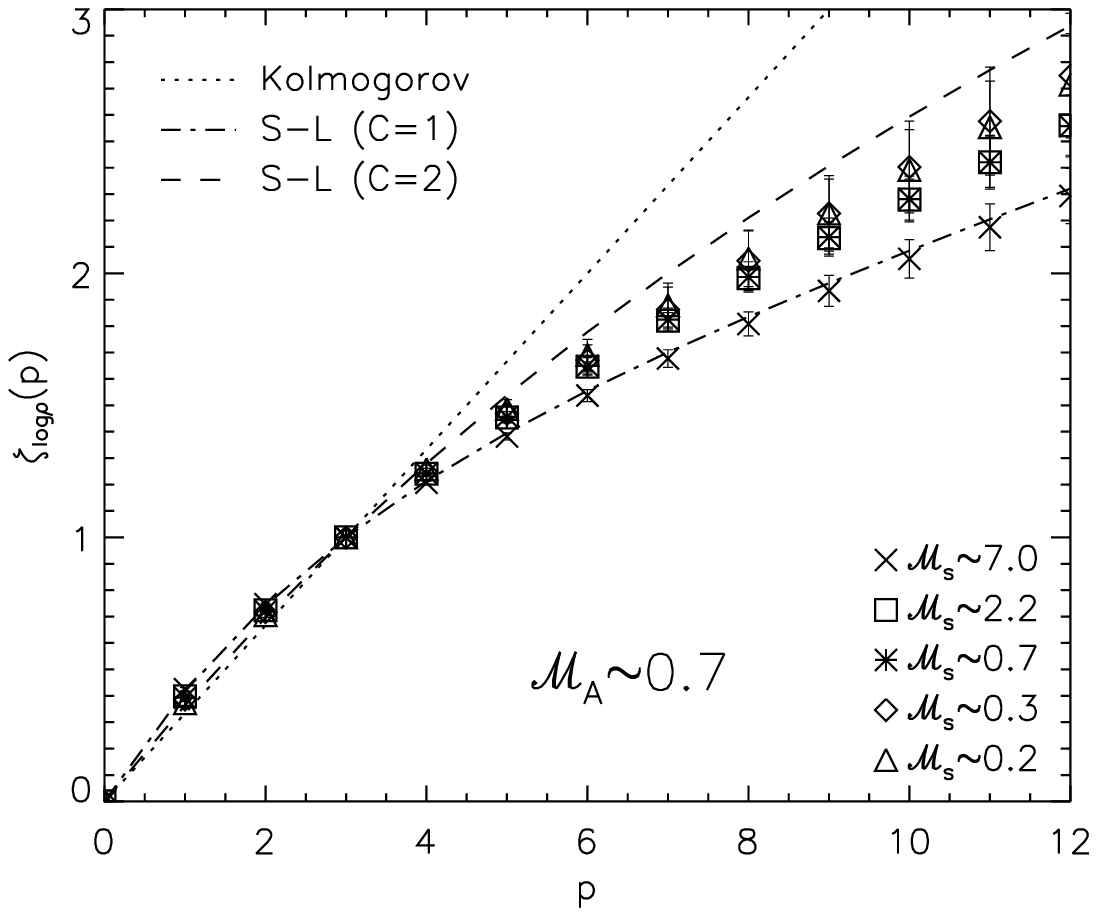}{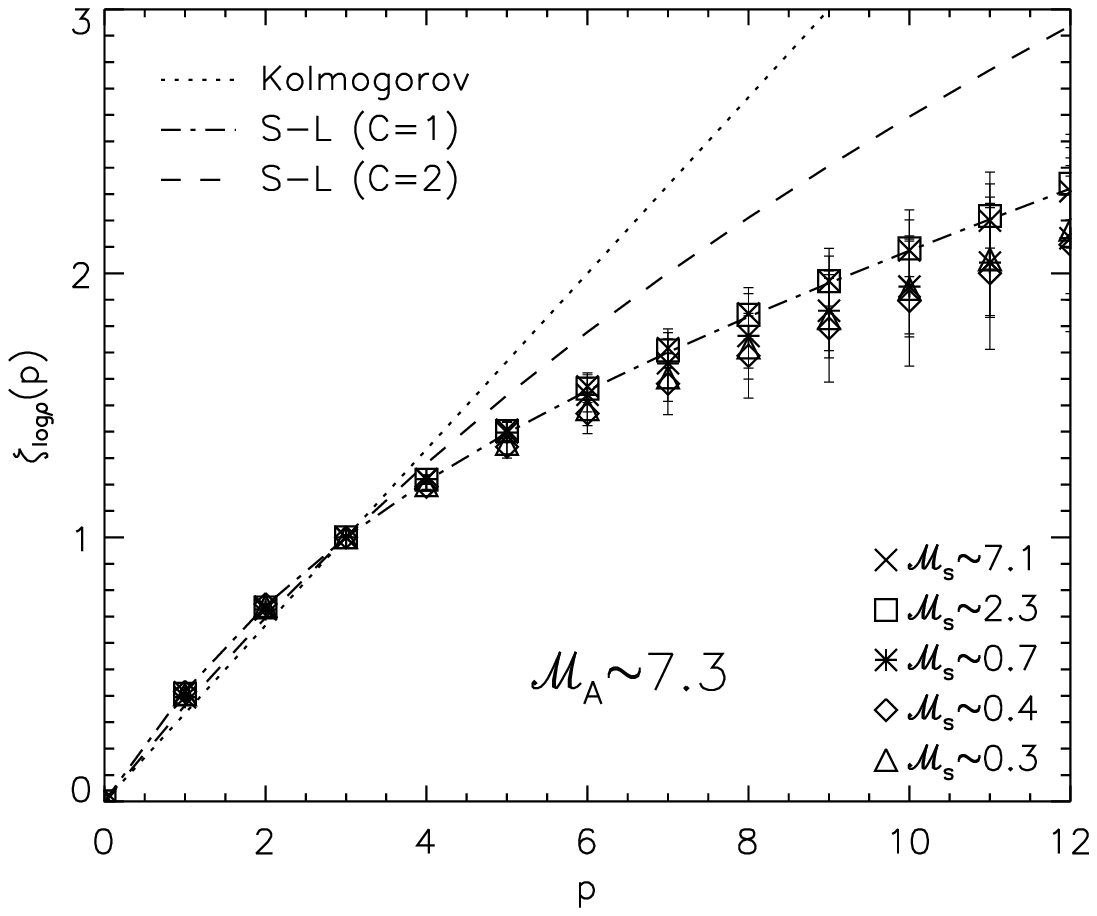}
 \caption{Scaling exponents for density ({\em top row}) and the logarithm of density ({\em bottom row}) normalized to the third-order exponent for models with ${\cal M}_A\sim0.7$ ({\em left column}) and ${\cal M}_A\sim7.3$ ({\em right column}). For density only the values for subsonic experiments are shown from ${\cal M}_s \sim 0.2$ ({\em triangles}) to $\sim 0.7$ ({\em stars}) for models with medium resolution. For the logarithm of density the values for subsonic ({\em triangles, diamonds, and stars}) and supersonic ({\em squares and crosses}) experiments are shown. Lines show theoretical scalings: dotted for Kolmogorov and dash-dotted and dashed for She-L\'{e}v\^{e}que with parameter $C=1$ and $2$, respectively ($g=3$ and $x=\frac{2}{3}$). \label{fig:dens_expons}}
\end{figure}

In Figure \ref{fig:dens_expons} we show the scaling exponents for density ({\em top row}) and the logarithm of density ({\em bottom row}) for models with ${\cal M}_A\sim0.7$ ({\em left column}) and ${\cal M}_A\sim7$ ({\em right column}). In the case of density ({\em top row}) we show only the subsonic models because of broken self-similarity of density structure functions in supersonic models. To overcome this problem one can use techniques filtering the high-density structures, which deform statistics the most. The application of such filtering has been discussed by BLC05. Nevertheless, we do not investigate this problem further in this article. When possible, we use extended self-similarity to obtain the exponents, applying the third-order structure function as a normalization. We perform such fitting procedures for all available time steps for each model. In this way we obtained the mean values of exponents and their standard deviation.

Comparing plots for sub- and superAlfv\'{e}nic models (Fig. \ref{fig:dens_expons}, {\em top left and top right, respectively}) we see that in the presence of a strong external magnetic field, density structures are less intermittent. Although the error bars are relatively large, the difference is undisputed. Moreover, the value of the parameter $C$ seems to be between one and two for the sub-Alfv\'{e}nic models with very low sonic Mach numbers and approximately equal to two when ${\cal M}_s \sim {\cal M}_A \sim 0.7$ (see Figure \ref{fig:dens_expons}, {\em top left}). When the magnetic field becomes weak, the intermittency of density grows. In addition, it grows somewhat with the sonic Mach number. The changes are smaller then the errors however.

In Figure \ref{fig:dens_expons} ({\em bottom row}) we present the scaling exponents calculated from the structure functions of the logarithm of density. After this transition, the scaling exponents can be more precisely determined for all models, even for supersonic ones. In turbulence with a strong magnetic field (Fig. \ref{fig:dens_expons}, {\em bottom left}) we see that intermittency behaves almost identically for subsonic models. We can restore the scaling of density even for supersonic motions. The value of $C$ is between one and two for models with ${\cal M}_s\lesssim 2.2$ and exactly two for the model with ${\cal M}_s\sim7.0$. The errors in the latter case are rather small. Our models with a weak magnetic field reveal more intermittent structure of $\log \rho$ as in the case of density (see Fig. \ref{fig:dens_expons}, {\em bottom right}). In these models, the scaling exponents suggest the dimension $C=2$. Although we see some small variations of exponents with the value of ${\cal M}_s$, they are smaller then the error bars.

What we can say about all the above models is that the structure of density is intermittent and the intermittency depends on the strength of the external magnetic field. One may note some changes in intermittency with the strength of sonic motions, but these changes are weak. In general, the supersonic motions generate more intermittent structures of density or the logarithm of density. The only possible exception is the logarithm of density, which has structure more intermittent with stronger sonic motions in the presence of a weak magnetic field, but with a larger uncertainty at the same time. Finally, in comparison to the density, the intermittency decreases when we use the logarithm of density.

\begin{figure}  
 \epsscale{0.9}
 \plottwo{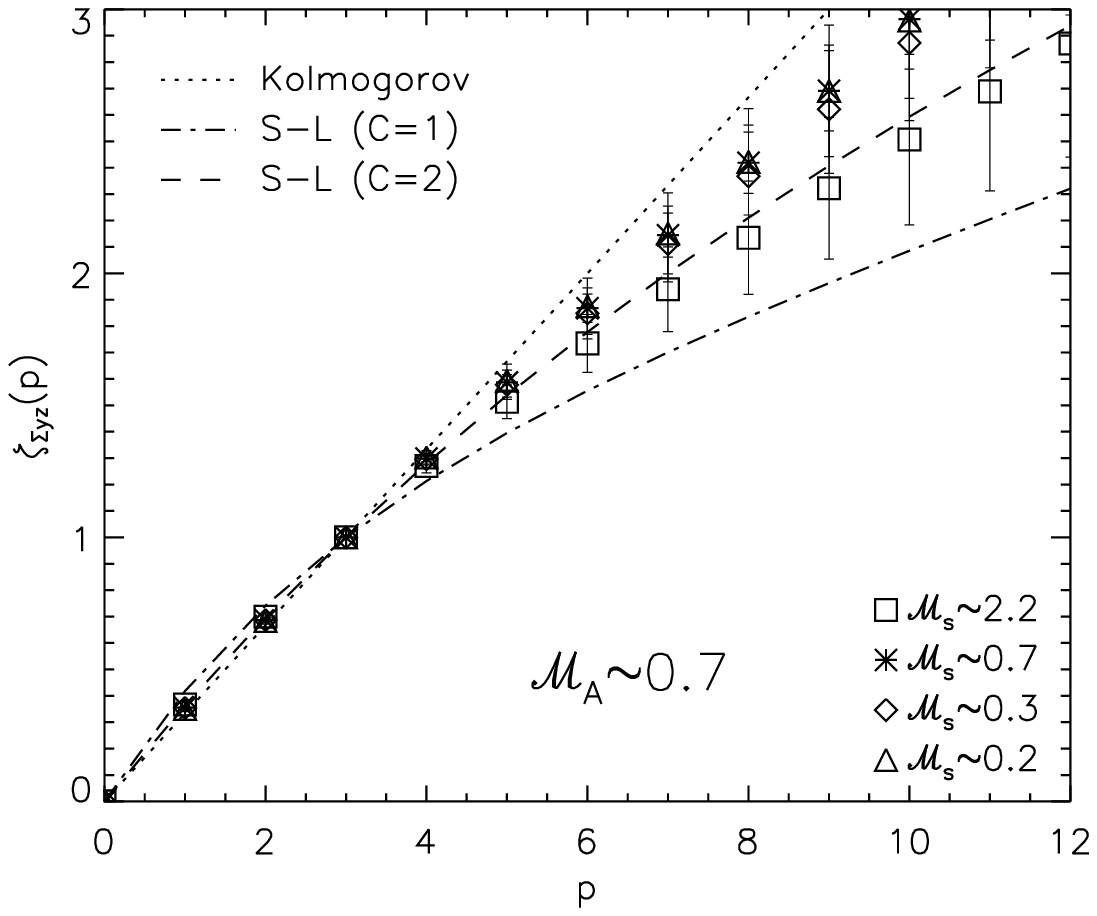}{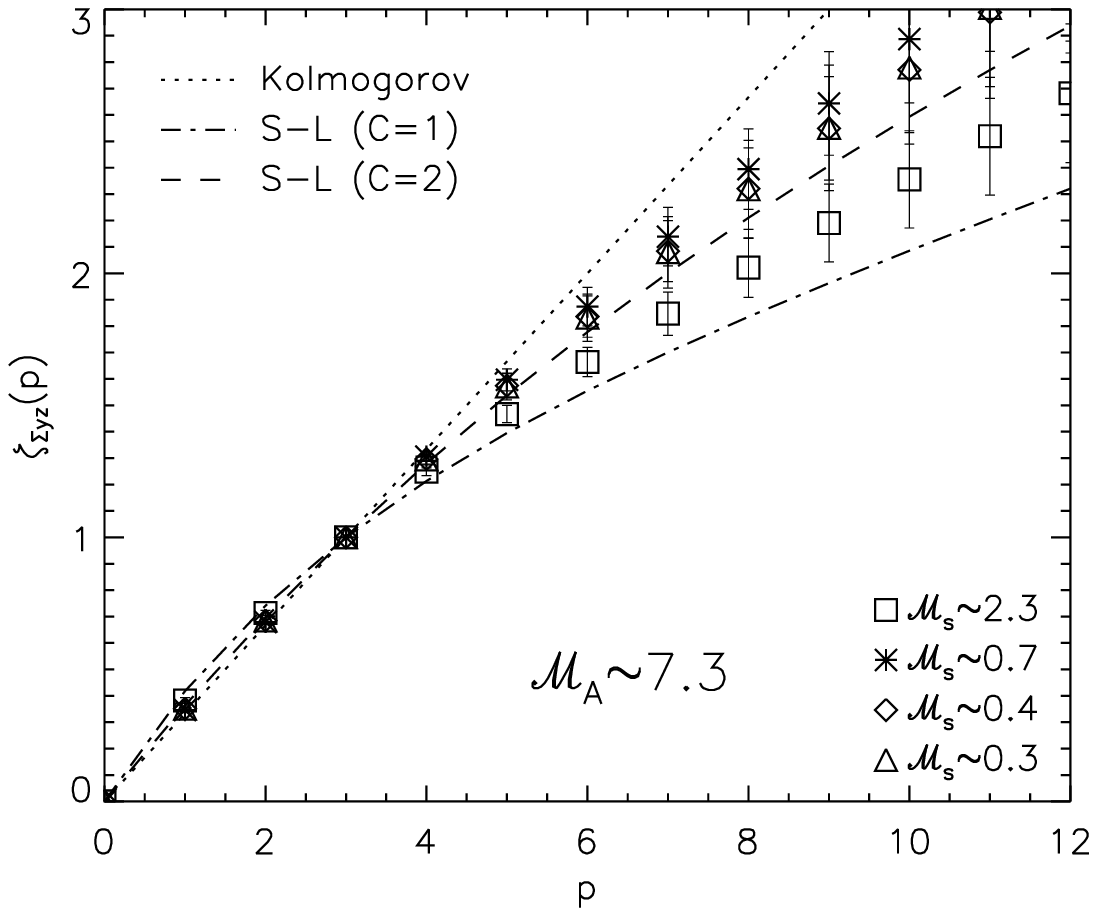}
 \plottwo{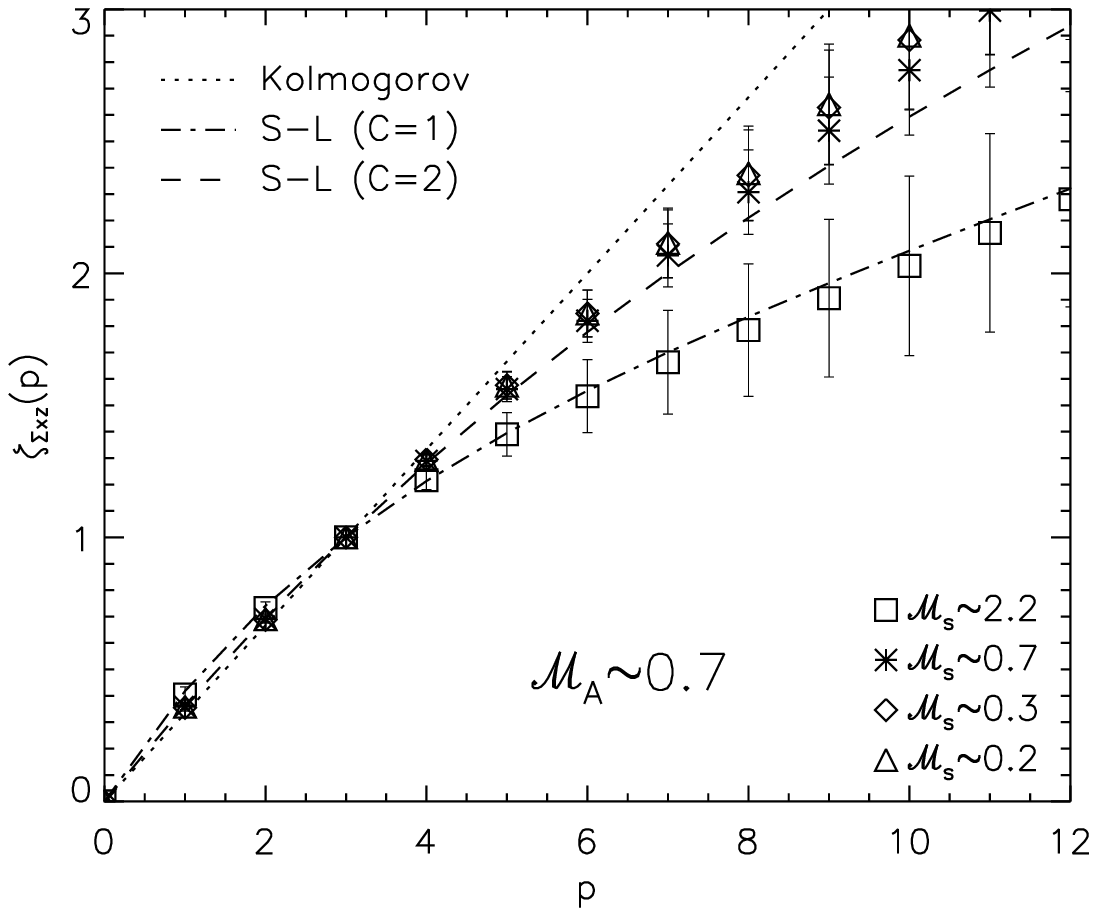}{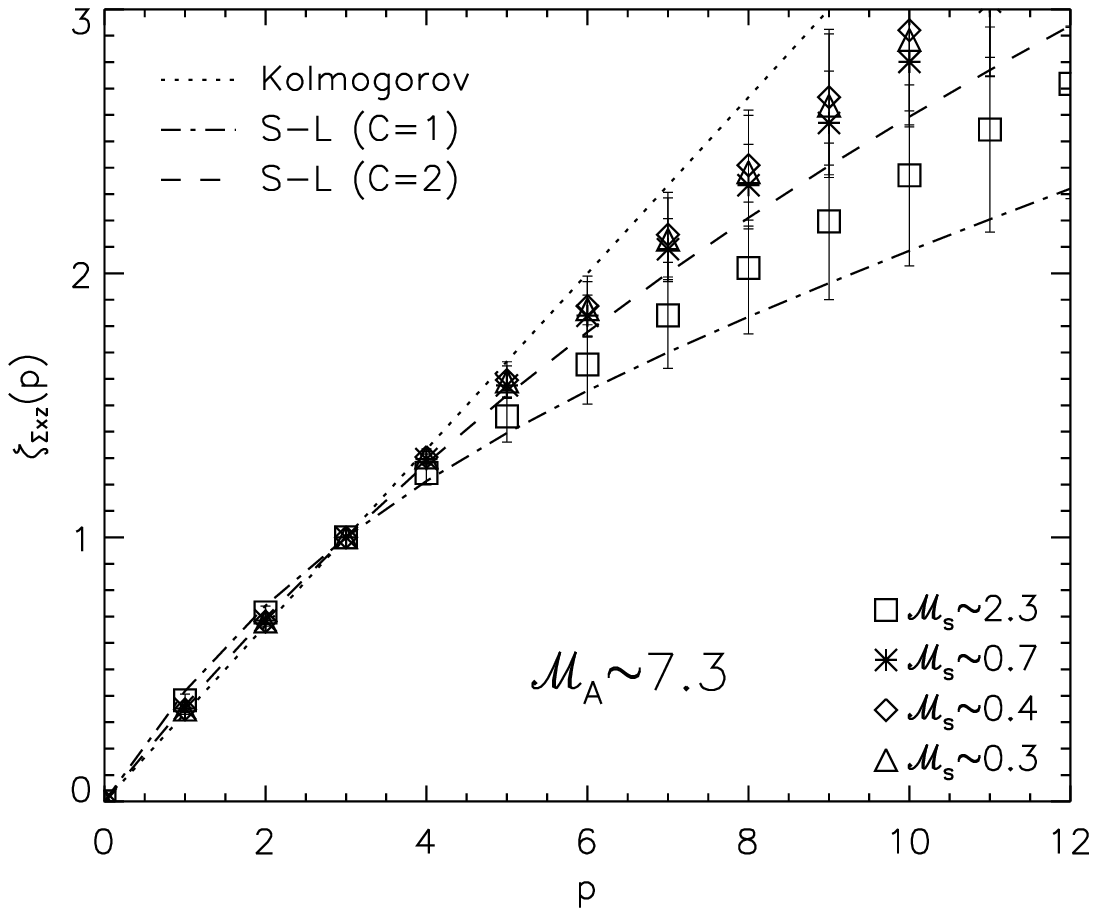}
 \caption{Scaling exponents of the column density integrated along and across the direction of $B_\mathrm{ext}$ ({\em top and bottom rows, respectively}) for experiments with ${\cal M}_A\sim0.7$ and $\sim7.3$ ({\em left and right columns, respectively}) for models with medium resolution. We use the third-order exponent to normalize. Lines show theoretical scalings: dotted for Kolmogorov and dash-dotted and dashed for She-L\'{e}v\^{e}que with parameter $C=1$ and $2$ respectively ($g=3$ and $x=\frac{2}{3}$). \label{fig:dens_int_expon}}
\end{figure}

\begin{figure}  
 \epsscale{0.9}
 \plottwo{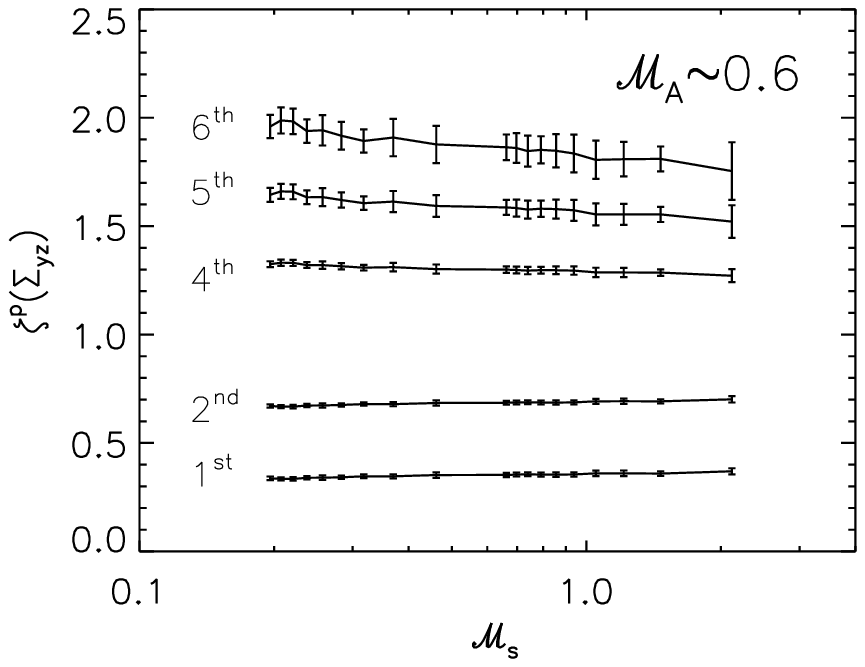}{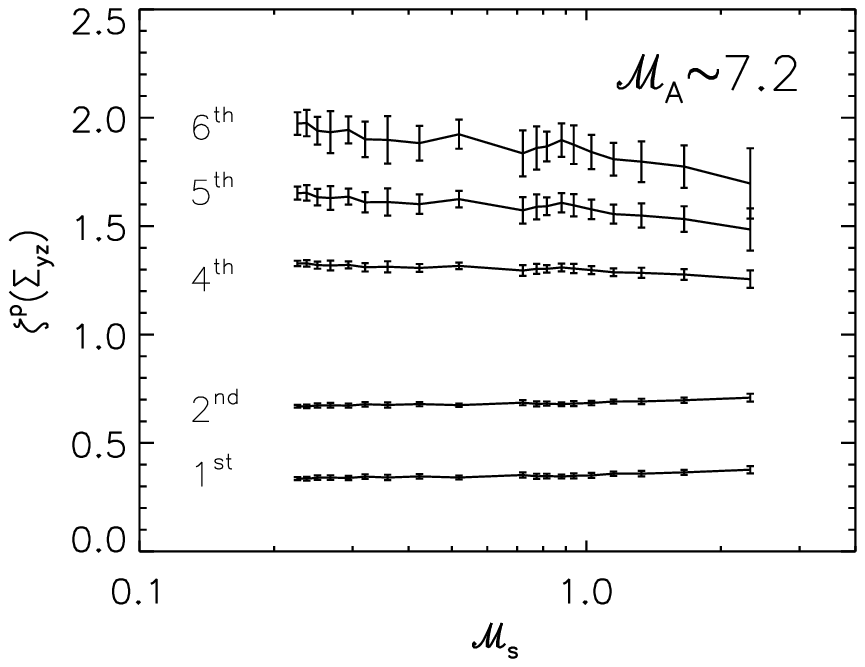}
 \plottwo{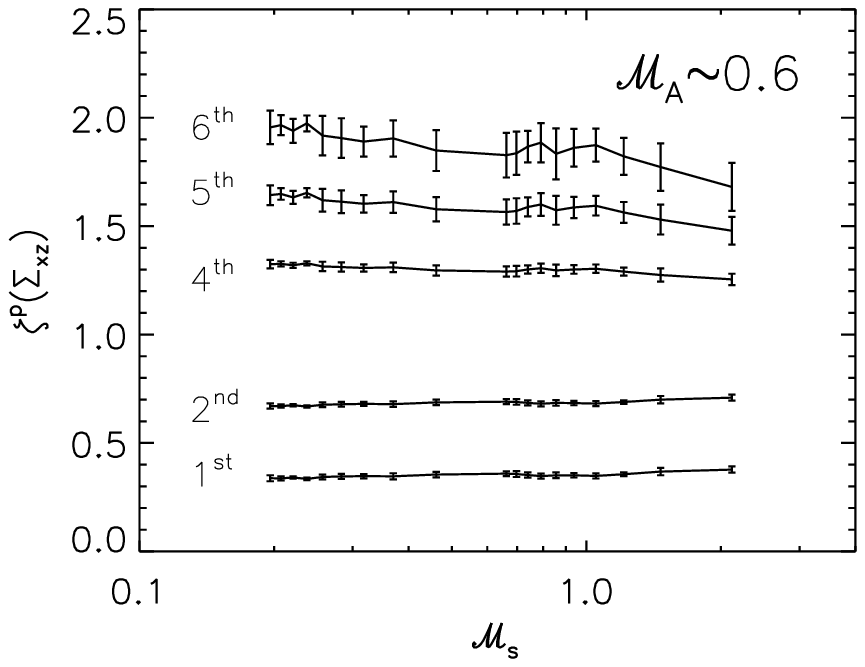}{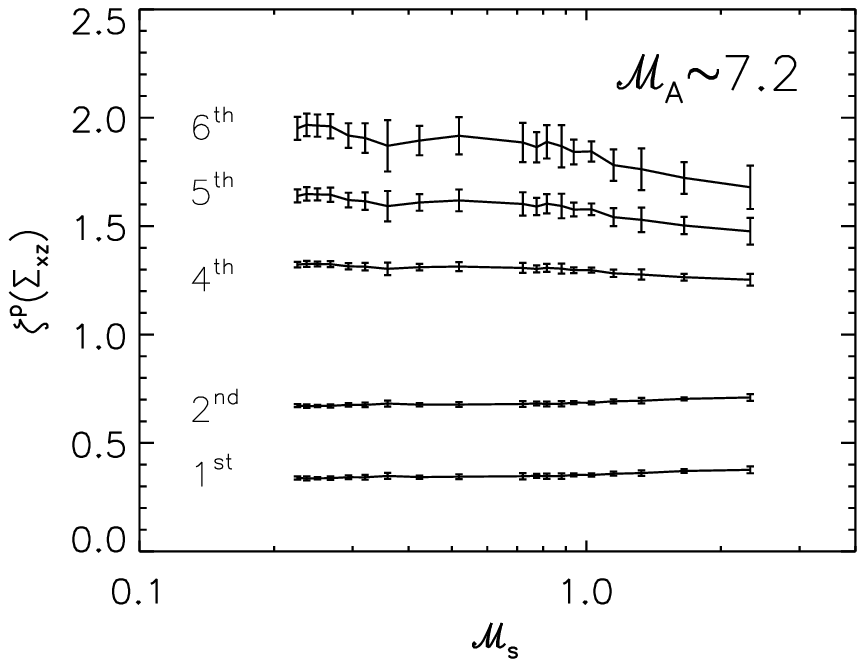}
 \caption{Scaling exponents of the structure functions of column density $\Sigma$ projected along ({\em top row}) and across ({\em bottom row}) the external magnetic field $B_\mathrm{ext}$ for sub-Alfv\'{e}nic ({\em left column}) and super-Alfv\'{e}nic ({\em right column}) turbulence for models with low resolution (128$^3$). We show the exponents for the structure functions from the first to the sixth order excluding the third one. \label{fig:col_dens_scal}}
\end{figure}

In Figure \ref{fig:dens_int_expon} we show the scaling exponents for column densities for different models. Although the estimation of exponents induce substantial uncertainties, we see a small change in intermittency with the sonic Mach number in all plots, especially in plots showing the column density integrated across $B_\mathrm{ext}$ in the models with a strong magnetic field. To investigate this change deeper we show how the intermittency of column density changes with sonic Mach number. In Figure \ref{fig:col_dens_scal} we present the scaling exponents of the structure functions of column densities $\Sigma$ for different experiments. As a result we obtain the relation of the scaling exponent and the sonic Mach number. The exponents are normalized to the third-order one, so Figure \ref{fig:col_dens_scal} shows exponents for the structure functions of the first, second, fourth, fifth, and sixth orders. We average the exponents for each model over 16 equally spaced moments in time $t\in [10,25]$, and we calculate their standard deviation. In this way we obtain also a measure of uncertainty. Figure \ref{fig:col_dens_scal} shows the exponents for column densities integrated along and across $B_\mathrm{ext}$. In all plots we see a systematic decrease of the exponents for orders larger than 3 and simultaneously an increase of the exponents for orders less than 3. This behavior signifies a progressive growth of the intermittency of density structure with sonic Mach number.

\section{Skewness and kurtosis}
\label{sec:moments}

The most straightforward statistical properties of density and column density distributions are the mean value and the variance. However, on a deeper level we are interested to know the degree of asymmetry of the distribution and its relation to the normal distribution. These two properties are measured by the skewness and kurtosis. The mean arithmetic value and the variance of the distribution of, for example, density $\rho$ are given by
\begin{equation}
 \bar{\rho} = \frac{1}{N}\sum_{i=1}^N{\rho_i} \quad \mathrm{and} \quad \sigma^2_{\rho} = \frac{1}{N-1}\sum_{i=1}^N(\rho_i-\bar{\rho})^2,
 \label{eqn:mean_variance}
\end{equation}
where $N$ is the number of samples or points of the mesh in the case of simulation data. The mean value is a less important property in our studies, so we do not consider it here; however, it is required to calculate higher moments. Variance measures the width of the distribution and is always positive by definition.

Skewness as a measure of the asymmetry is defined through the third-order statistical moment,
\begin{equation}
 \gamma_{\rho} = \frac{1}{N} \sum_{i=1}^N {\left( \frac{\rho_i-\bar{\rho}}{\sigma_{\rho}} \right)^3}
 \label{eqn:skewness}
\end{equation}
The skewness for a normal distribution is zero. Negative values of the skewness indicate distribution that is skewed in the left direction, and positive values for the skewness indicate the distribution that is skewed in the right direction. By skewed left, for instance, we mean that the left tail is longer compared to the right tail.

Kurtosis is a measure of whether the quantity has a distribution that is peaked or flat in comparison to the normal distribution. That is, data sets with high kurtosis tend to have a distinct peak near the mean, decline rather rapidly, and have heavy tails. Data sets with a low kurtosis tend to have a flat top near the mean rather than a sharp peak. A uniform distribution would be the extreme case. Kurtosis is defined in a similar way to skewness, but using the fourth-order statistical moment,
\begin{equation}
 \beta_{\rho} = \frac{1}{N} \sum_{i=1}^N {\left( \frac{\rho_i-\bar{\rho}}{\sigma_{\rho}} \right)^4} - 3.
 \label{eqn:kurtosis}
\end{equation}
Positive kurtosis indicates a ``peaked'' distribution and negative kurtosis indicates a ``flat'' distribution.

\begin{figure}  
 \epsscale{0.9}
 \plottwo{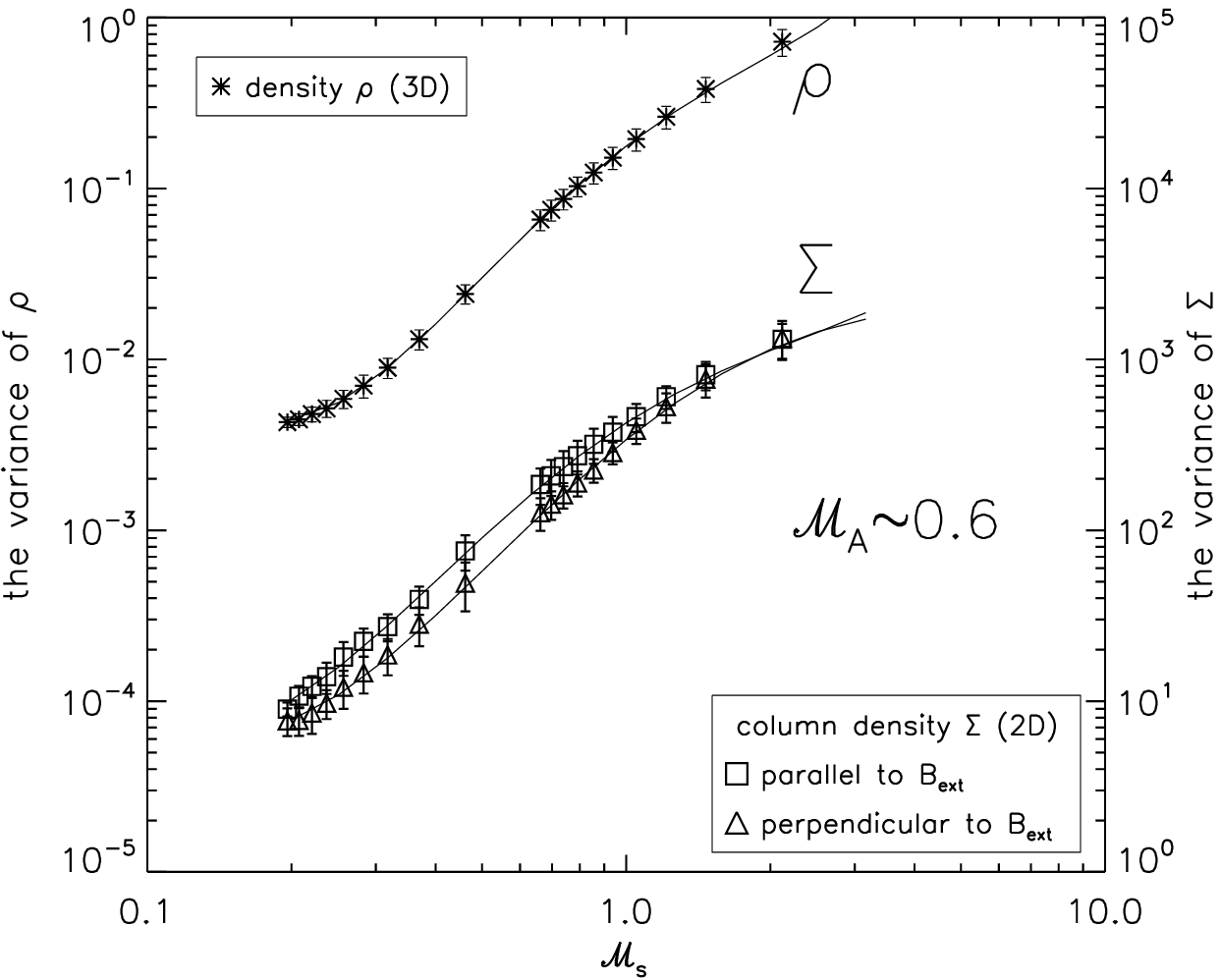}{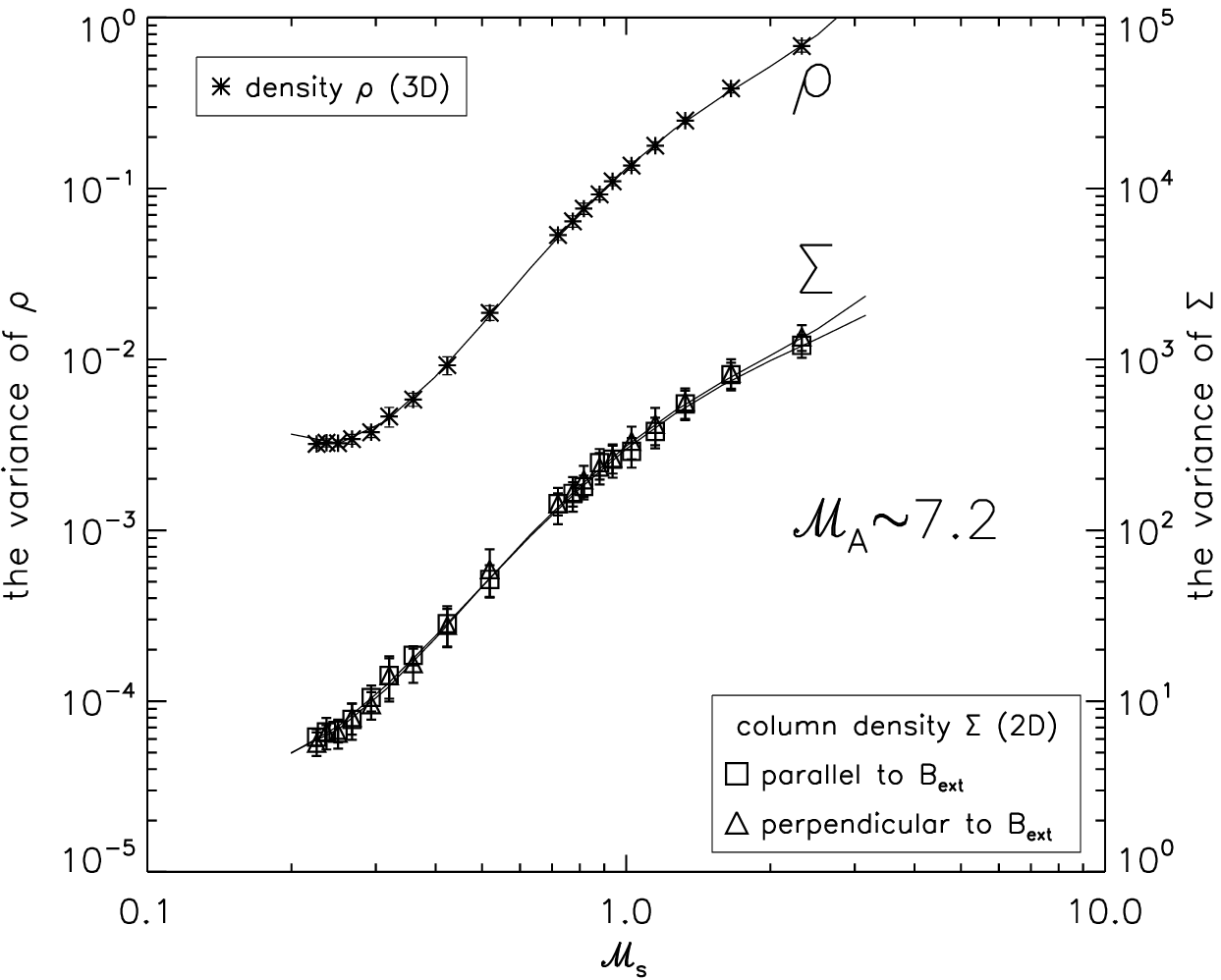}
 \plottwo{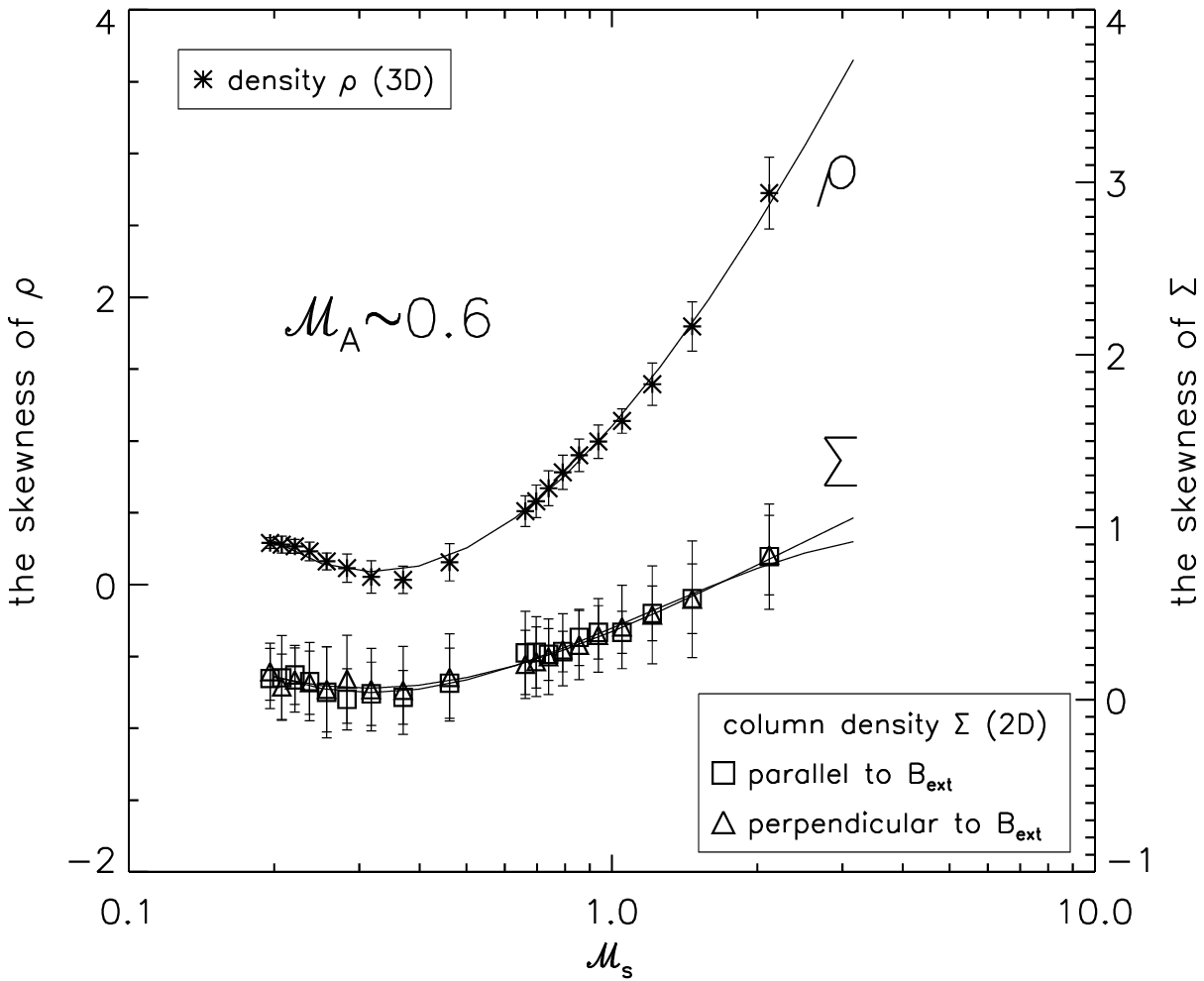}{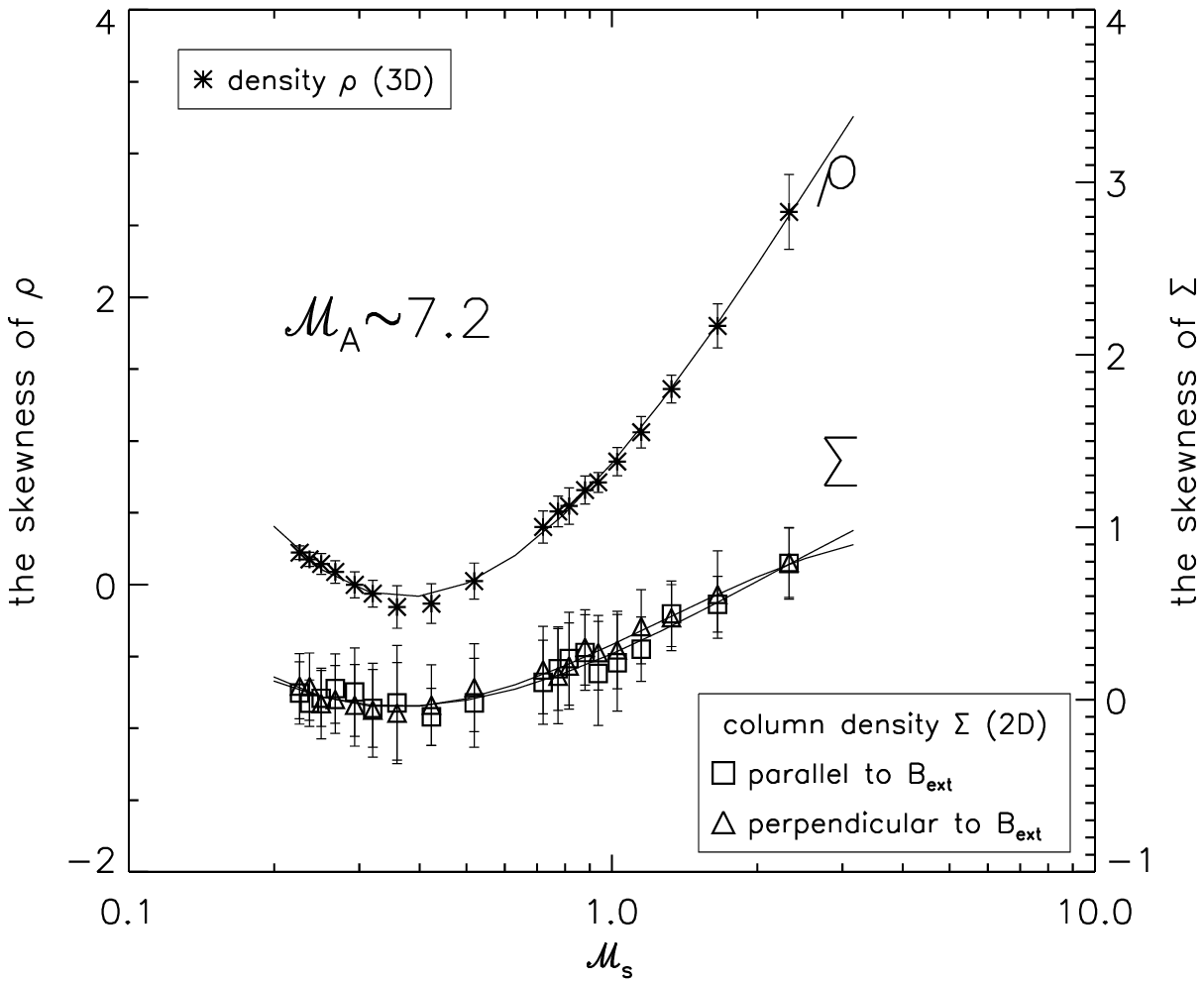}
 \plottwo{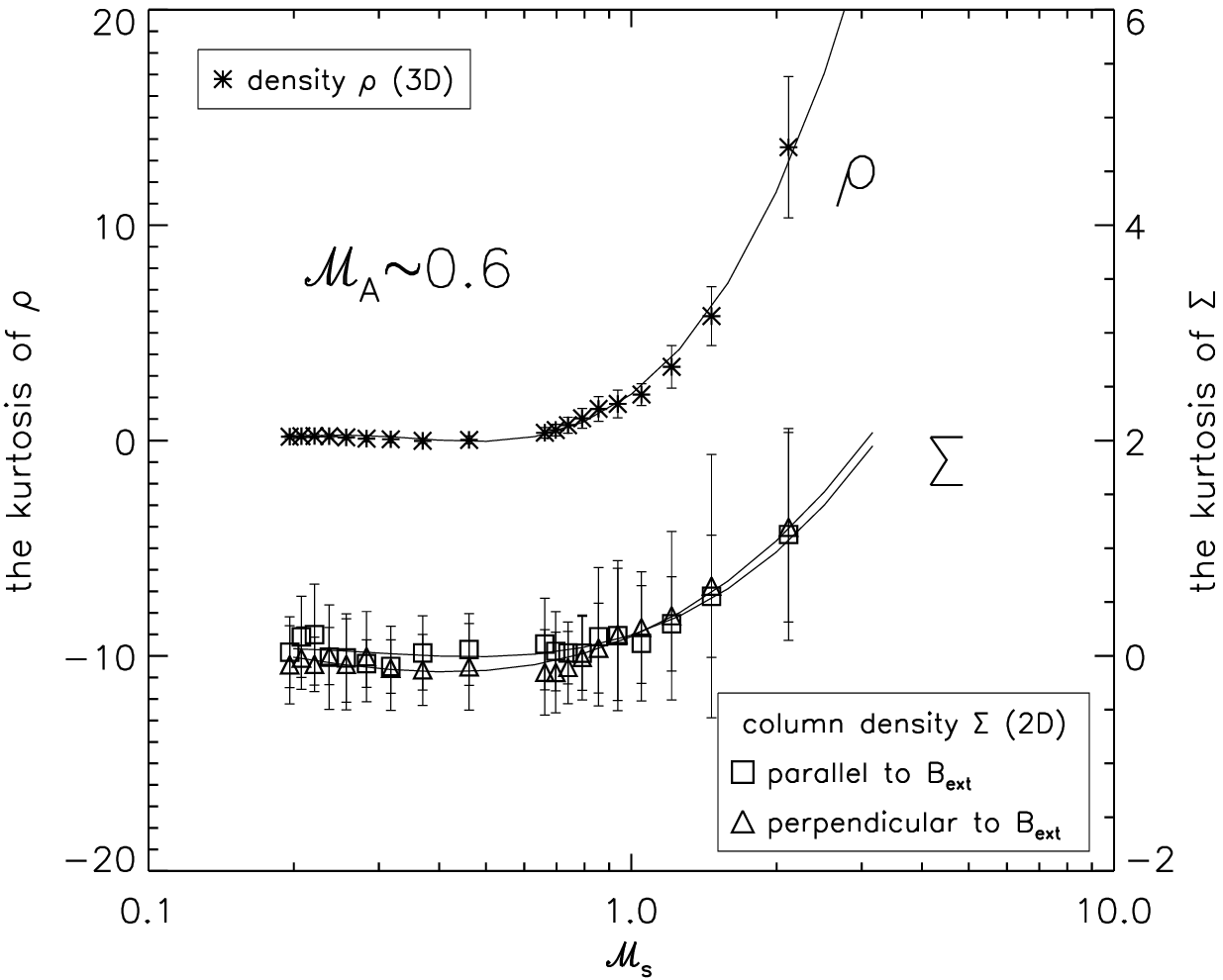}{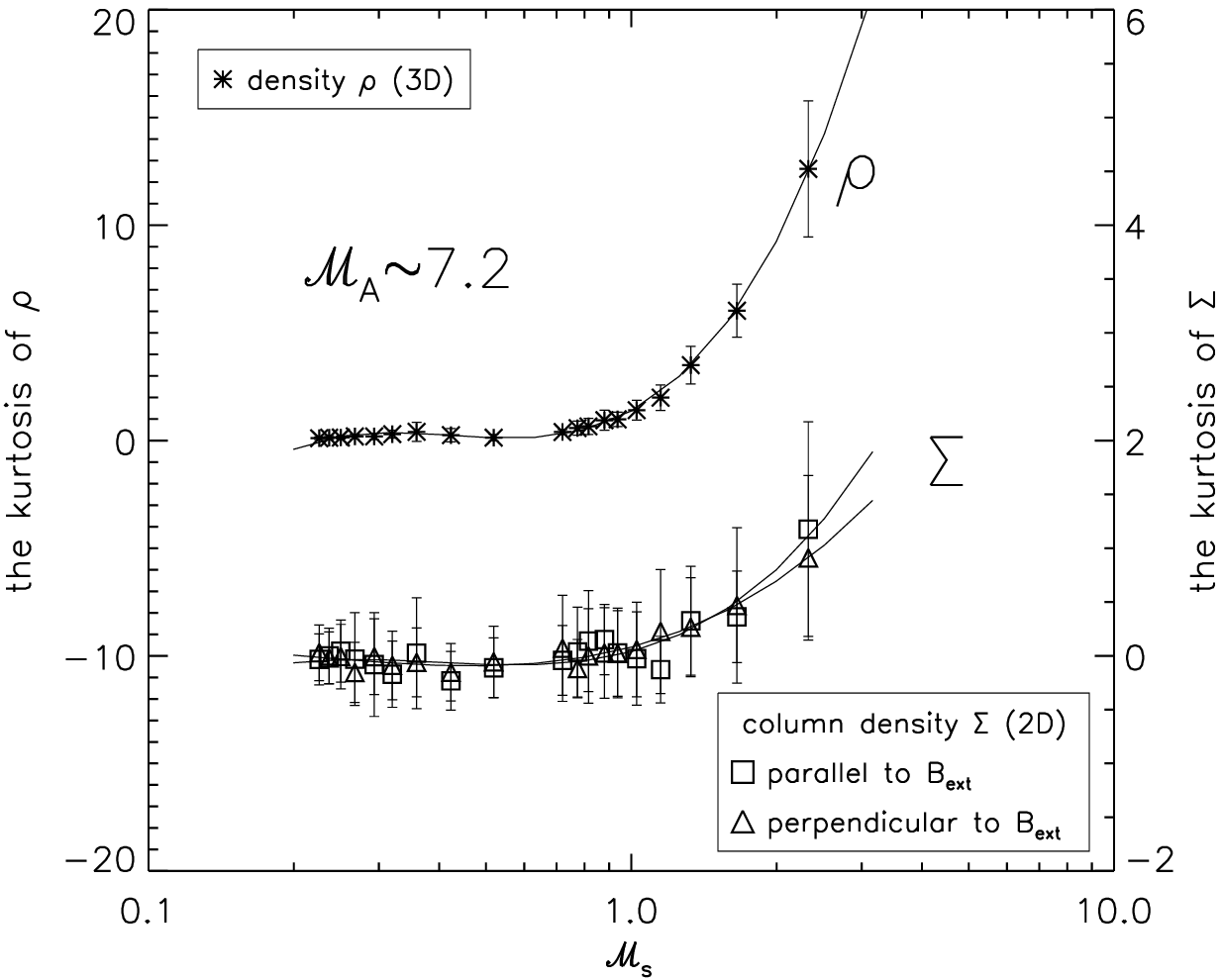}
 \caption{Statistical measures describing the distribution of density $\rho$ and column densities $\Sigma$ projected along and across the external magnetic field $B_\mathrm{ext}$ for models with ${\cal M}_A\sim0.6$ ({\em left column}) and ${\cal M}_A\sim7.2$ ({\em right column}) for models with low resolution (128$^3$). Plots in the top row show variances of the density distribution. In the middle row we show skewness and in the bottom row kurtosis. In each plot we combine lines for densities with a corresponding scale on the left axis of the graph, and lines for the column densities with the scale on the right axis of the graph. Solid lines signify a polynomial of fourth order fitted to the points. \label{fig:dens_moments}}
\end{figure}

In Figure \ref{fig:dens_moments} we show the variance, skewness and kurtosis ({\em top, middle, and bottom rows, respectively}) for densities and column densities for all low-resolution models (models with resolution 128$^3$, see Table \ref{tab:models}). From such a set of models we obtain the dependence of the moments on the sonic Mach number. Error bars are determined using the standard deviation of the results for each model from all subsequent steps for times $t\in[5,25]$.

\begin{figure}  
 \epsscale{0.9}
 \plottwo{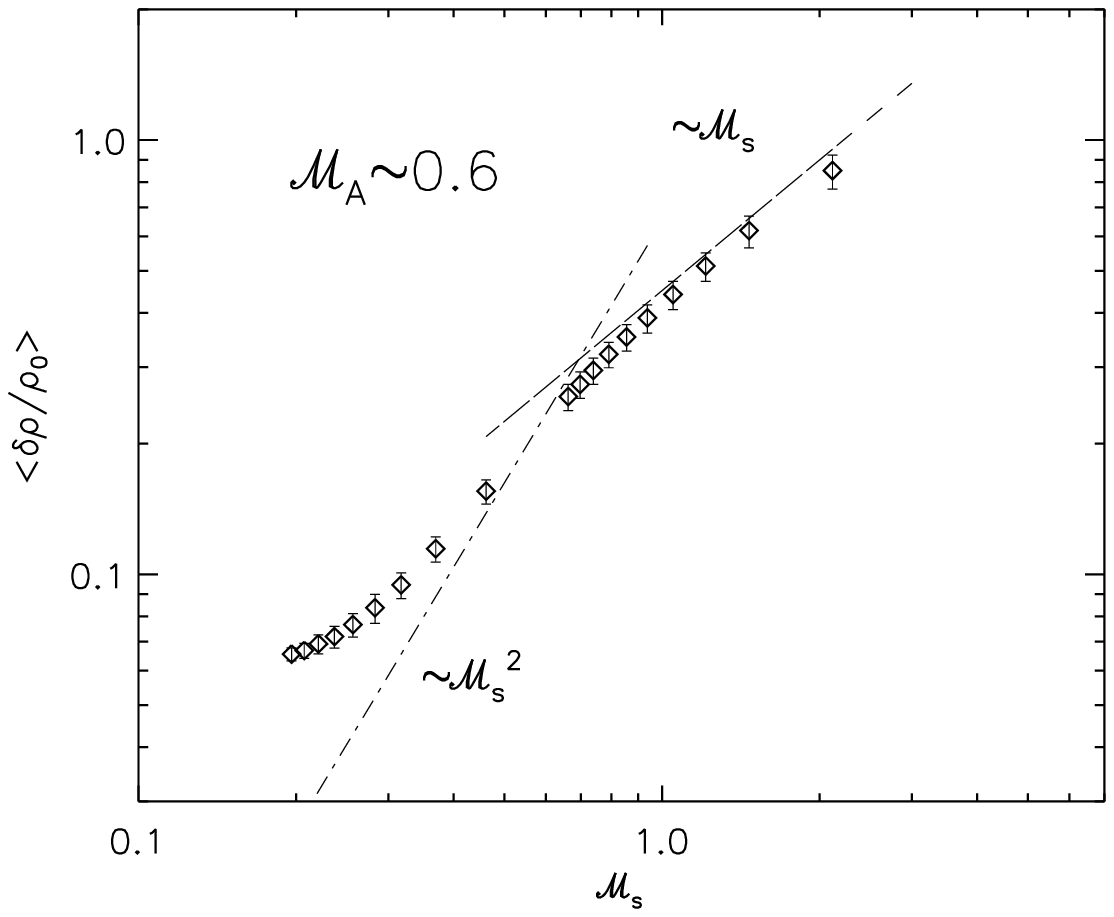}{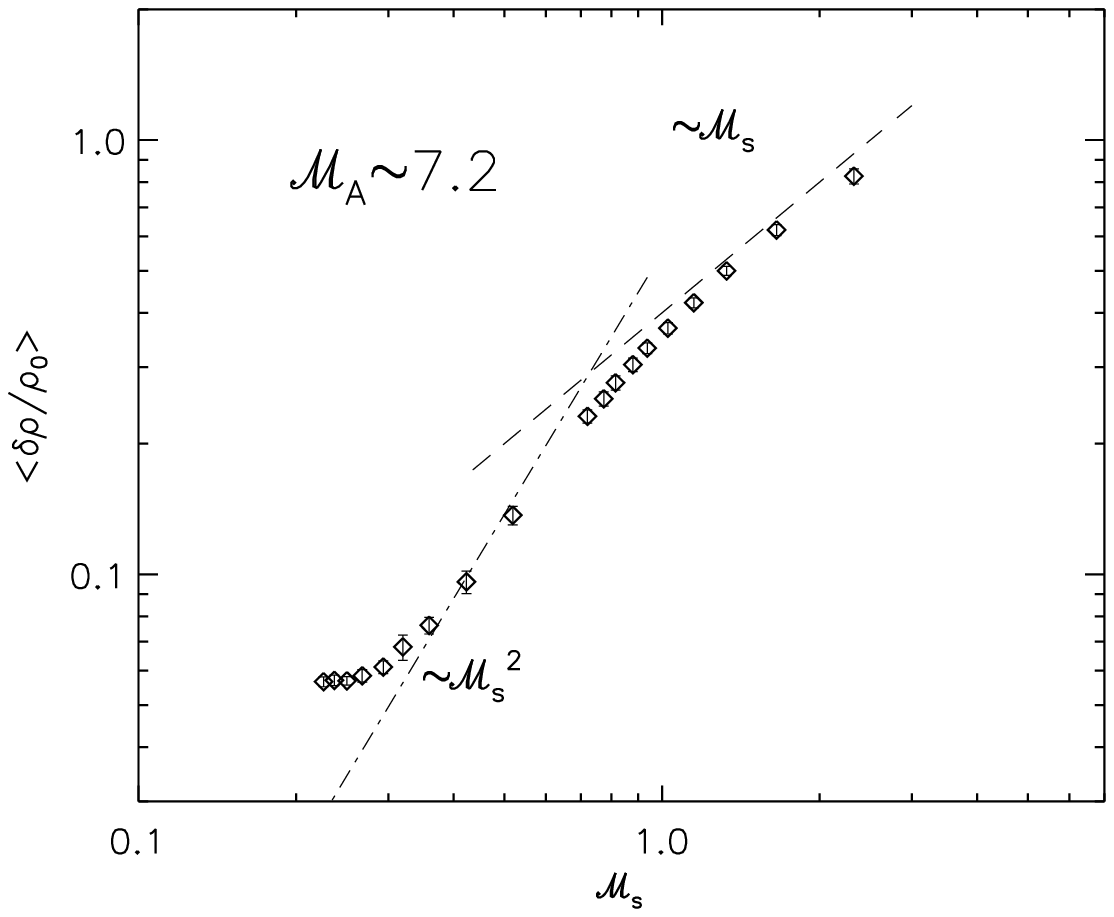}
 \caption{Relation between the mean value of the standard deviation of density fluctuation $\langle \delta \rho/\rho_0 \rangle$ and the sonic Mach number ${\cal M}_s$ for models with low resolution. Two lines show the analytical relations derived for magnetic-pressure-dominated turbulence ($\sim {\cal M}_s$, {\em dashed line}) and the gas-pressure-dominated turbulence ($\sim {\cal M}_s^2$, {\em dot-dashed line}). \label{fig:var_mach}}
\end{figure}

The variance of density fluctuations has been studied extensively, so its relation to ${\cal M}_s$ is well known \cite[e.g.]{nordlund99,ostriker01,cho03}. The variance of density grows with the sonic Mach number and consequently with the broadening of the PDF. The same behavior is observed in the variance of column densities. Lines for variances in Figure~\ref{fig:dens_moments} look very similar for models with strong and weak magnetic fields. However, the comparison of the variances for column densities integrated along and across the external magnetic field reveals a difference in its values, a sort of split, in the case of the sub-Alfv\'{e}nic models (see Fig. \ref{fig:dens_moments}, {\em top left}). In the case of super-Alfv\'{e}nic turbulence, there is no split at all. Lines perfectly overlap. The latter is natural, as the super-Alfv\'{e}nic turbulence easily tangles the external magnetic field. According to \cite{cho03}, the relations between the variance of density fluctuations and the sonic Mach number are $\delta \rho/\rho_0 \sim {\cal M}_s$ when the magnetic pressure dominates and $\delta \rho/\rho_0 \sim {\cal M}_s^2$ when the gas pressure dominates. The former case is observed when ${\cal M}_s \gg {\cal M}_A$. Indeed, in Figure~\ref{fig:var_mach} we see that the mean standard deviation of the density fluctuations $\langle \delta \rho/\rho_0 \rangle$ scales with ${\cal M}_s$ almost linearly when ${\cal M}_s > {\cal M}_A$. When ${\cal M}_s < {\cal M}_A$, the relation is much steeper. A similar behavior is observed in the case of super-Alfv\'{e}nic turbulence (Fig.~\ref{fig:var_mach}, {\em right}) although the gas pressure dominates in all models. For models with a very small value of the sonic Mach number, the relation is different. The value of $\langle \delta \rho/\rho_0 \rangle$ depends much less on ${\cal M}_s$.

Studies by \cite{passot01} showed the dependence of density fluctuations on the Alfv\'{e}nic Mach number, which is different for different modes. When ${\cal M}_{A}$ is small, the slow mode dominates with the relation $B^2 \sim \rho$. The fast mode starts to contribute more significantly in the generation of density fluctuations with the relation $B^2 \sim \rho^2$ when ${\cal M}_{A}$ is large. A split in the $\sigma^2_{\Sigma}$--${\cal M}_s$ relation for column densities parallel and perpendicular to $B_\mathrm{ext}$ in the sub-Alfv\'{e}nic turbulence (Fig. \ref{fig:dens_moments}, {\em top left}) can be explained by a correlation of clumps and filaments of the density structure with the direction of the mean magnetic field resulting from the action of the slow mode as a main contributor to the generation of density fluctuations. In the super-Alfv\'{e}nic models, the variance of the column density is independent of the direction of the LOS (compare to Fig.~\ref{fig:dens_moments}, {\em top right}). Plots of variances show that the strength of sonic motions can be directly measured from the variance of intensity in the maps of molecular clouds.

More interesting information is provided by the two other statistical measures: skewness and kurtosis. In Figure \ref{fig:dens_moments} ({\em middle row}) we see that asymmetry of density and column density distributions grows with the value of ${\cal M}_s$, but only if ${\cal M}_s\gtrsim0.5$. When the sonic Mach number falls below this value, the asymmetry of the distribution starts to grow again. The skewness of column density varies in the same way with the minimum placed exactly at the same value of ${\cal M}_s$. In all cases, the skewness has positive values, which is interpreted as an asymmetry of the distribution into the direction of higher densities. The skewness is rather insensitive to the strength of the external magnetic field. The kurtosis, presented in Figure \ref{fig:dens_moments} ({\em bottom row}), oscillates around zero for models with highly subsonic motions. This suggests an almost perfect Gaussian distribution for density and column density. It is also in agreement with the fact that if density is perturbed weakly, its distribution is normal. The growth of kurtosis with ${\cal M}_s$ and its positive values signify a distribution ``peaked'' more than the standard normal distribution. All three measures reveal strong dependence on the sonic Mach number. Moreover, the utility of the above measures for the column density follows from the similarity to the corresponding measures of 3D density.

\section{Additional considerations}
\label{sec:considerations}

\subsection{How does the cloud boundary influence the statistics?}
\label{sec:boundary}

\begin{figure}  
 \epsscale{0.32}
 \plotone{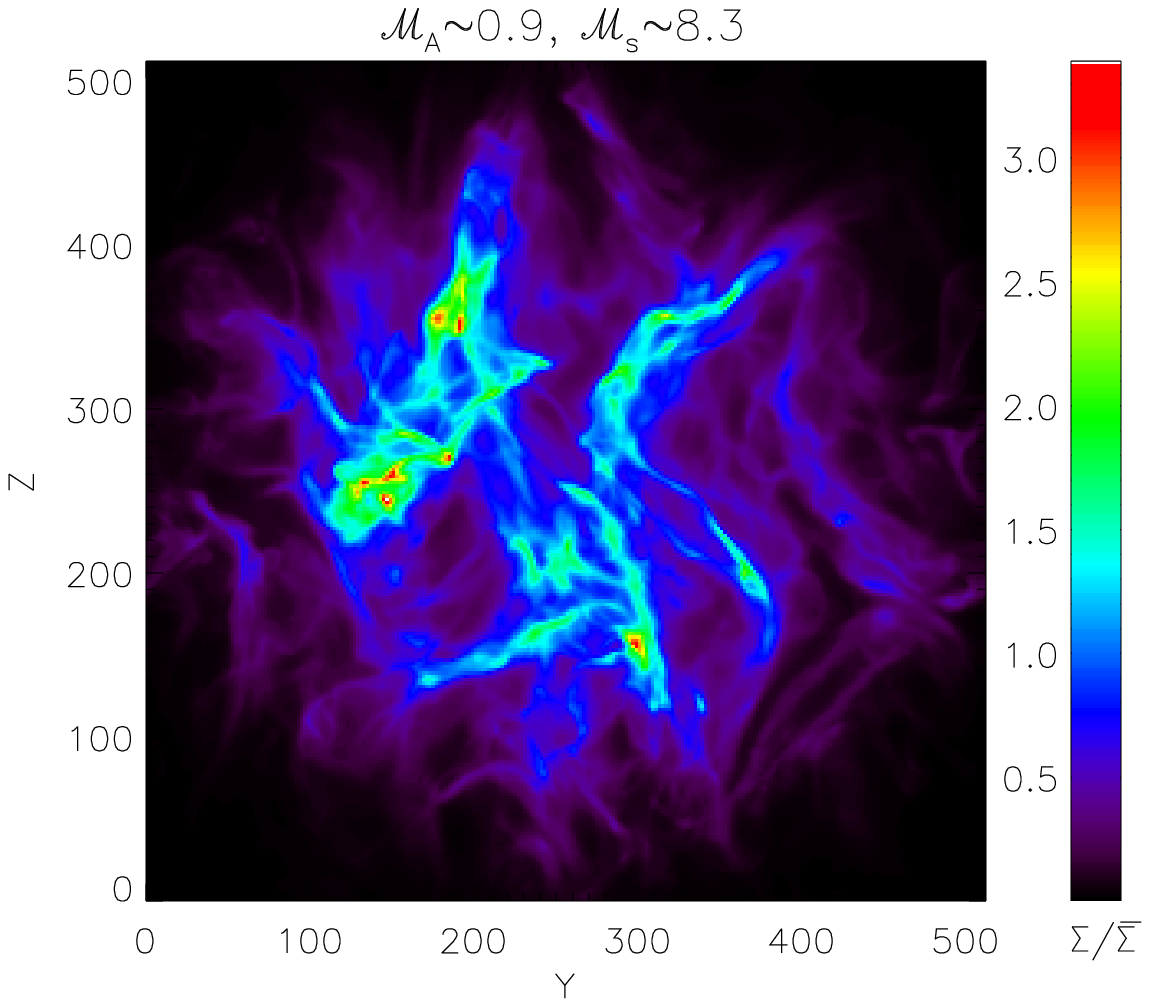}
 \plotone{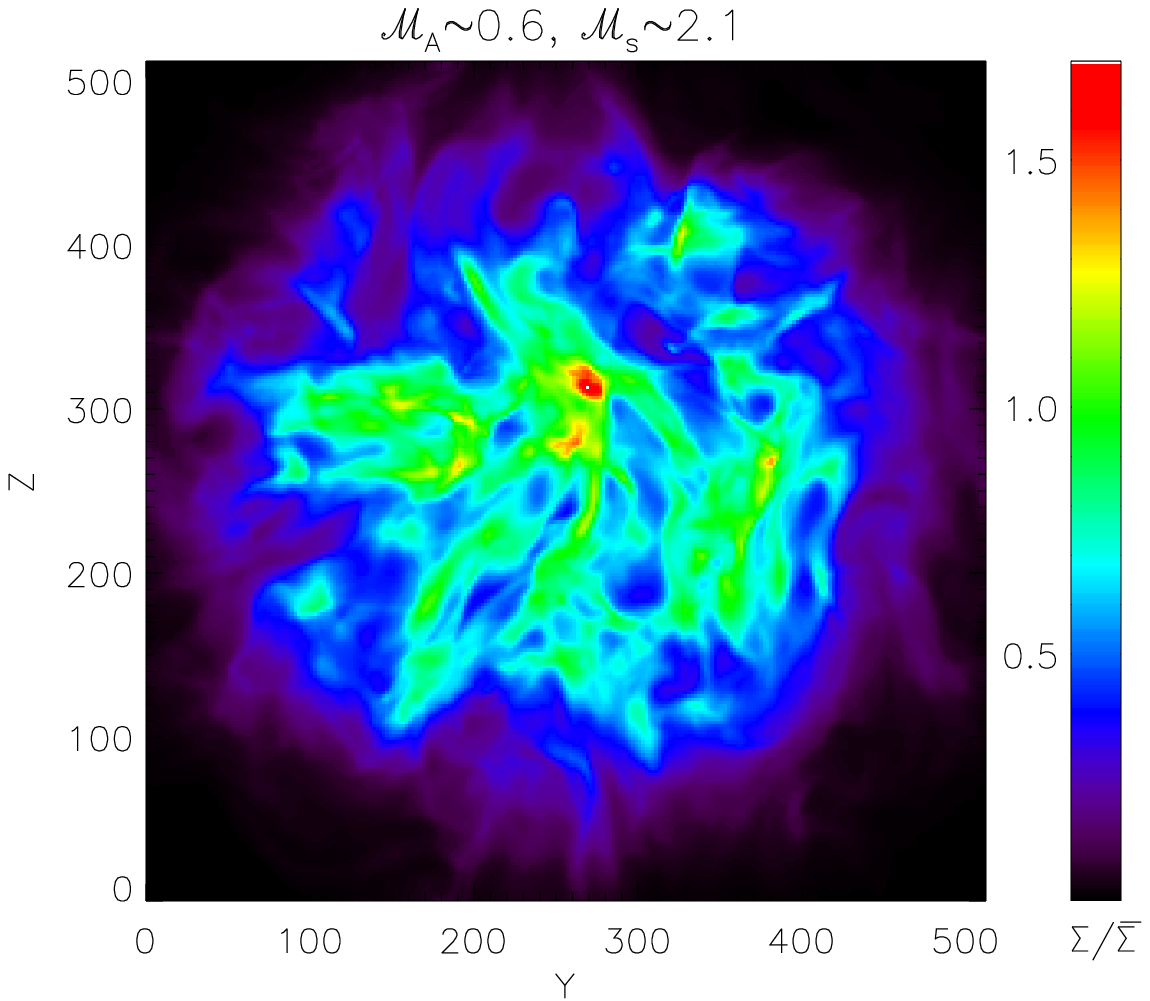}
 \plotone{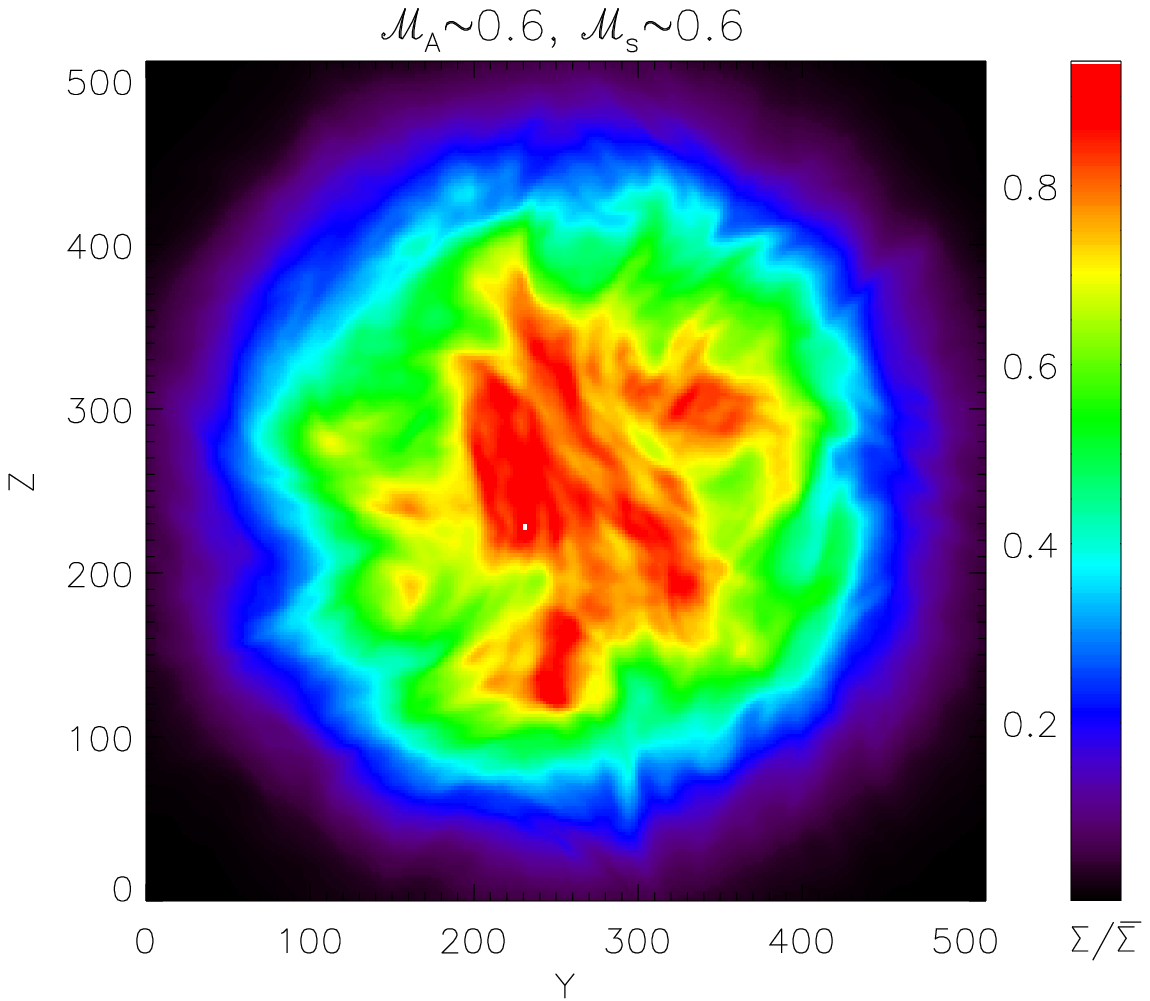}
 \plotone{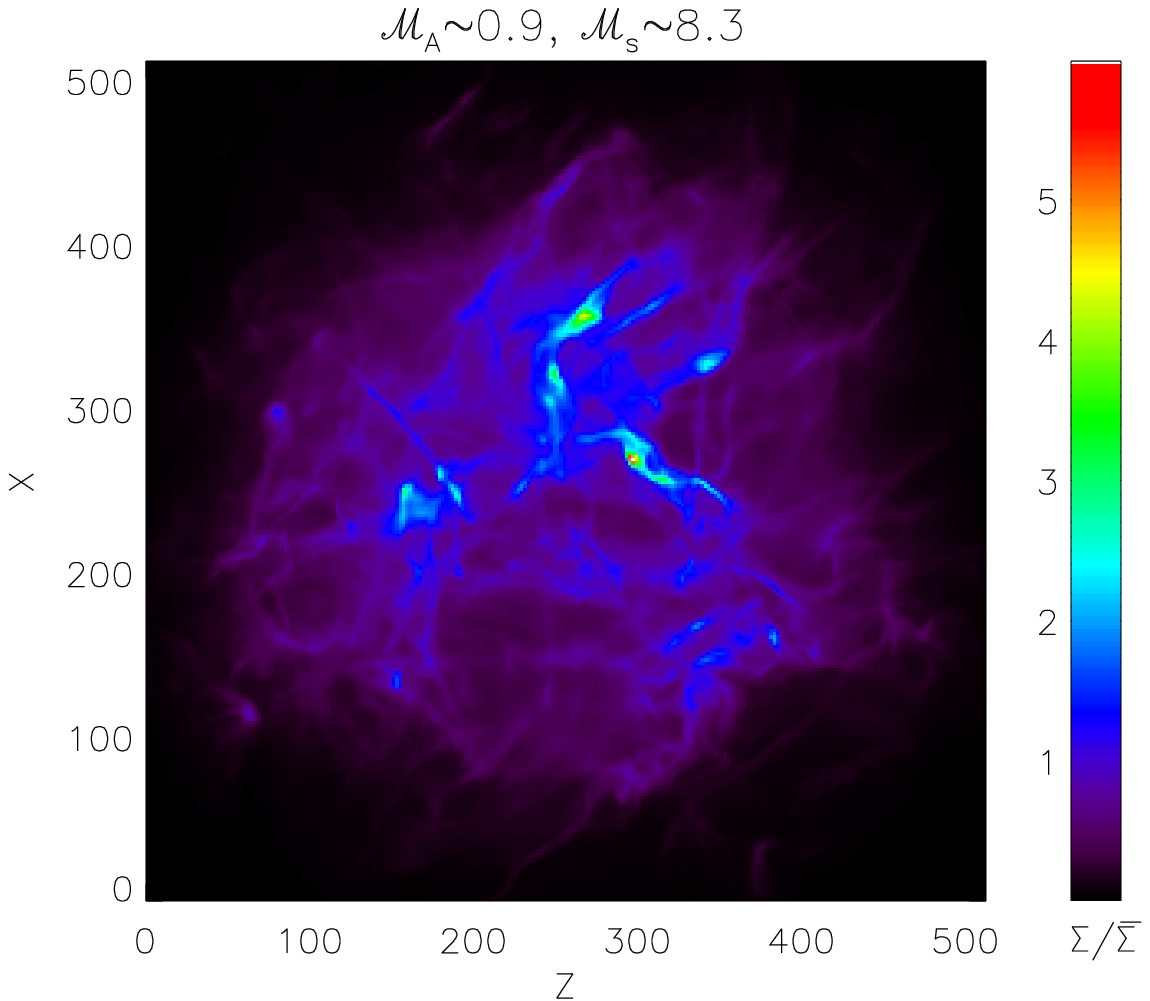}
 \plotone{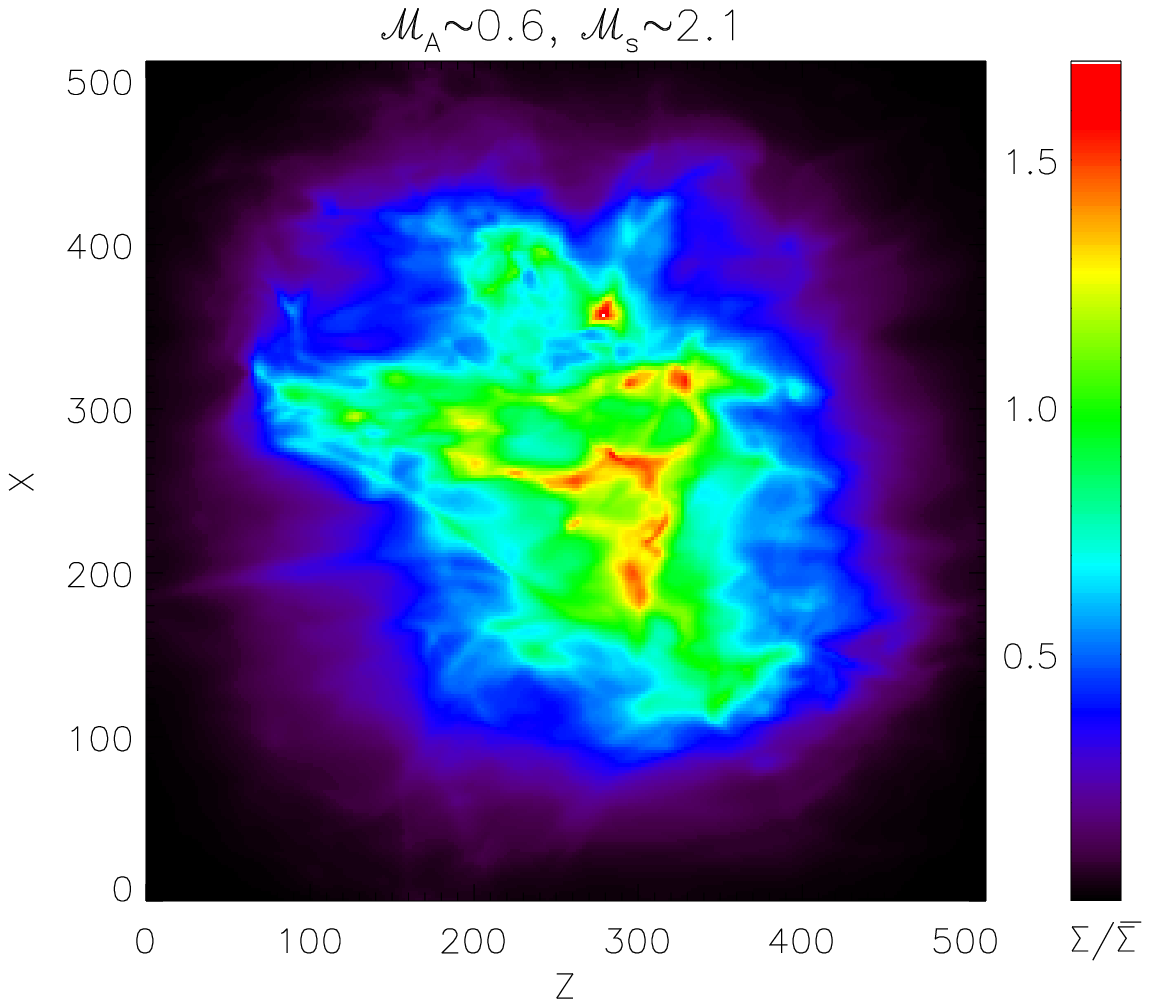}
 \plotone{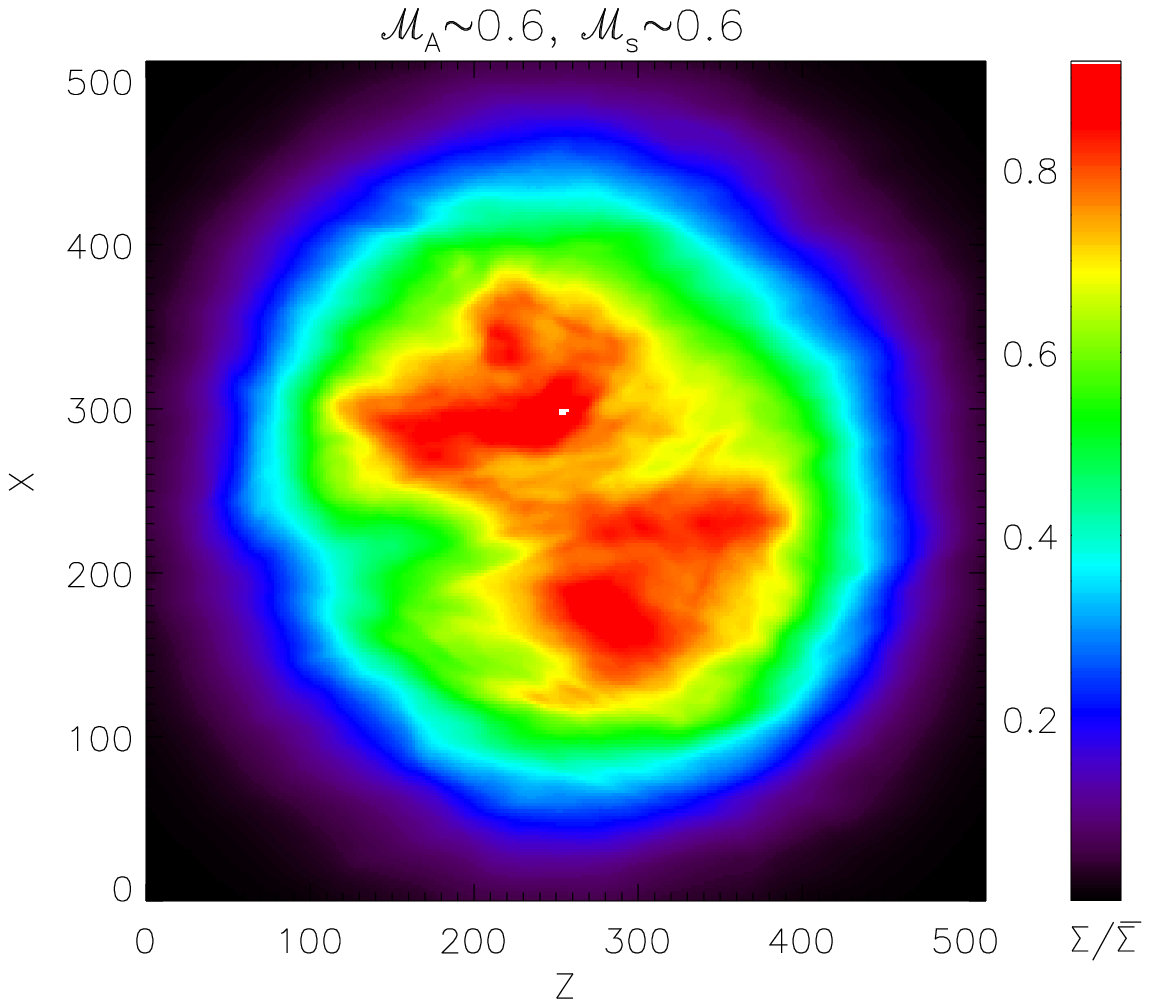}
 \caption{Examples of generated synthetic maps for high-resolution models with the same strength of the external magnetic field and different sonic Mach numbers (${\cal M}_s=8.3$, $2.1$, $0.6$ {\em for left, middle and, right columns, respectively}). Plots in the top row show column densities integrated along $B_\mathrm{ext}$, and in the bottom row, densities are integrated across $B_\mathrm{ext}$. \label{fig:synth_maps}}
\end{figure}

To connect our studies with column densities available from observations, we transform our density cube to a cloudlike shape. This operation has been done by a simple multiplication of the density by a 3D function, which is uniform within a ball of radius $R$ and decaying outside $R$ with a Gaussian profile. In the next step we calculated column densities from the resulting data and performed an analysis in the same way as was previously done. In this simple way we obtain the synthetic map of a molecular cloud. We should emphasize here that the results obtained from the synthetic observations of this type are mostly applicable to supersonic turbulence. The restriction is due to the fact that the density fluctuations in subsonic turbulence are small in comparison to the mean cloud profile, so the two-point statistics may be dominated by this large-scale distribution. Therefore, in this section we narrow down our studies only to the supersonic experiments. Examples of maps obtained after applying the above procedure are presented in Figure \ref{fig:synth_maps}. The change in structure is visible ``by eye''. For highly supersonic motions (Fig.~\ref{fig:synth_maps}, {\em left column}), the column density exposes filamentary structure, which is created by strong shocks. The overall shape of the cloud is difficult to determine. In the middle column of Figure~\ref{fig:synth_maps}, for medium sonic Mach numbers, the overall shape of the cloud starts to appear. The maps still show filaments but in a much more compact form with many spongelike elements. For subsonic turbulence, the cloud looks like a ball of cotton. We can recognize some whirled streaks. In this case, the mean shape of the cloud clearly dominates.

\begin{figure}  
 \epsscale{0.9}
 \plottwo{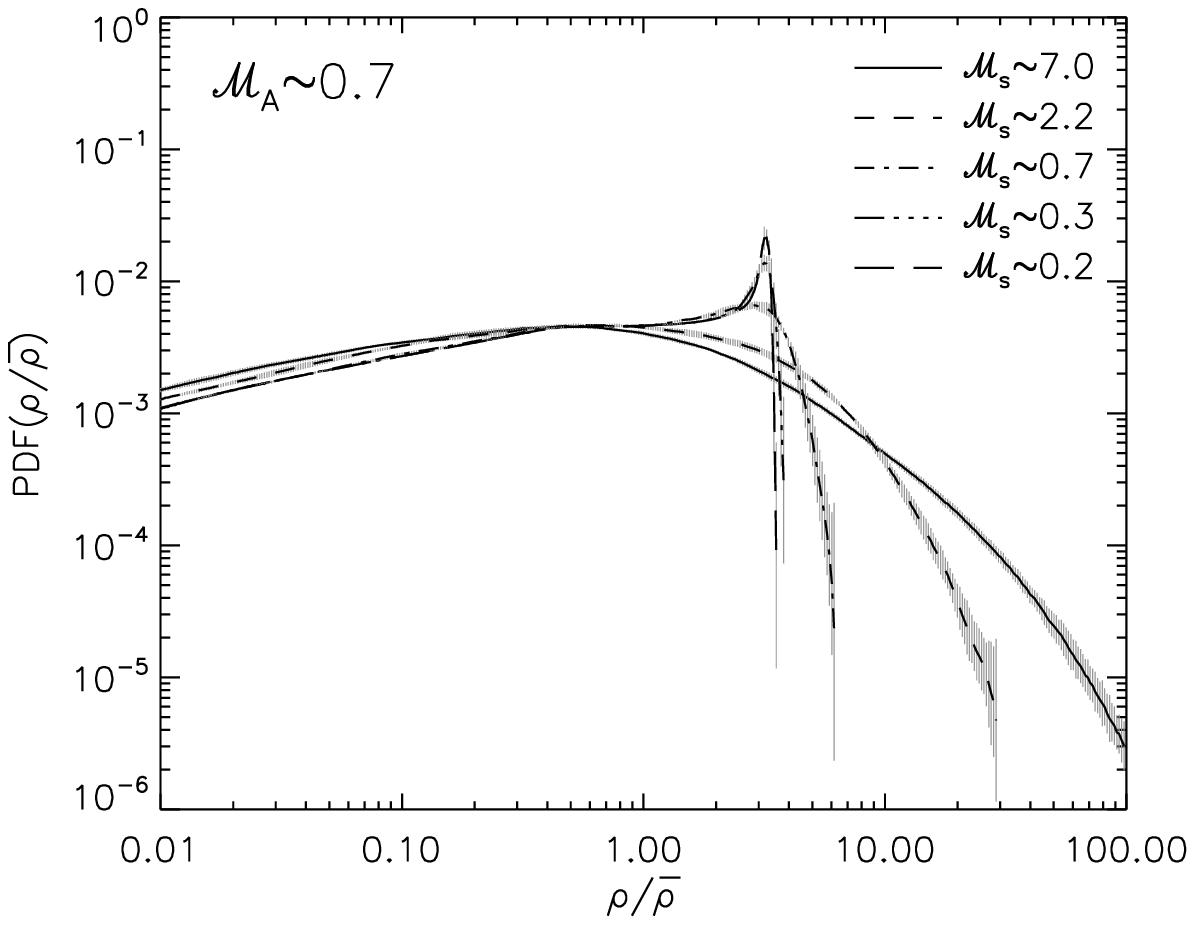}{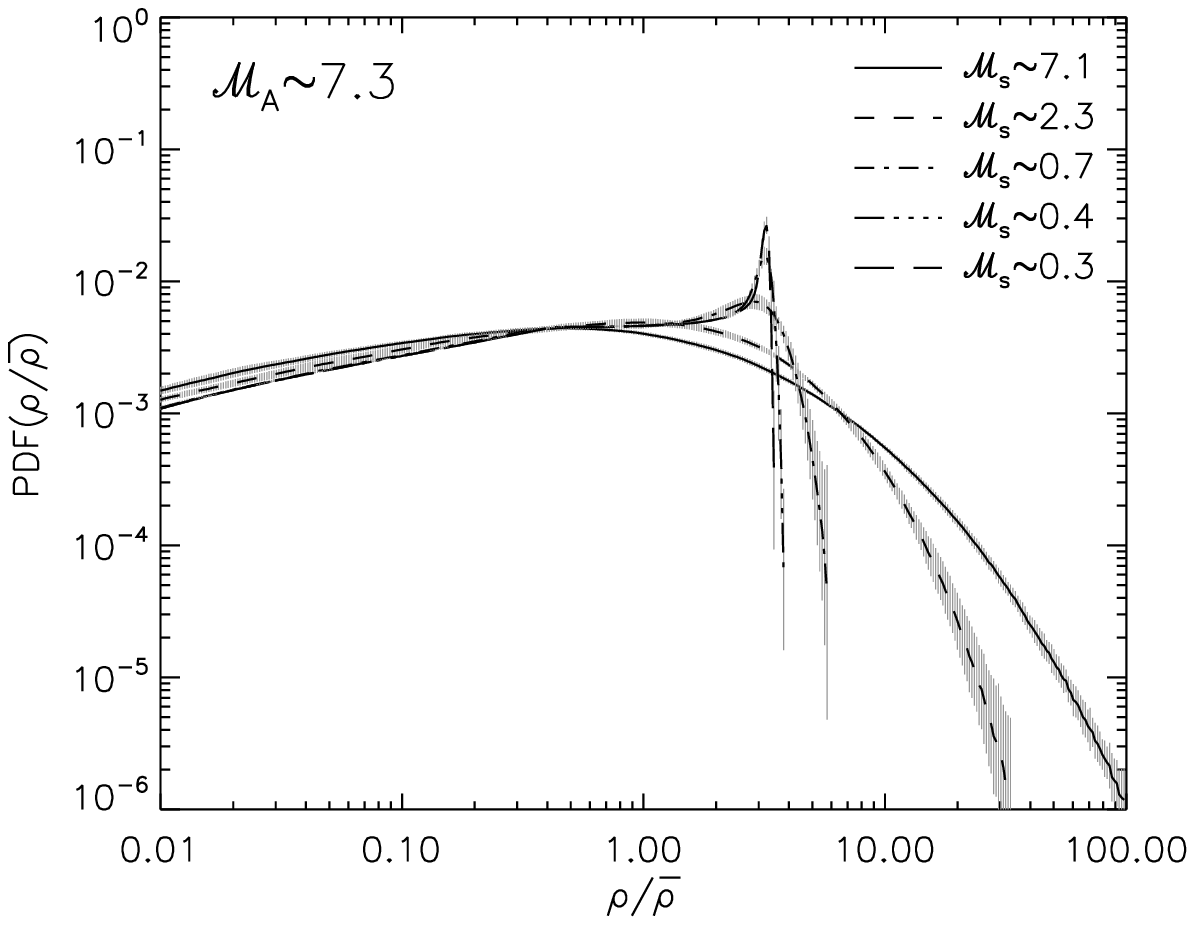}
 \plottwo{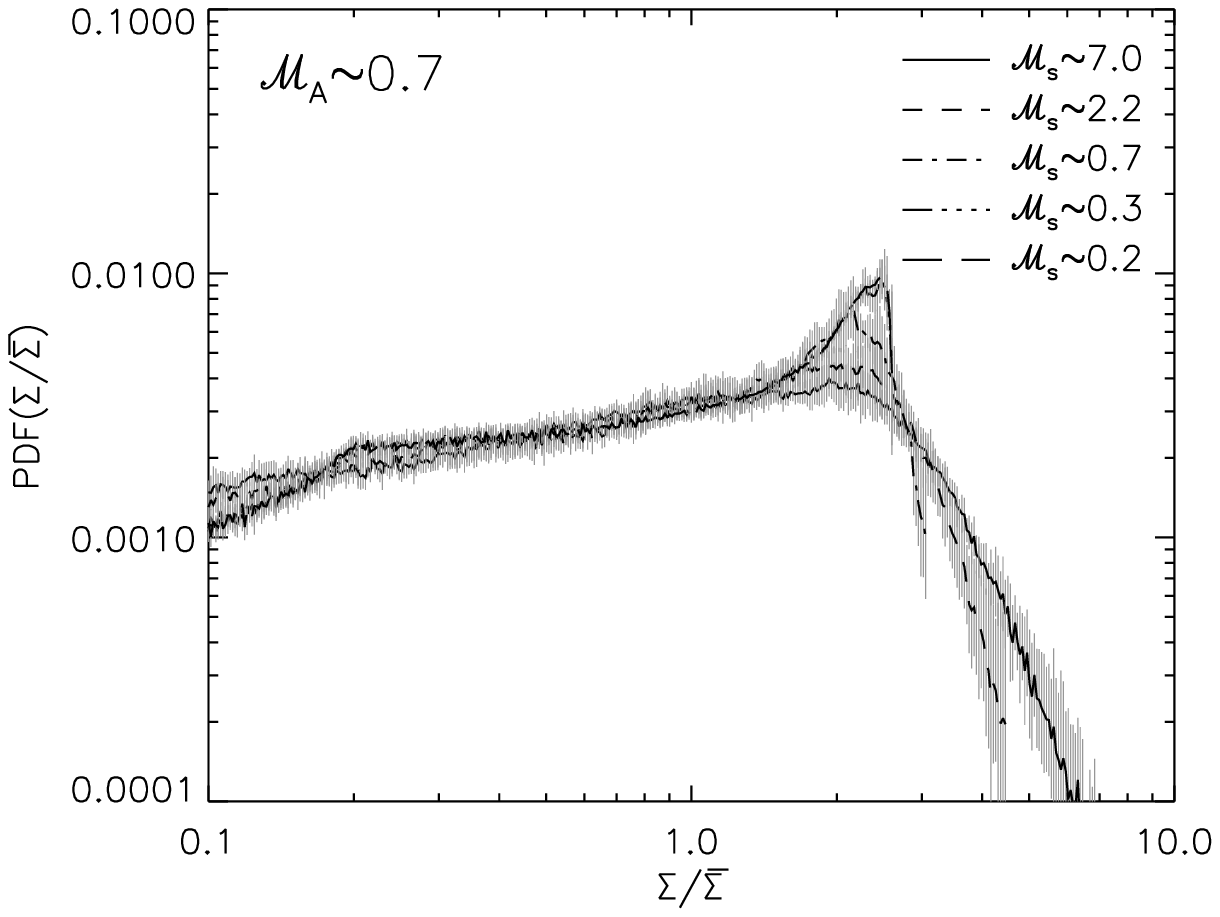}{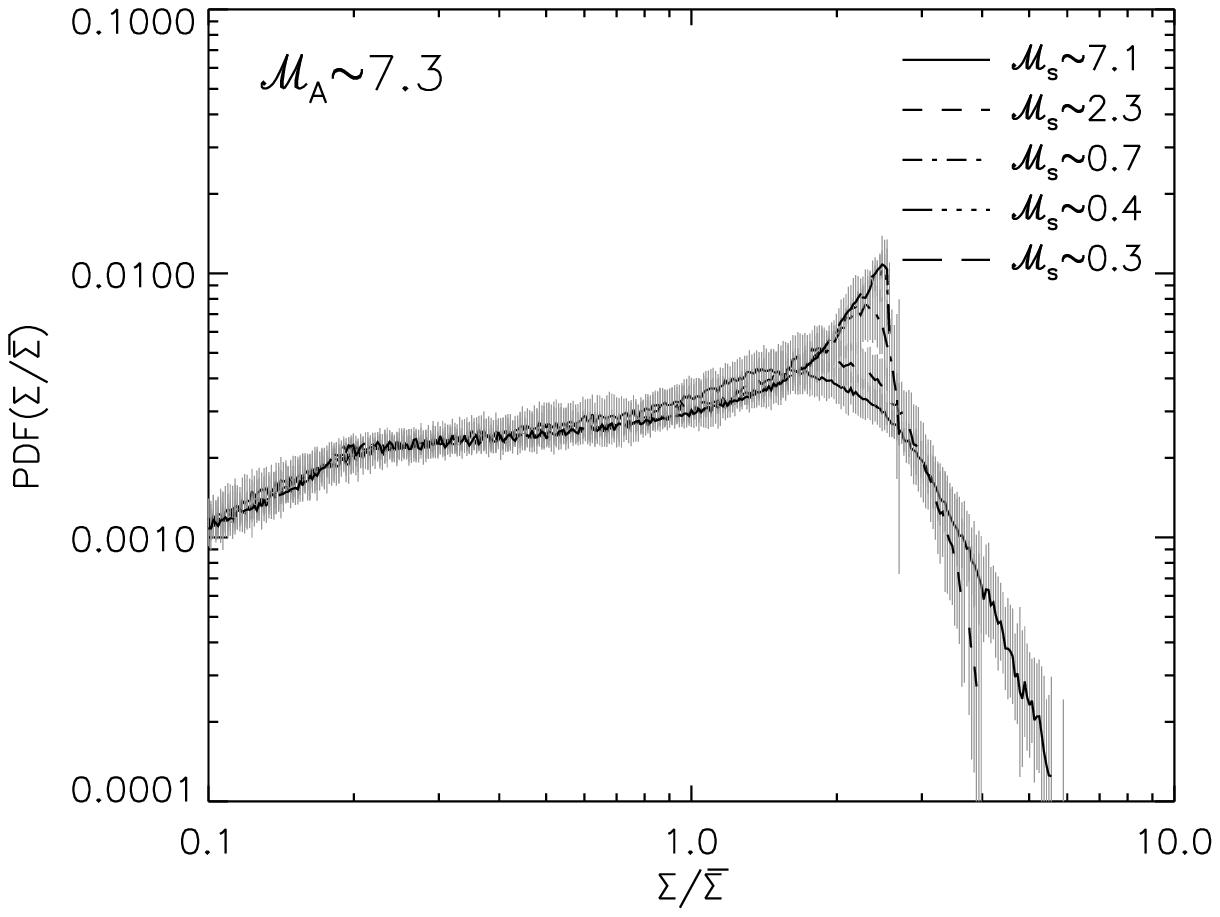}
 \plottwo{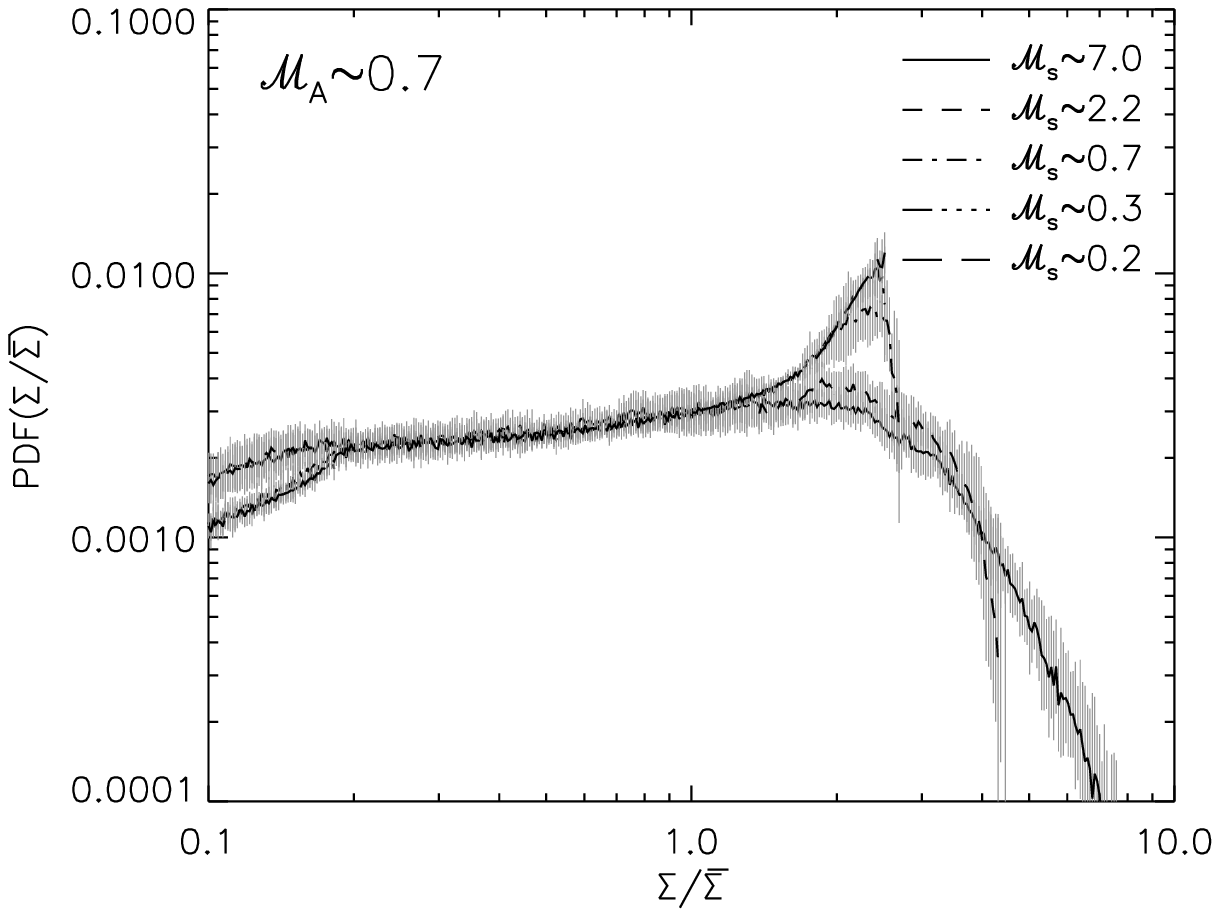}{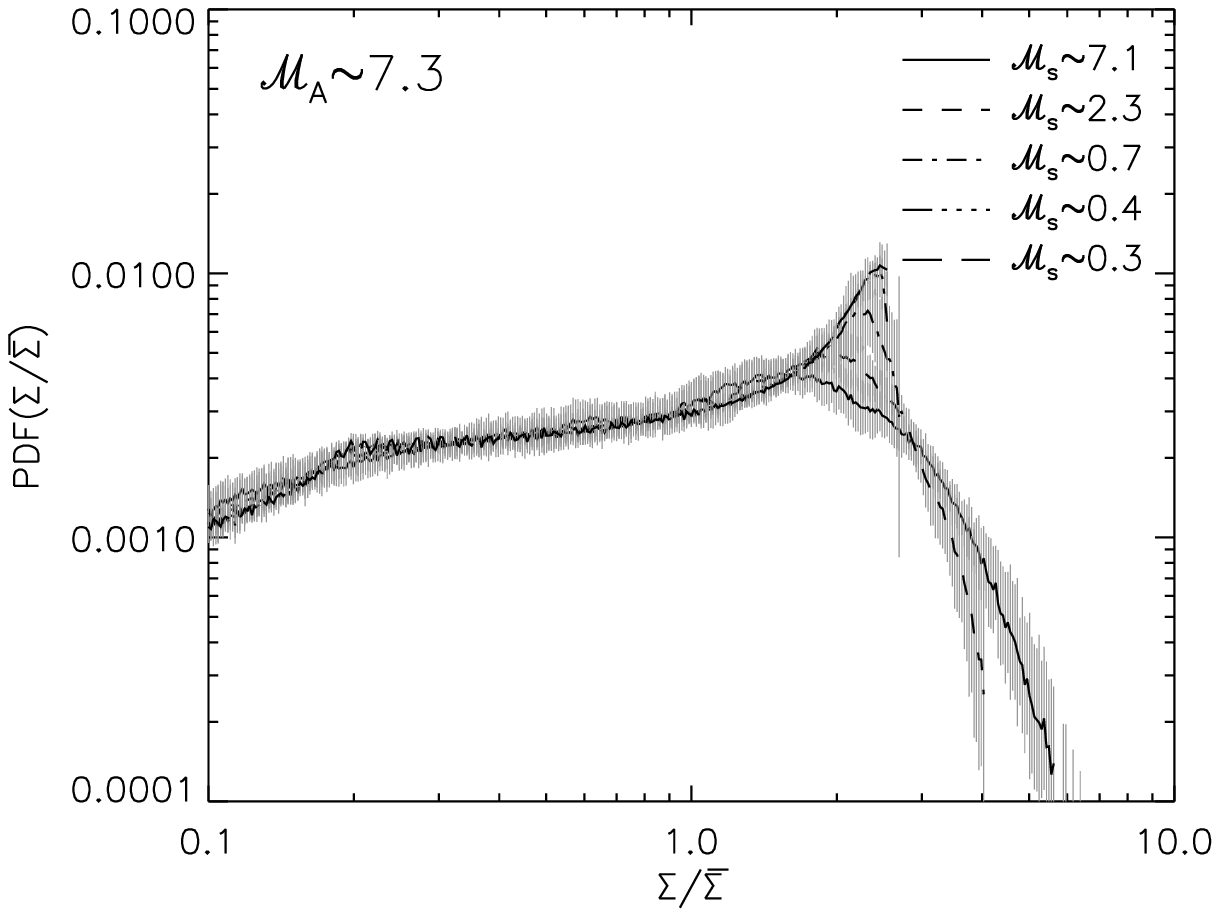}
 \caption{PDFs of density $\rho$ shaped in the form of a cloud ({\em top row}) and with its column densities $\Sigma$ projected along and across the external magnetic field $B_\mathrm{ext}$ ({\em middle and bottom rows, respectively}) for sub-Alfv\'{e}nic and super-Alfv\'{e}nic turbulence ({\em left and right columns, respectively}). We used models with medium resolution here. \label{fig:dens_sph_pdfs}}
\end{figure}

To answer the question concerning the influence of boundaries on the statistics, we start by presenting the PDFs of column densities (Figure \ref{fig:dens_sph_pdfs}). All models have distributions very different from those presented in Figure \ref{fig:dens_pdfs}. The PDFs are very far from the symmetry with respect to the mean value. The shape of the low-density part is influenced mainly by the boundaries of the cloud. The boundary of a cloud is an interface where the density decreases from the mean value of the interior of the cloud to the density of the ambient medium, so this explains why the shape of the low-density part looks very similar in all models independent of the Mach number. However, if we take a look at the high-density part of the PDFs, we see that here the shape does depend on the sonic Mach number. Because the densities of the ambient medium are too small to disturb this part of the PDF, it is clear the these are formed only by the internal structure of the cloud. Thus, for our cloud model, the PDFs from the higher density part of the cloud are more reliable.

\begin{figure}  
 \epsscale{0.9}
 \plottwo{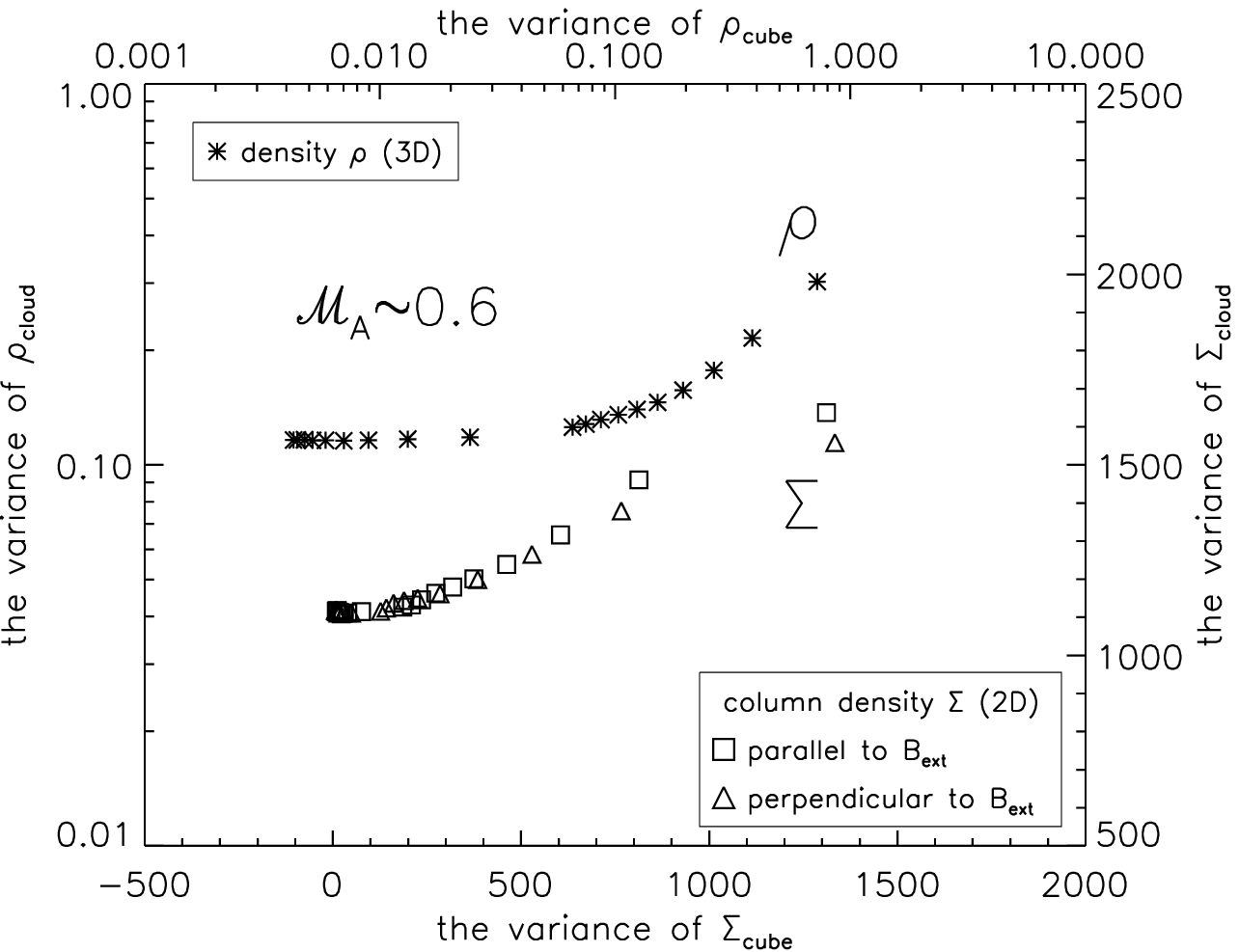}{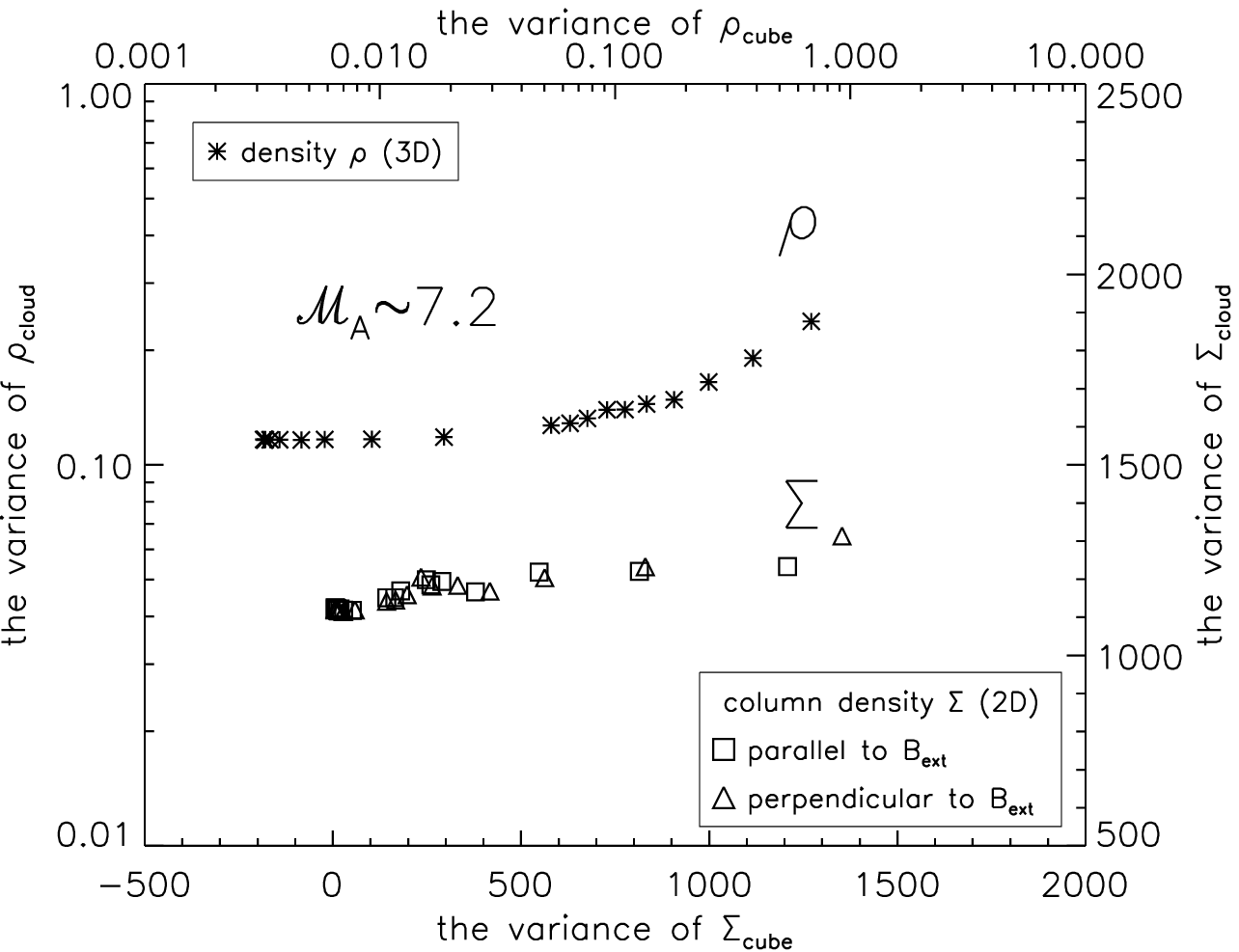}
 \plottwo{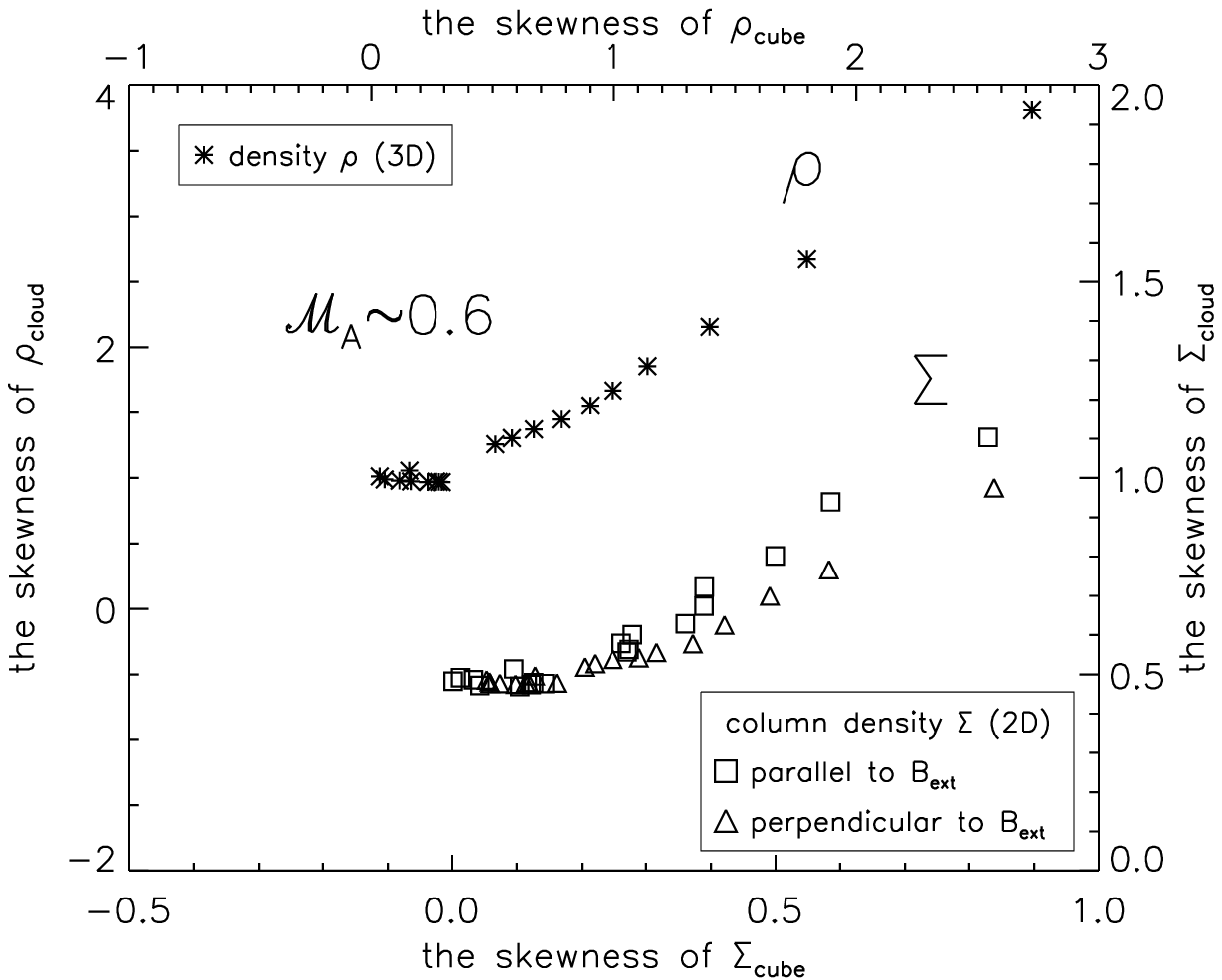}{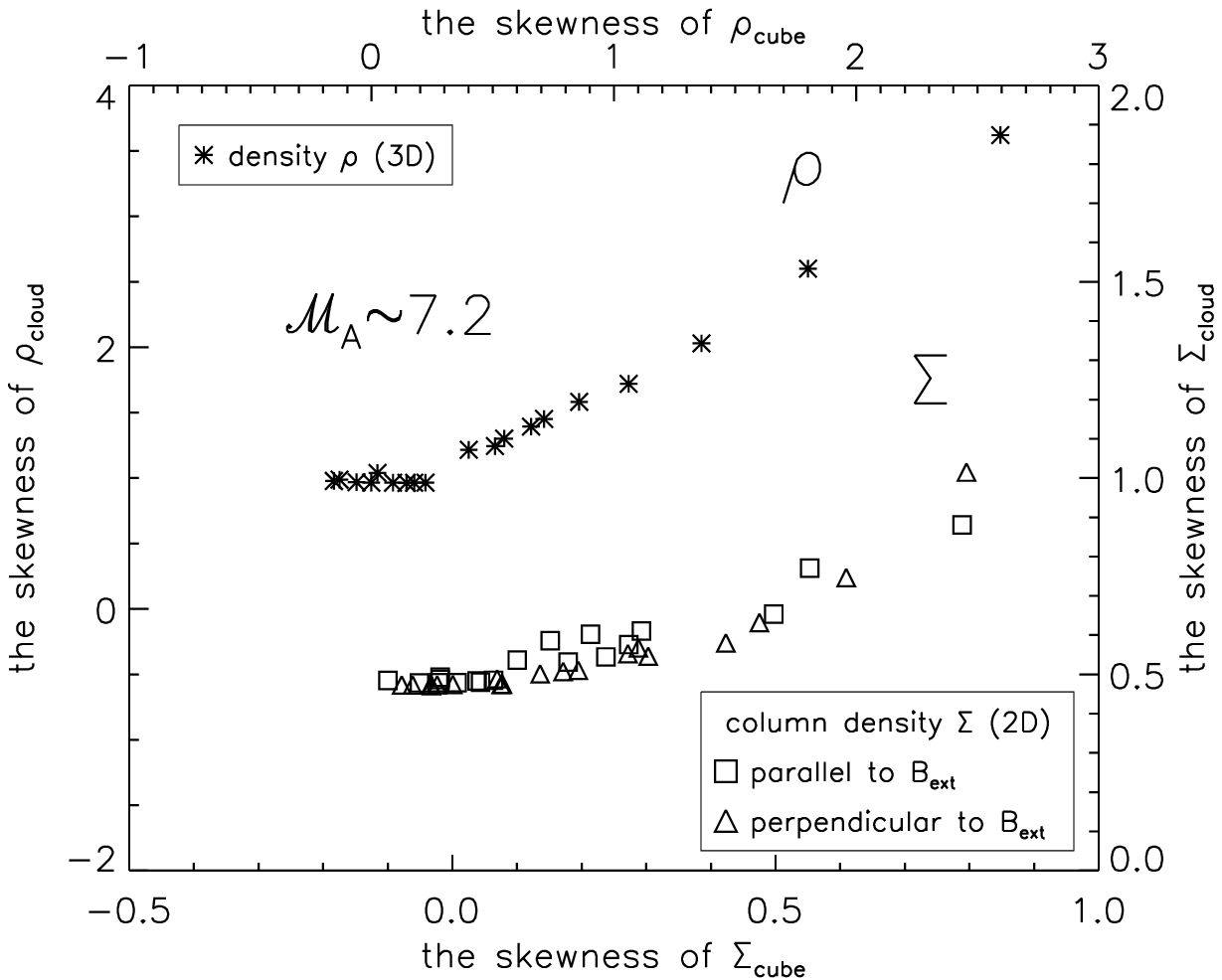}
 \plottwo{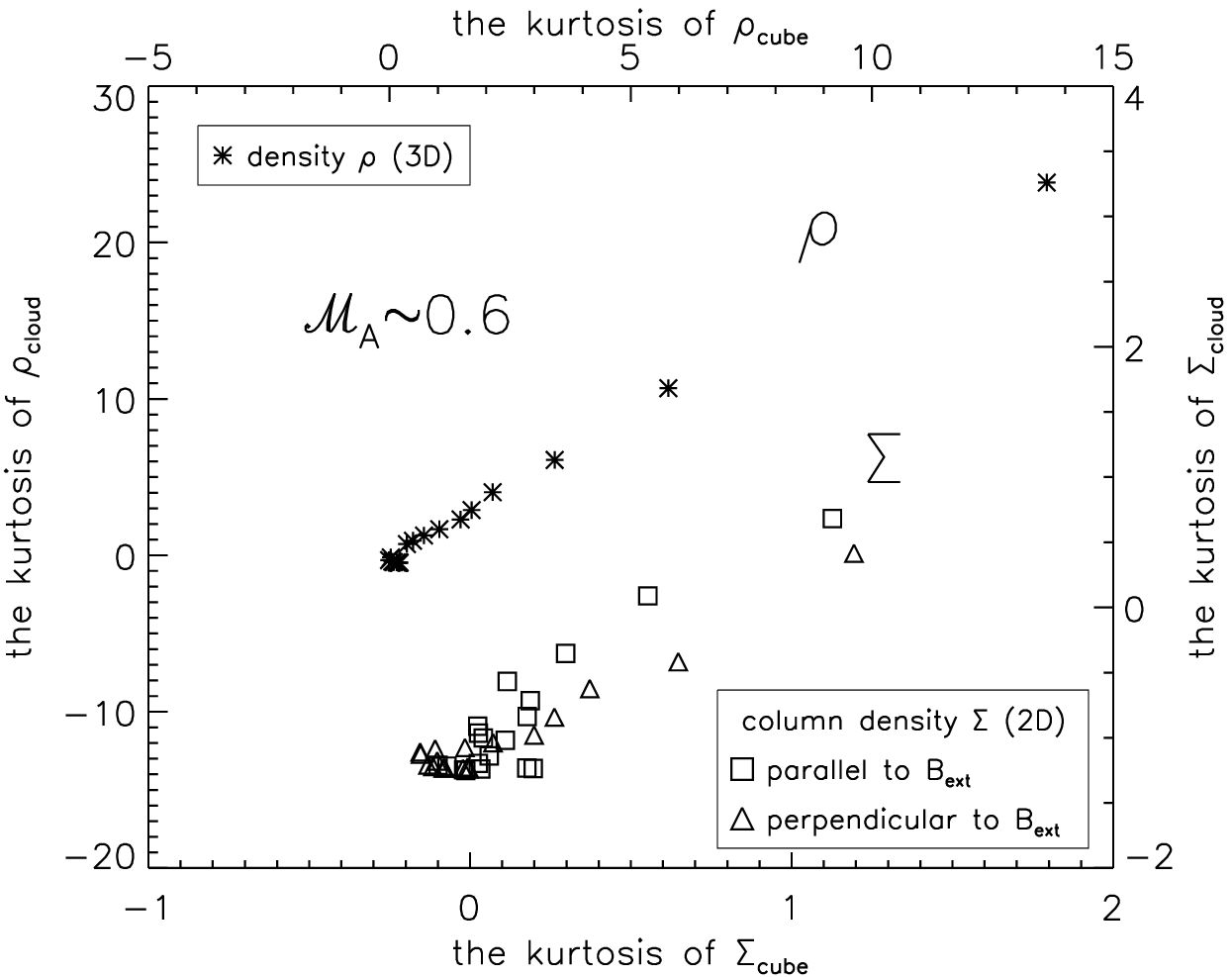}{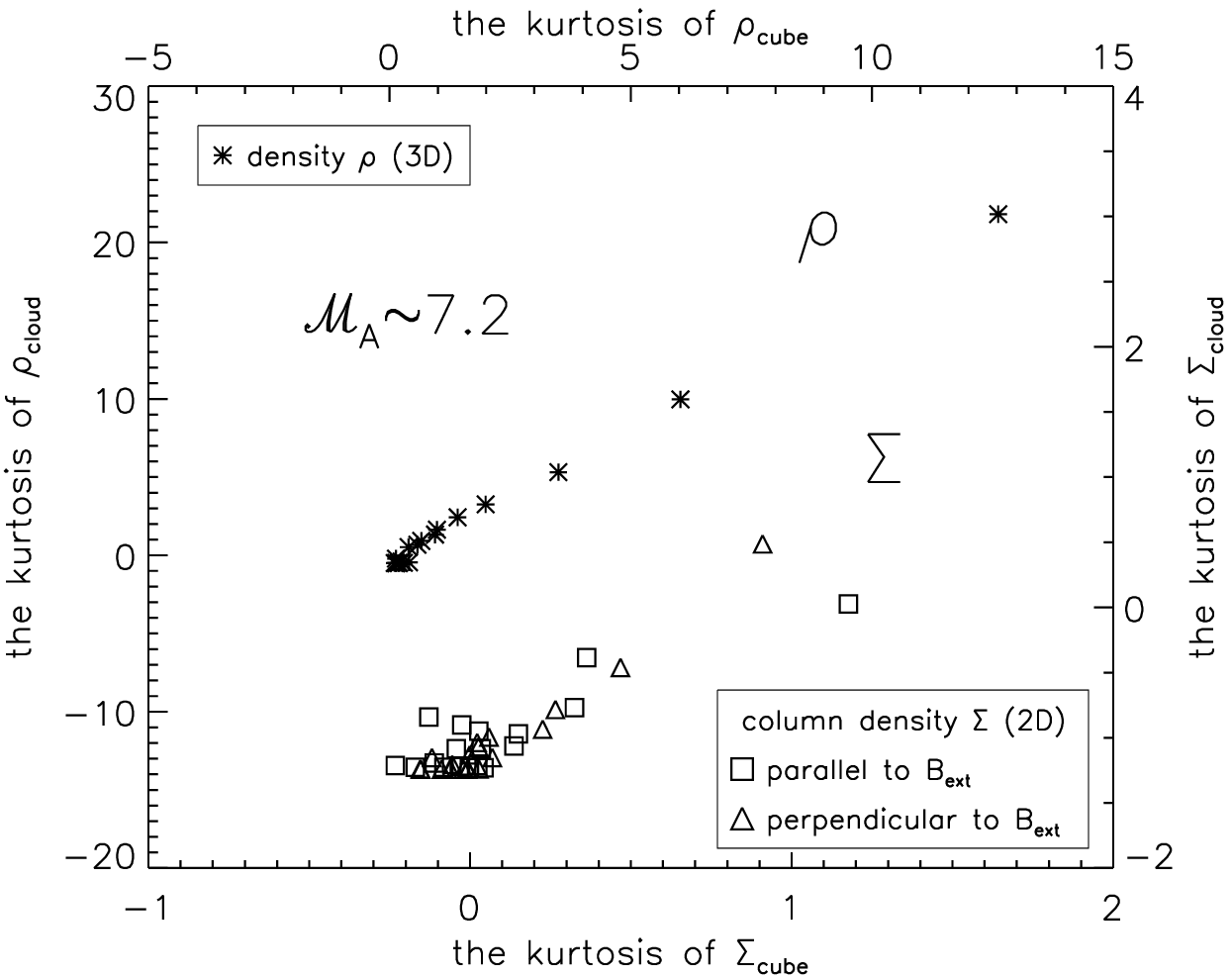}
 \caption{Comparison of the variance, skewness, and kurtosis ({\em top, middle and bottom rows, respectively}) of the cloudlike-shaped density plotted as functions of the unshaped full cube densities. In the left column we present results for sub-Alfv\'{e}nic turbulence, while the right column contains results for super-Alfv\'{e}nic turbulence. Each plot consists of the moments for models with different sonic Mach numbers for density and column densities integrated along and across the external magnetic field. The low-resolution models are used here. \label{fig:dens_sph_moments}}
\end{figure}

In Figure \ref{fig:dens_sph_moments} we show three statistical moments of cloudlike-shaped densities as functions of moments for the original full cube data. In the six plots in Figure \ref{fig:dens_sph_moments} we show variances ({\em top row}), skewnesses ({\em middle row}) and kurtosises ({\em bottom row}) for sub-Alfv\'{e}nic ({\em left column}) and super-Alfv\'{e}nic ({\em right column}) experiments. We plot the values of moments obtained from models with different sonic Mach numbers as a function of moments for the same models but with densities modulated by the Gaussian distribution. The boundary does not influence significantly the moments if the relations presented in the plots of Figure \ref{fig:dens_sph_moments} are linearly increasing. On the contrary, the boundary dominates if values of the moments for $\rho_\mathrm{cloud}$ or $\Sigma_\mathrm{cloud}$ do not change with the moments for $\rho_\mathrm{cube}$ and $\Sigma_\mathrm{cube}$. In this way we can see how the boundary influences the variance, skewness, and kurtosis of density and column density.

Taking a closer look into the relation between the moments of 3D density (lines plotted using stars) we see that the linearity of the relation grows with the order of the statistical moment, which means that the boundary contributes the most in variance, less in skewness, and in the case of kurtosis, the contribution is almost negligible (see Fig. \ref{fig:dens_sph_moments}, {\em top, middle, and bottom rows, respectively}). It means that if we want to avoid the influence of the boundaries, we should use the kurtosis as a measure of the sonic Mach number in turbulence with a proper scaling factor. However, we are mainly interested in the moments of 2D data, which are available from observations. Lines corresponding to the moments of column densities (plotted using triangles and squares) reveal much higher dispersion and irregularity. Nevertheless, we are still able to determine regions where the relation between statistical moments of column density calculated from cloudlike-shaped density and moments of unmodified column density is linear (see Fig.~\ref{fig:dens_sph_moments}). The skewness of $\Sigma$ could also be a measure of the strength of the external magnetic field, because its relations, seen in middle left plot of Figure \ref{fig:dens_sph_moments}, are different for densities integrated along and across $B_\mathrm{ext}$ (lines plotted using squares and triangles, respectively). We do not see such differences in the results of models with weak $B_\mathrm{ext}$ (see fig.~\ref{fig:dens_sph_moments}, {\em middle right}). The relations for kurtosises in the case of $\Sigma$ seem to be least helpful. The points are strongly agglomerated, with only a few points, corresponding to the highest sonic Mach numbers, creating a straight line. As a result of these agglomerations, the relations can only be used in the case of highly supersonic turbulence.

\begin{figure}  
 \plottwo{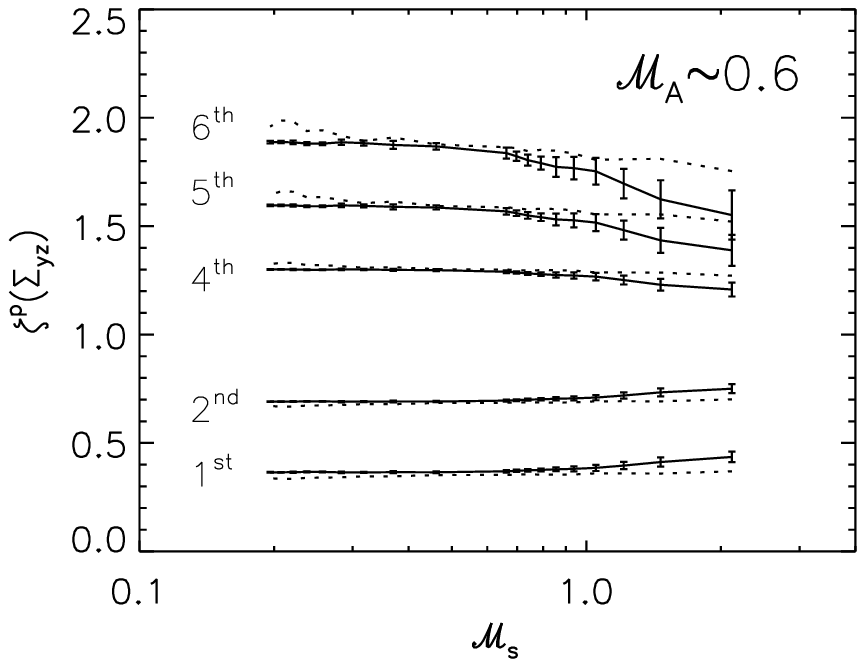}{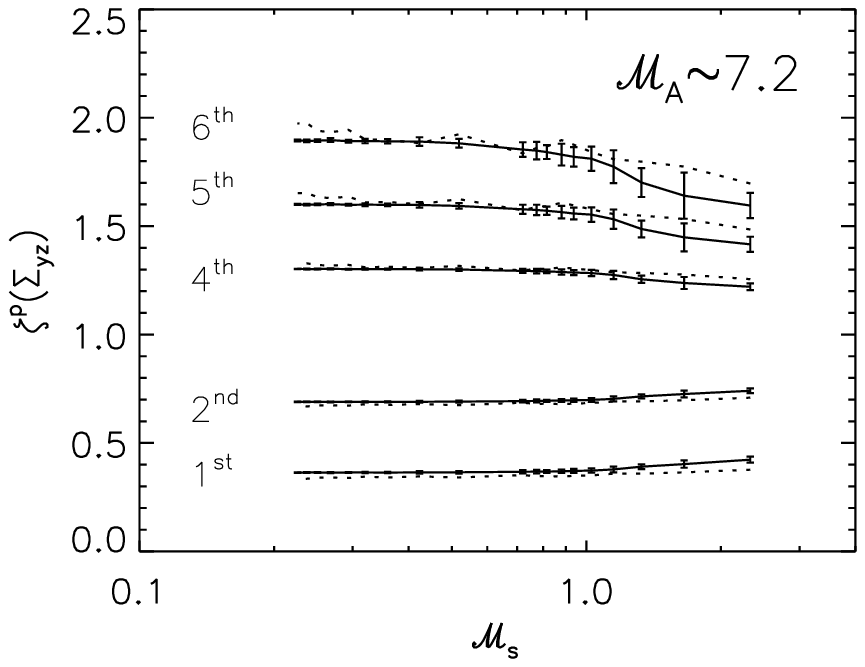}
 \caption{Scaling exponents of the structure functions of column density obtained from densities shaped in the form of a cloud. Plots show exponents of the structure functions of first to sixth order excluding the third one, which is taken as a normalization. For comparison the dotted lines show the corresponding values of exponents for the original density, i.e. not shaped in the form of a cloud (the same as in Figure \ref{fig:dens_int_expon}). The low-resolution models are used here as well. \label{fig:dens_sph_expon}}
\end{figure}

Figure \ref{fig:dens_sph_expon} shows two plots for a relation between the exponents of the structure functions of column densities for the sub- and super-Alfv\'enic turbulence ({\em left and right plots, respectively}). these plots suggest a strong dependence of intermittency on ${\cal M}_s$. It seems to be even stronger than in the plots in Figure \ref{fig:col_dens_scal} due to the presence of a boundary. As previously described, in models with low values of ${\cal M}_s$, the contribution in the structure functions mainly comes from the large-scale structure of the cloud. In this case, the exponents are almost independent of ${\cal M}_s$ (see the subsonic part of the plots in Figure \ref{fig:dens_sph_expon}). However, when the sonic motions become stronger, they are able to produce a high contrast in density structures, enough to dominate in the structure functions. Thus, we are able to study the intermittency from the intensity maps only in turbulence with sufficient values of ${\cal M}_s$. However, to overcome this problem in the case of subsonic turbulence, we should restrict analogous analysis only to the intensity maps of the interior of the cloud. In both cases, we should treat the statistics obtained from the intensity maps very carefully, including the boundary effect as much as possible.

\subsection{How do the statistics of density depend on resolution?}
\label{sec:resolution}

One could ask how our results depend on the resolution of our calculations. Undoubtedly, in models with a higher resolution the small-scale structure of density is better resolved. We have seen in previous sections that these small-scale structures can significantly affect the spectra and structure functions. In this section we examine to what extent our results depend on numerical resolution.

\begin{figure}  
 \plottwo{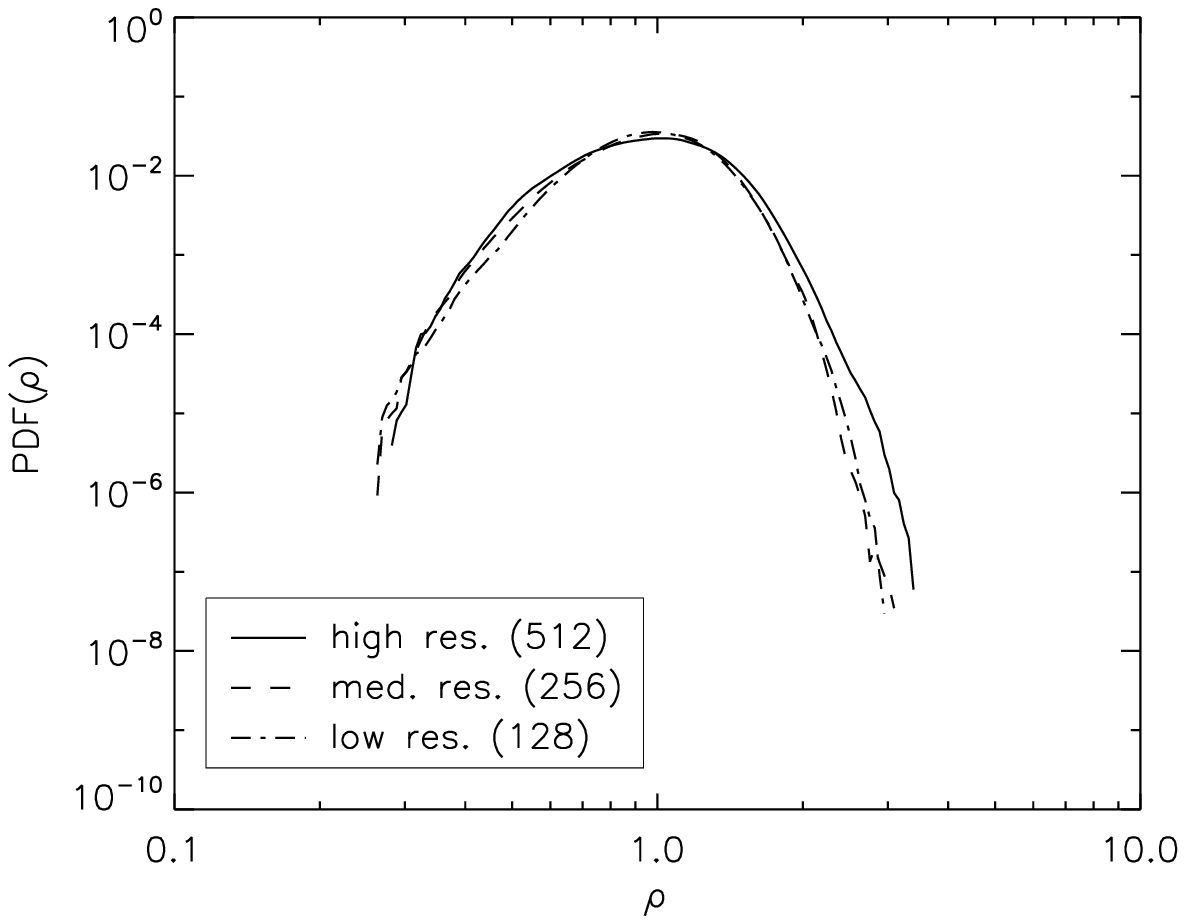}{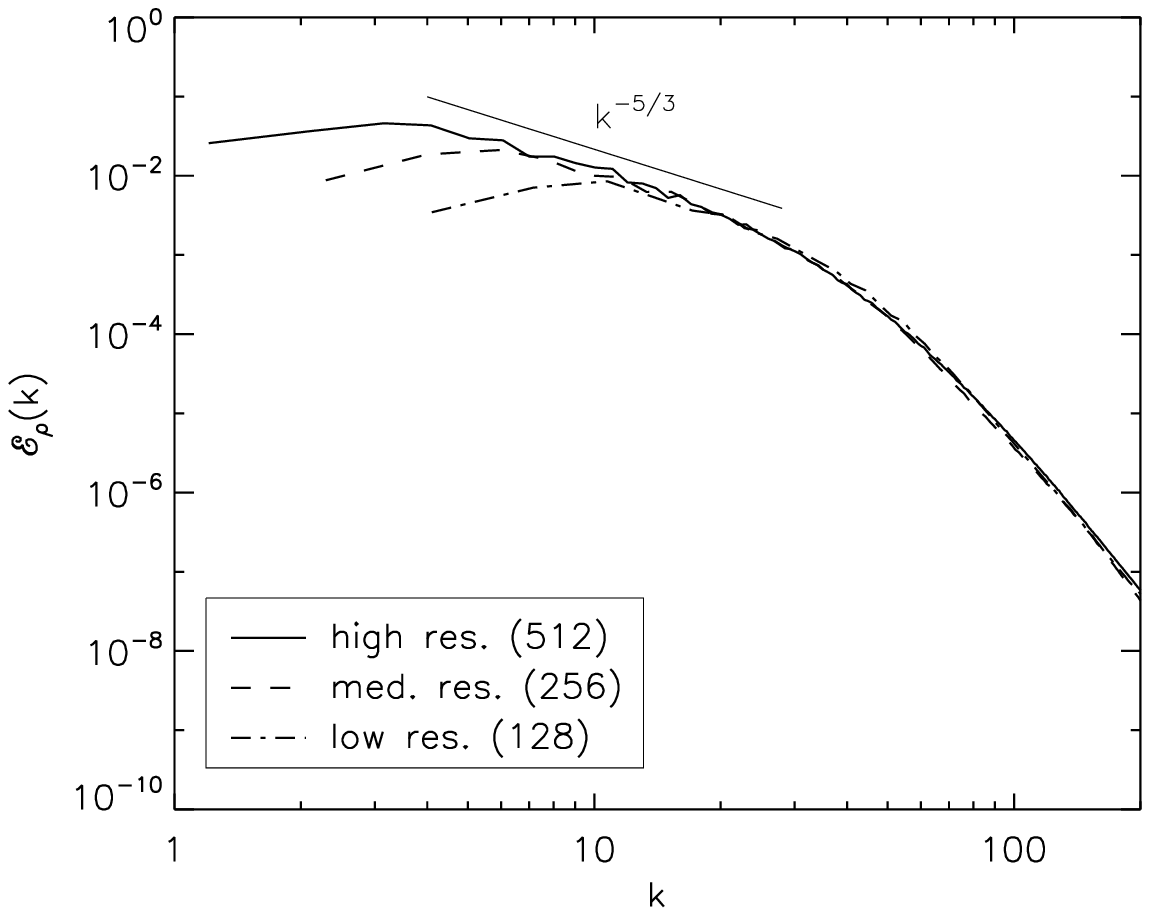}
 \caption{PDFs for density ({\em left}) and power spectra of density ({\em right}) for experiments with ${\cal M}_s\sim0.7$ and ${\cal M}_A\sim0.7$ for three different resolutions: 128$^3$, 256$^3$, and 512$^3$. Spectra in the right plot have been shifted to make the dissipative parts coincide. \label{fig:reso_pdfs_spectra}}
\end{figure}

In Figure \ref{fig:reso_pdfs_spectra} we show PDFs and spectra of density for experiments with the same physical conditions, but for three different resolutions: 128$^3$, 256$^3$, and 512$^3$. As a reference example we show only models with ${\cal M}_s\sim0.7$ and ${\cal M}_A\sim0.7$. Plotted functions for the low and middle resolutions are the result of averaging over several time steps to get the results independent of time variations. For the highest resolution we have only a single snapshot in time. Figure \ref{fig:reso_pdfs_spectra} ({\em left}) suggests that the distribution of density is almost insensitive to the resolution. We see very similar shapes and values of lines. The only visible difference is a tail at higher densities for the high-resolution experiment, which can be a result of time-dependent fluctuations of the highly compressed small-scale structure. It could also be a result of better resolved shocks due to the smaller numerical diffusivity in the case of the high-resolution model, which smooths the density compression at shocks less, resulting in a higher density contrast. A progressive growth in the numerical diffusivity is especially well seen in the spectra of density (see Fig. \ref{fig:reso_pdfs_spectra}, {\em right}). Due to a high diffusivity, the inertial range is very short for the low-resolution model, but its slope is consistent with slopes for higher resolution experiments. Almost perfect covering of the dissipation ranges for all resolutions in the right plot of Figure \ref{fig:reso_pdfs_spectra} shows that the numerical dissipation acts in the same way on the small-scale density structure independent of the resolution. The resolution changes only the length of the inertial range.

\begin{figure}  
 \plotone{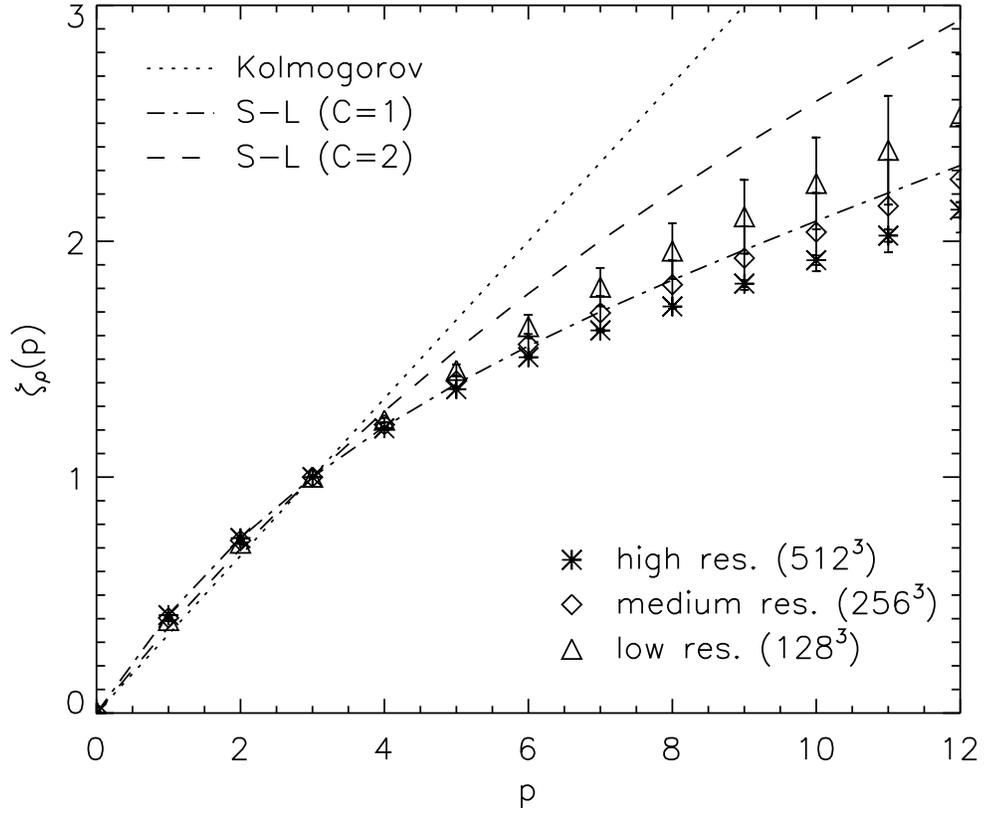}
 \caption{Scaling exponents for density for models with ${\cal M}_s\sim0.7$ and ${\cal M}_A\sim0.7$ for three different resolutions: 128$^3$, 256$^3$, and 512$^3$. Lines show theoretical scalings: dotted for Kolmogorov and dash-dotted and dashed for She-L\'{e}v\^{e}que with parameter $C=1$ and $2$, respectively ($g=3$ and $x=\frac{2}{3}$). \label{fig:reso_expons}}
\end{figure}

Figure \ref{fig:reso_expons} shows the scaling exponents for density for three different resolutions. To estimate exponents and errors we calculate the mean value and standard deviation from results for all available time snapshots at $t\ge 5$ for a given experiment. Figure \ref{fig:reso_expons} suggests the growth of the intermittency of density with resolution. The intermittency is more sensitive to the resolution. It is especially well seen at higher orders $p$. Simultaneously, we see bigger errors for lower resolutions. But bigger errors could also indicate that the intermittency of structure is oscillating stronger over time. An interplay of forces creating small-scale and dense structures through shock formation and diffusion (also numerical) and destroying such structures could establish an equilibrium of some sort when the development of turbulence reaches a stationary state. In this kind of analysis, there are also statistical restrictions due to the sampling size effects. Structure functions are reliable to some order depending on the resolution of numerical models. In our studies we have the size of analyzed samples ranging from about 2$\cdot$10$^6$ to over 10$^8$. In such a case, the maximum order should be for $p=8$. This explains the increase of the error bars. Our conclusion is that we should expect a somewhat higher intermittency of density in experiments with higher resolution.

In some sense our study sends a warning to the low-resolution studies of the She-L\'{e}v\^{e}que scaling. We see that while the extended self-similarity may be fulfilled for the low-resolution studies, their higher order scalings can still be different from higher resolution runs.

\subsection{How do the statistics of density depends on time averaging?}

In our analysis we used 16 snapshots of data uncorrelated in time for the low-resolution models and from eight up to 17 snapshots for the medium-resolution models depending on the model.\footnote{The accurate number of snapshots in each model is given in Table \ref{tab:models}.} Our results show a relatively significant magnitude for the errors (see Figures \ref{fig:dens_expons}, \ref{fig:col_dens_scal}, and \ref{fig:dens_moments}) understood here as the measure of the time variance of the analyzed quantities. Continuous interaction between waves results in creation of the structures. We observe smooth transitions from one structure to another. The process is stochastic. It means that as with all stochastic processes, it should not be considered for one particular snapshot, but a relatively large number of snapshots should be taken into account. For instance, if we had used only one snapshot, we would come to the wrong conclusions about whether the intermittency is higher or lower for a particular case or whether the statistical moments scale differently with ${\cal M}_s$. This is especially important for measures with a relatively high level of uncertainty like scaling exponents or statistical moments for the column densities. In addition, in numerical simulations the time step is defined by the maximum characteristic speeds in the system. These speeds depend on the local amplitudes of the velocity and magnetic field, which can be different for different models. In this case, a small shift in time at which we compare snapshots for different models would result in completely different conclusions for measures strongly varying in time. In this way, the analysis of turbulence for only small time averaging produces results that could be very unreliable.

\section{Topology of density structures}
\label{sec:topology}

To study the topology of density structures, we applied the genus statistics that are successfully used in the cosmic microwave background studies \citep[for example]{gott86}. For a given density threshold we define an isosurface of density separating higher and lower density regions and evaluate the genus of this contour. Roughly speaking, the genus of an isosurface is the number of holes in this surface minus the number of isolated regions for a given threshold. This difference changes with the change of the threshold. We plot the genus of these contours as a function of the density threshold.

To perform genus statistics we use a Fortran algorithm published by \cite{weinberg88}. From the analysis of the genus of isosurfaces of density, we can extract information about the density distribution and about the topology of structures that are created by turbulence. The results and plots in this section refer to the medium-resolution models.

\begin{figure}  
 \plottwo{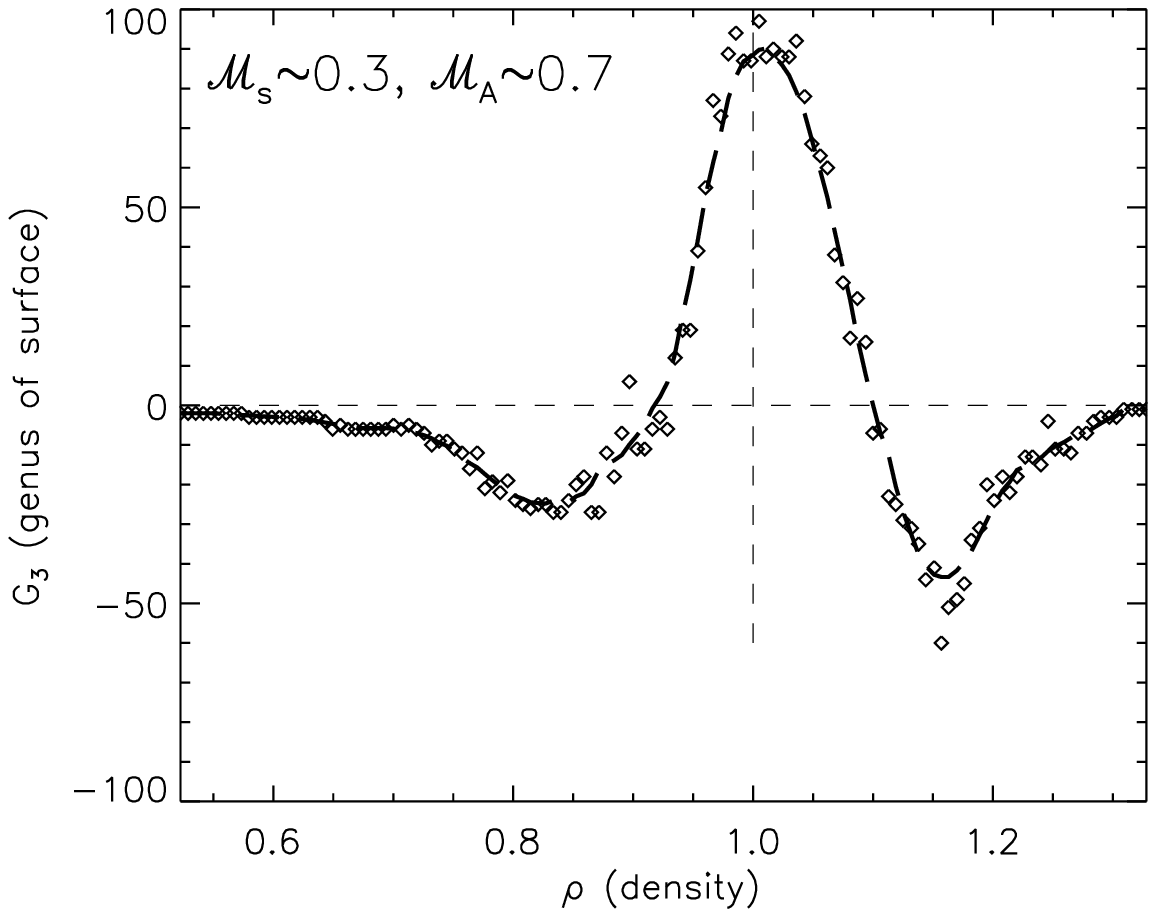}{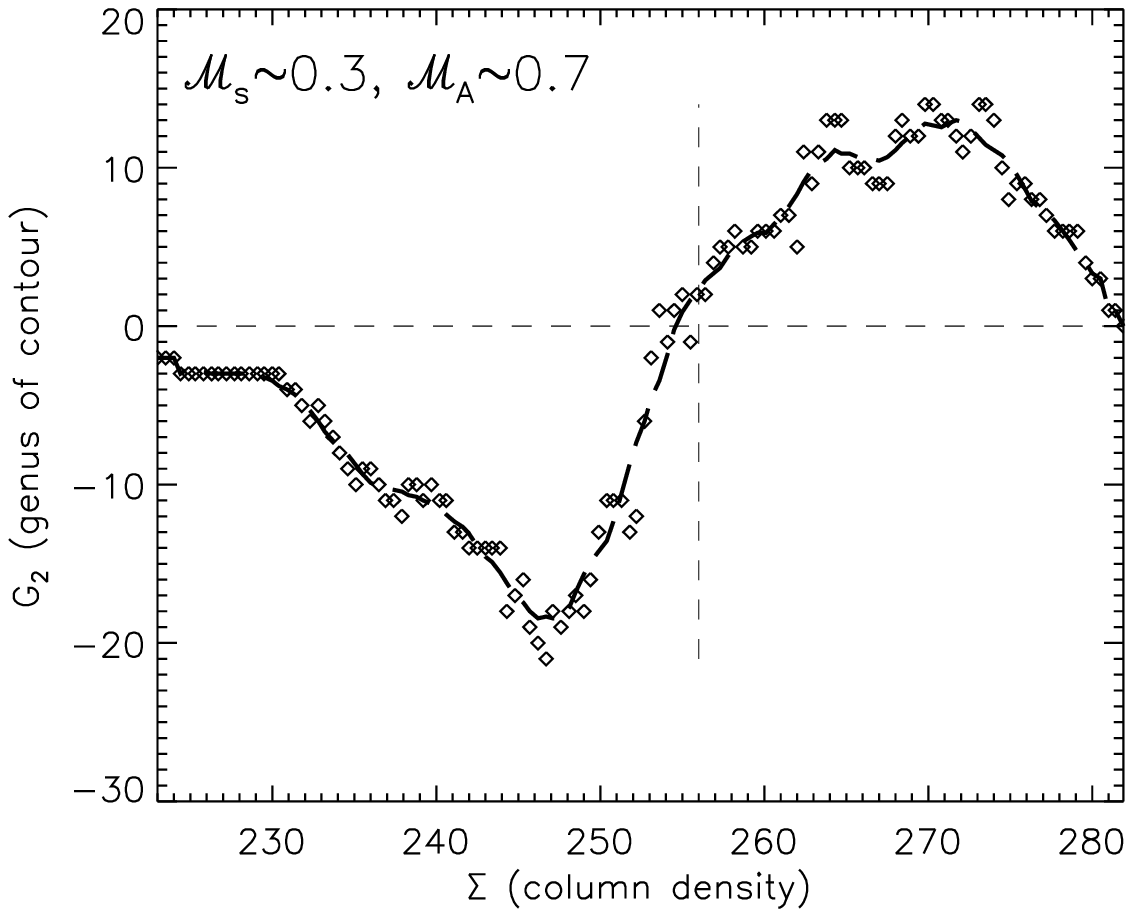}
 \plottwo{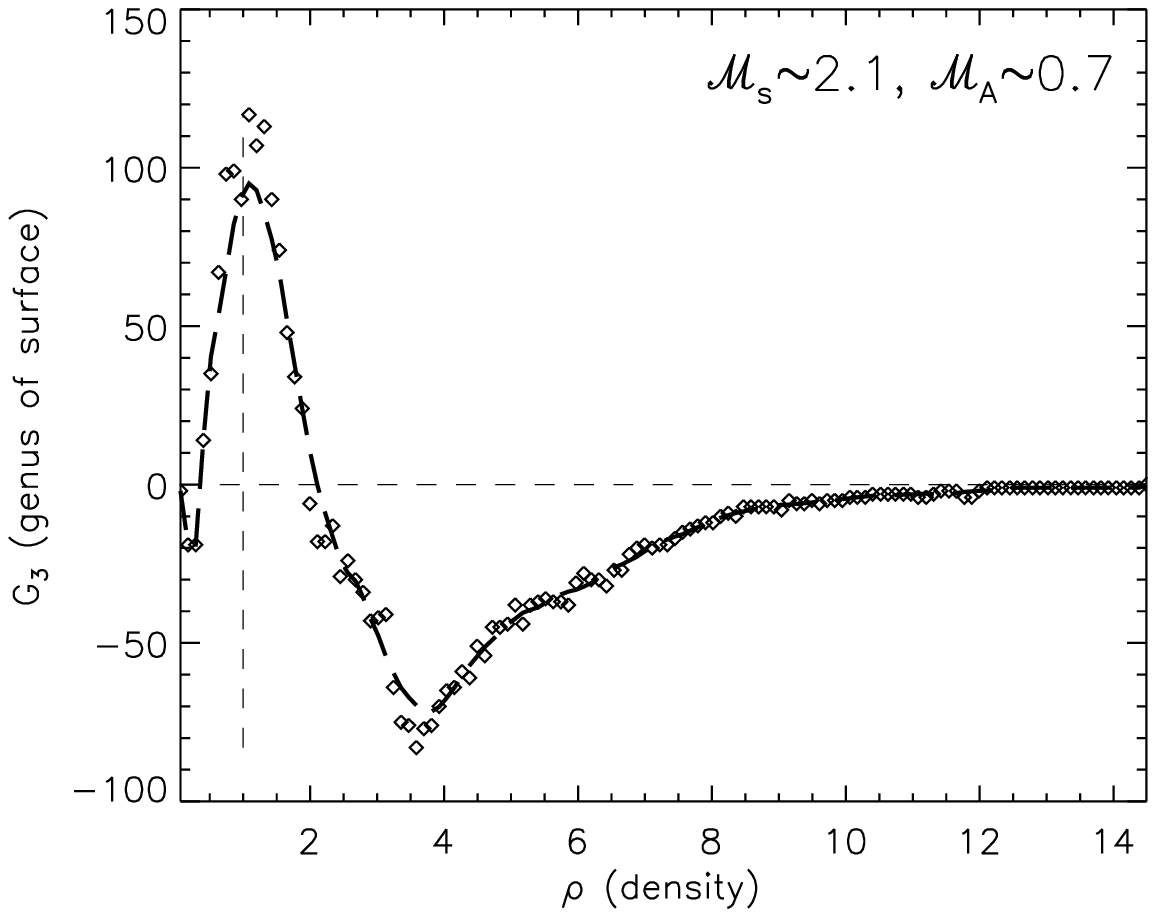}{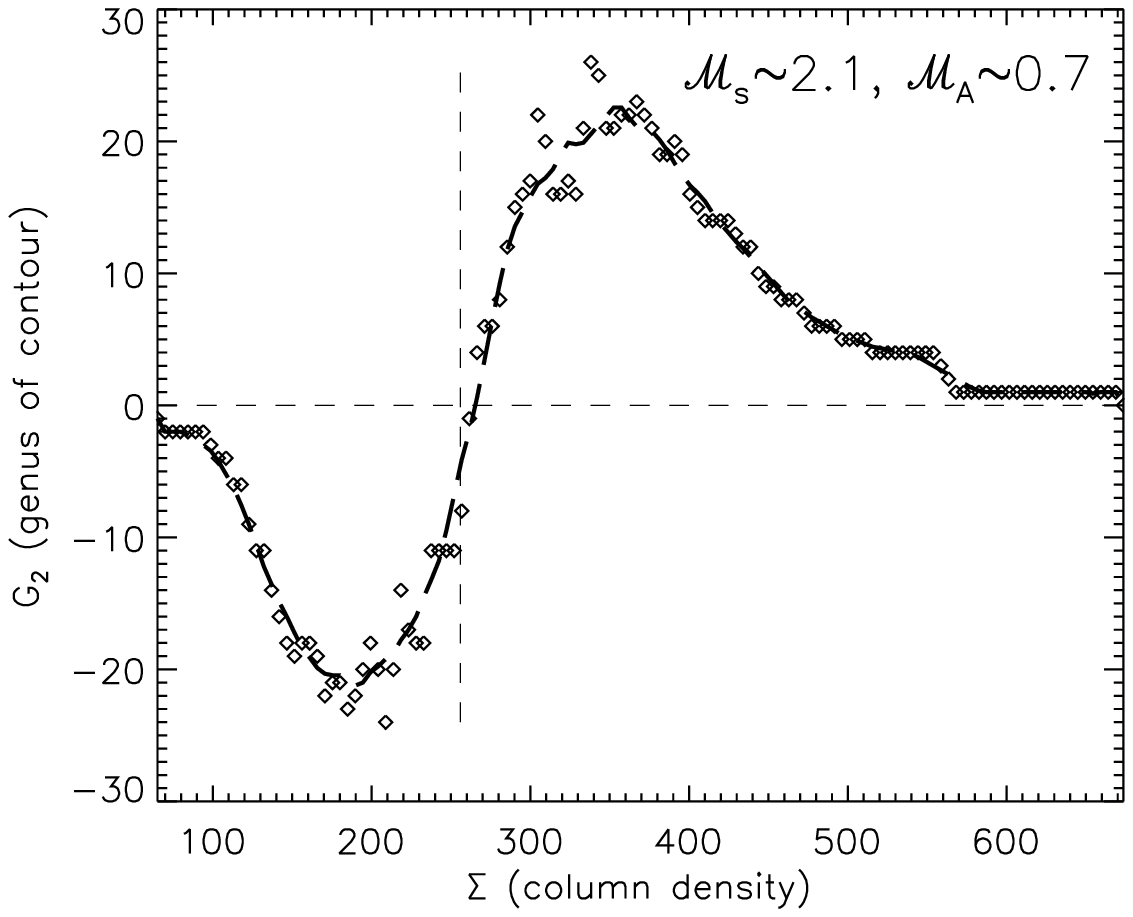}
 \plottwo{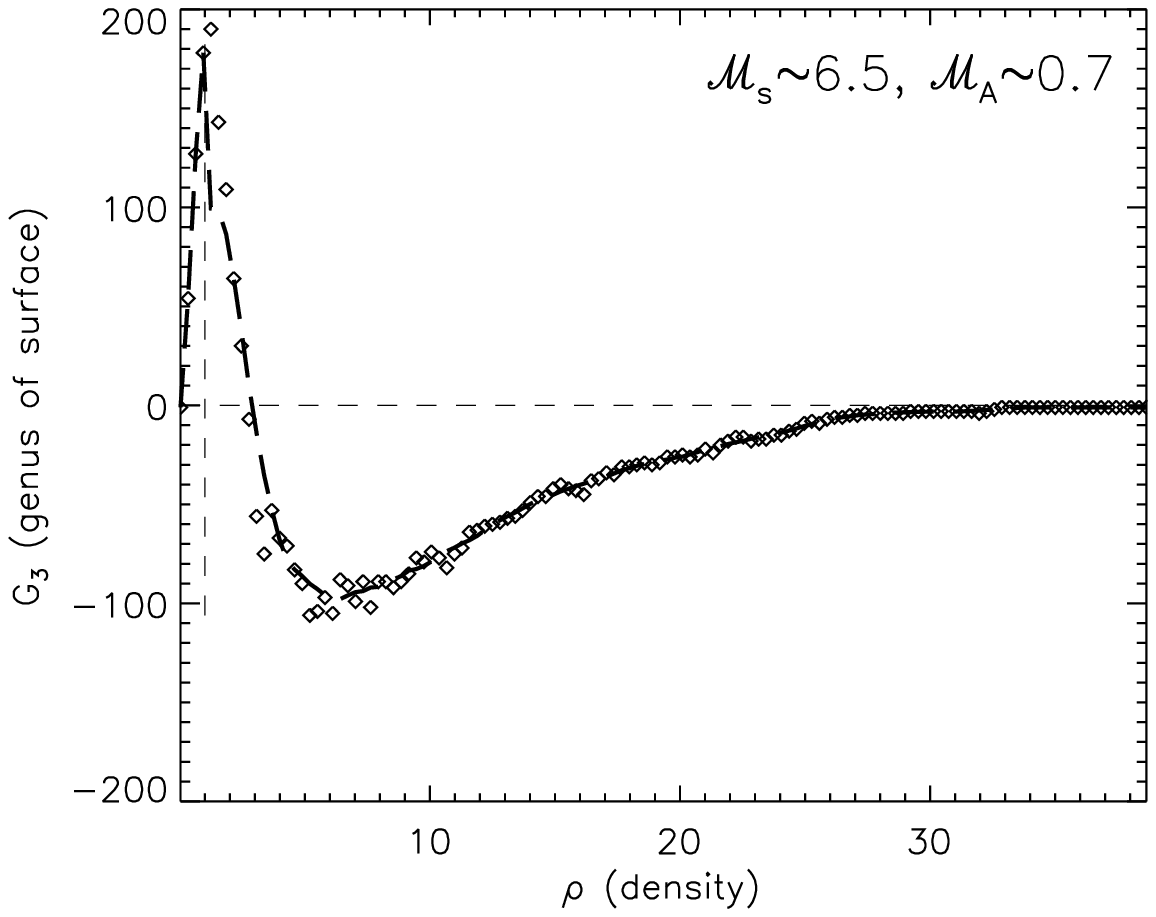}{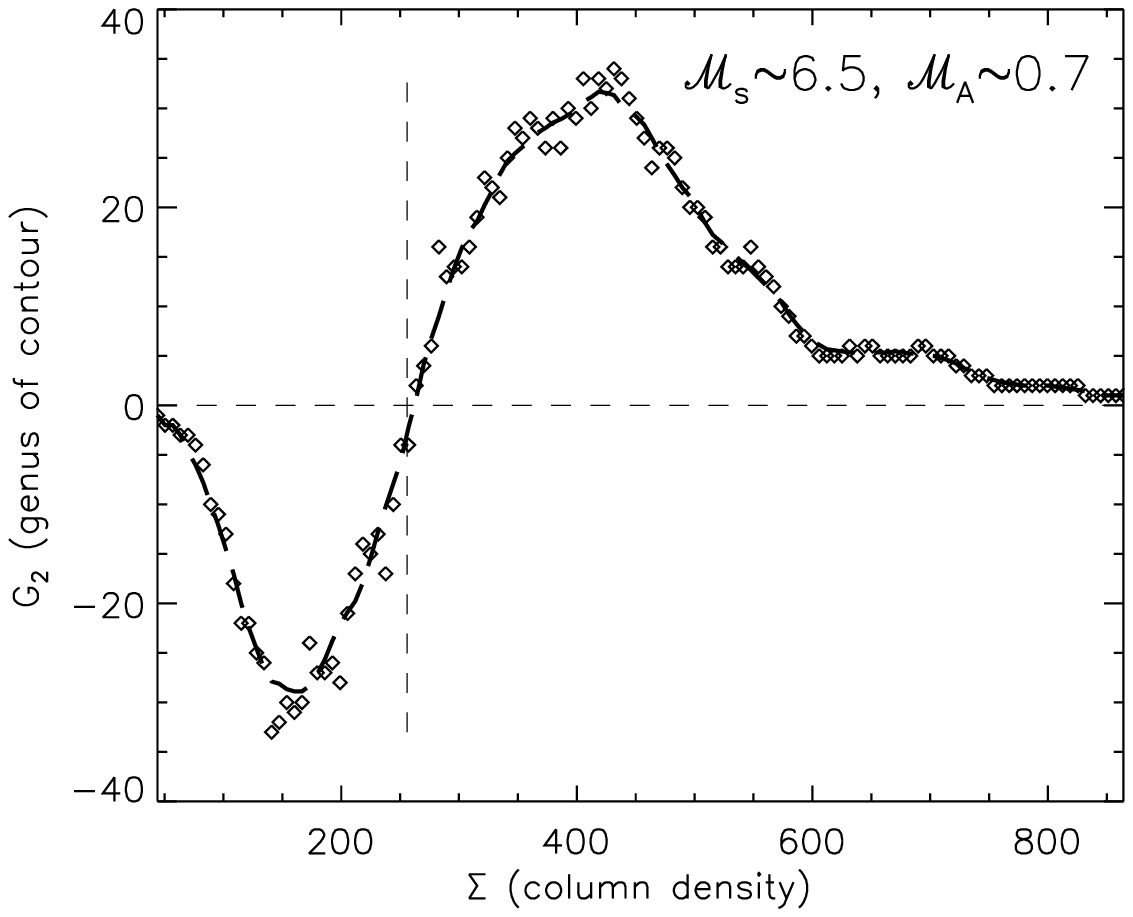}
 \caption{Topology of the density structures ({\em left column}), and topology of the density integrated along the magnetic field ({\em right column}). Subsonic models have symmetric distributions of the genus line, and the supersonic models have much longer tails of higher densities (see text for explanation). In all plots, vertical dashed lines designate the position of the mean density value. \label{fig:genus_statistics}}
\end{figure}

In Figure \ref{fig:genus_statistics} ({\em left column}) we present the distribution of genus for three different sonic Mach numbers, one for the subsonic case (Fig. \ref{fig:genus_statistics}, {\em top left}) and two for the supersonic cases (Fig. \ref{fig:genus_statistics}, {\em moddle left and bottom left}). For the subsonic case in which the distribution is almost symmetrical with respect to the mean density value, the shape of the genus line suggests a spongelike density structure with occasional overdense clumps or under-dense holes. We observe a somewhat higher number of higher amplitude dense regions than rarefactions, because on the high-density side of the distribution, the line decreases to highly negative values (down to about -80). Correspondingly the low distribution of rarefactions reaches a value of only -50. For supersonic models with ${\cal M}_{s}$ around 2.5 or 7.0, the genus line looks much different. Although the structure seems to be also spongelike (the distribution in the vicinity of the mean density value is broad and highly positive), the low-$\beta$ turbulence produces an extended tail in the high-density direction. This suggests the existence of many highly dense singular structures. It may be evidence for the formation of strong shocks, which perturb and compress density locally in small volumes (see BLC05). Density in these regions grows to a value 25 times larger than the mean value in the case of ${\cal M}_s =7$ turbulence. In connection with this we observe the disappearance of the low-density tail in the distribution, which means that underdense holes are short lived.

We can ask the question about whether the genus statistics of integrated density along one chosen axis can give us adequate information about the 3D topology of the density structure. With such a relation, we could interpret maps of column densities obtained from observations of the diffuse ISM and molecular clouds. In Figure \ref{fig:genus_statistics} ({\em right column}) we show the genus statistics of density integrated along the $X$-axis (or along the external magnetic field, which is also directed along the $X$-axis). We applied the 2D version of the algorithm for genus computations \citep{melott90}. A 2D genus is interpreted as the number of isolated 2D high-density regions minus the number of isolated 2D low-density regions. While the 3D genus is symmetric about the mean value, the 2D version is antisymmetric. For the high-$\beta$ case (Fig. \ref{fig:genus_statistics}, {\em top right}) the genus distribution is very symmetrical, and we only see one rarefied hole structure with column density lower than 180. This corresponds to the Gaussian distribution of the integrated density for this model. For the low-$\beta$ cases the structure is extended more into high-density structures, like in the 3D case of the genus line, and we do not observe any symmetry there. Although the number of more dense regions is somewhat lower than the less dense regions, the maximum is shifted to the right. We can say that all the dense clumps have densities above 300, while holes have densities below 150.

\begin{figure}  
 \plottwo{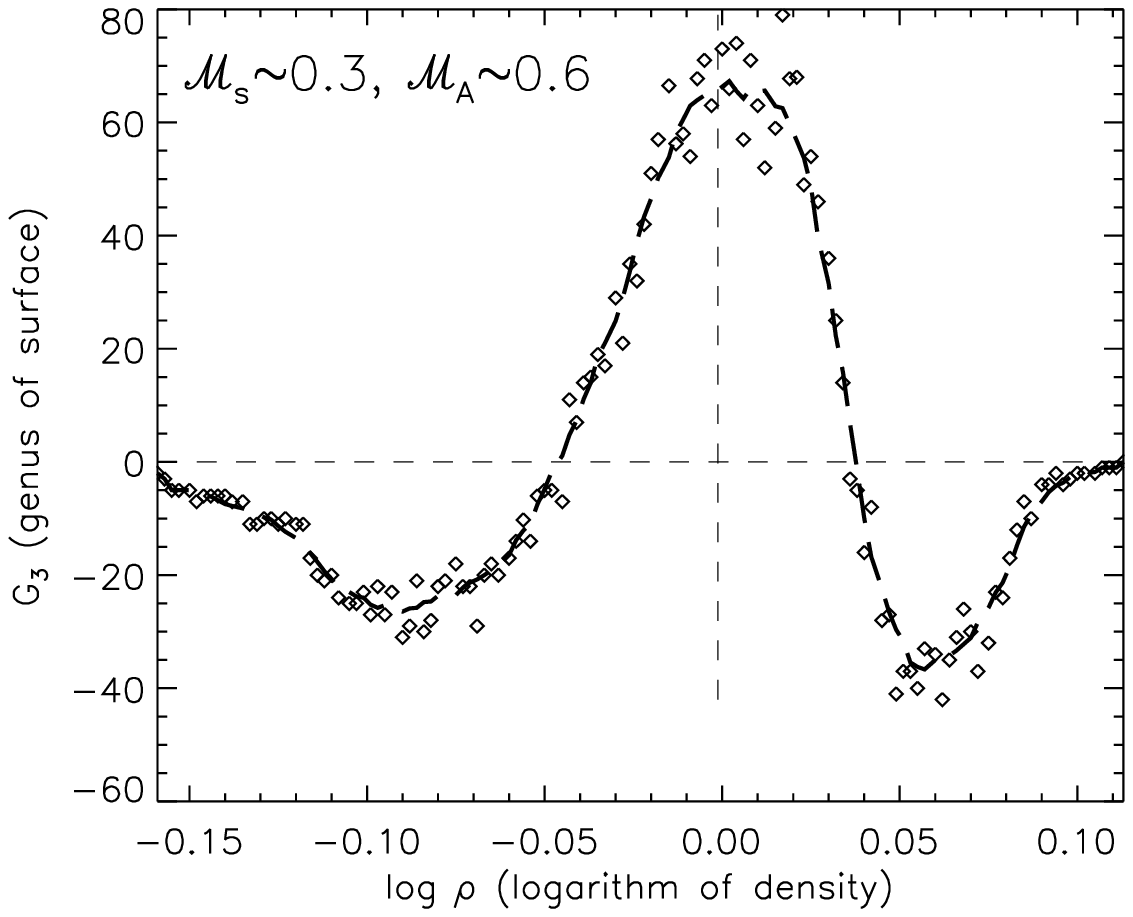}{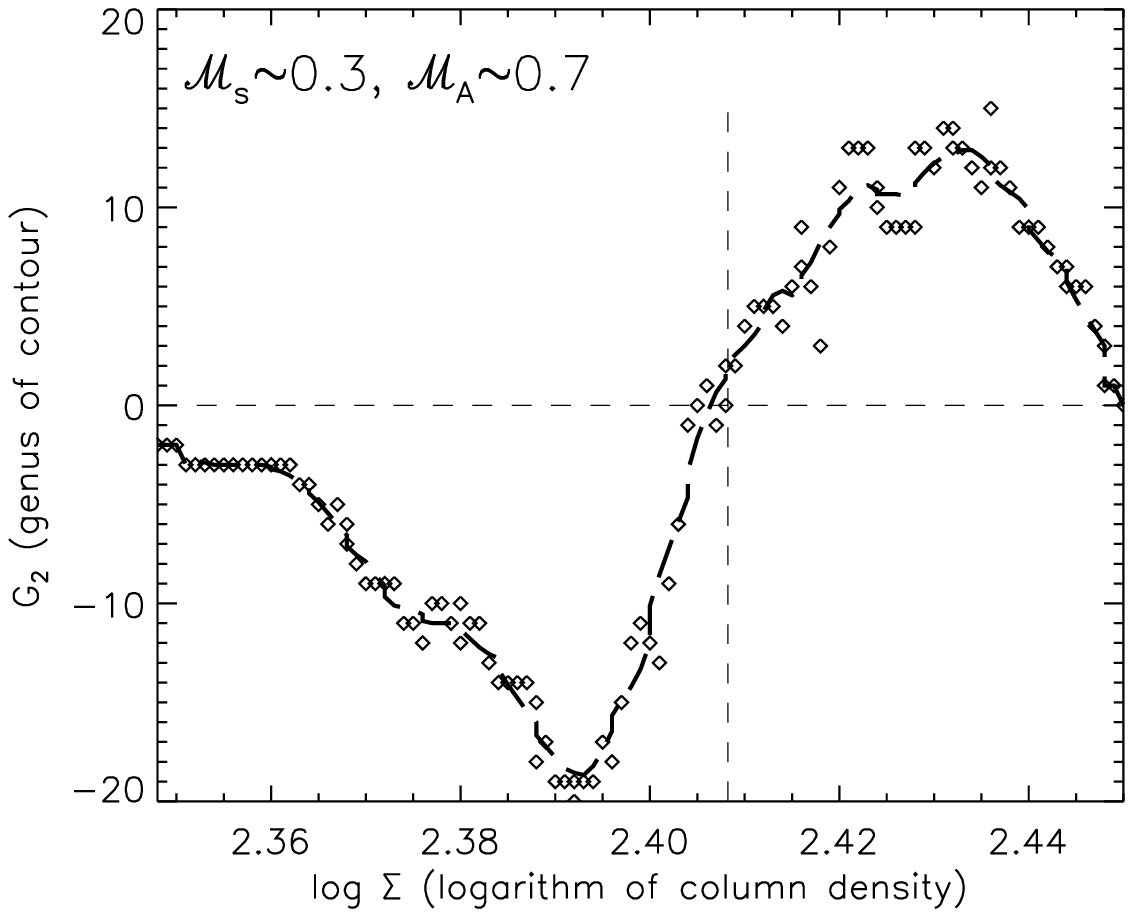}
 \plottwo{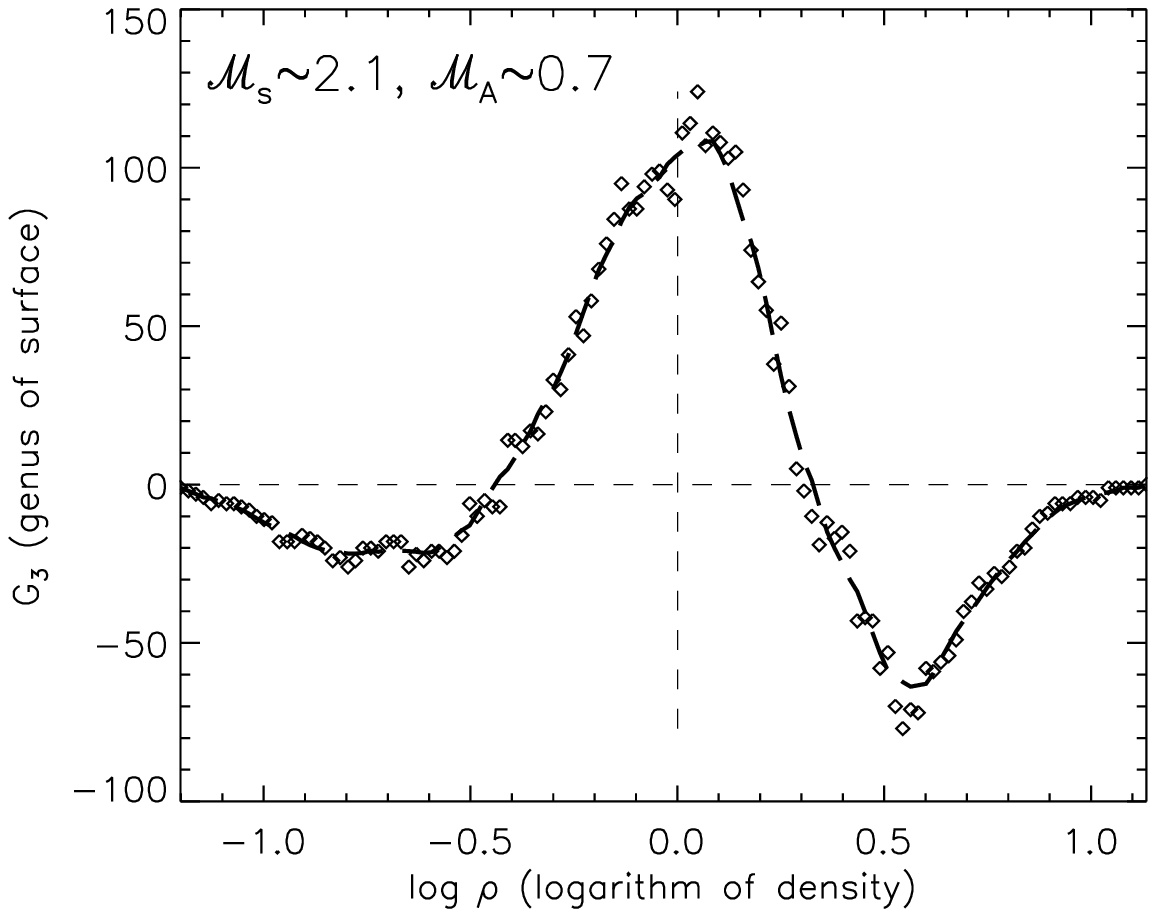}{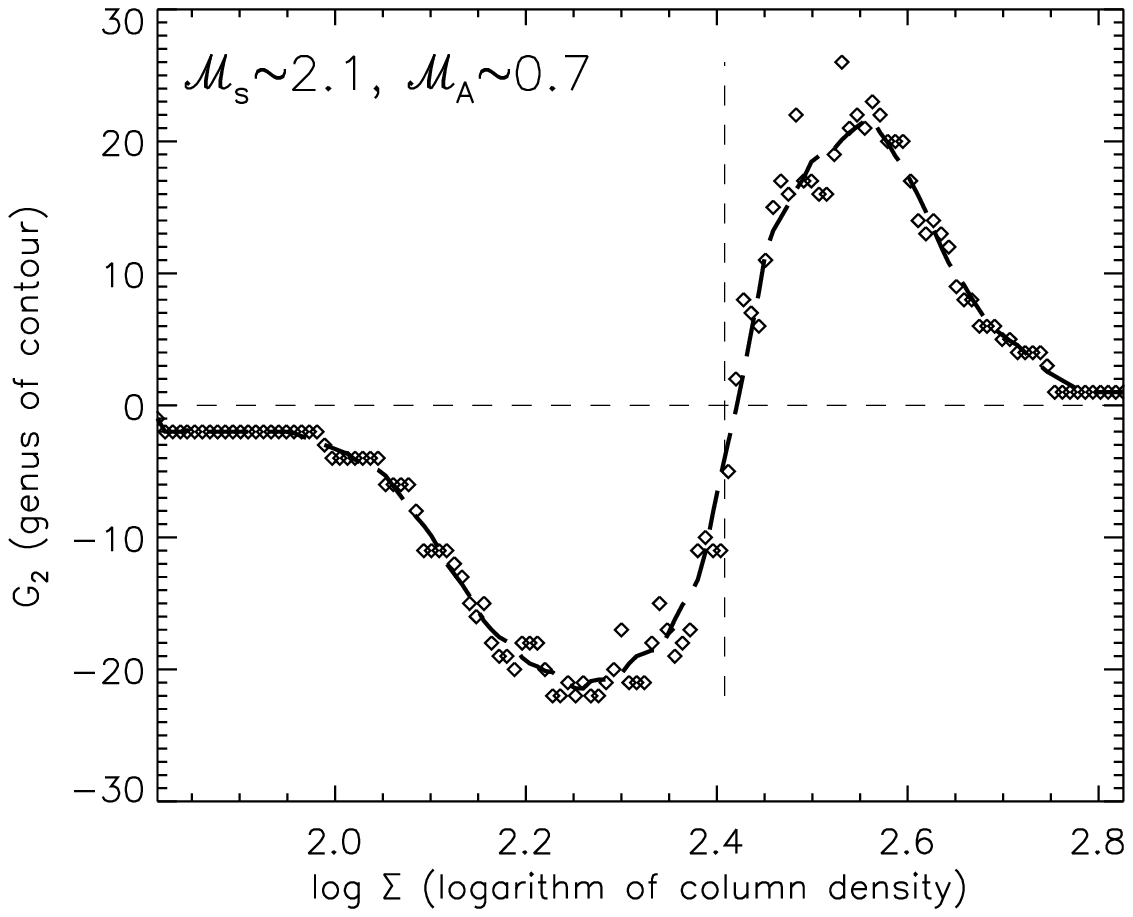}
 \plottwo{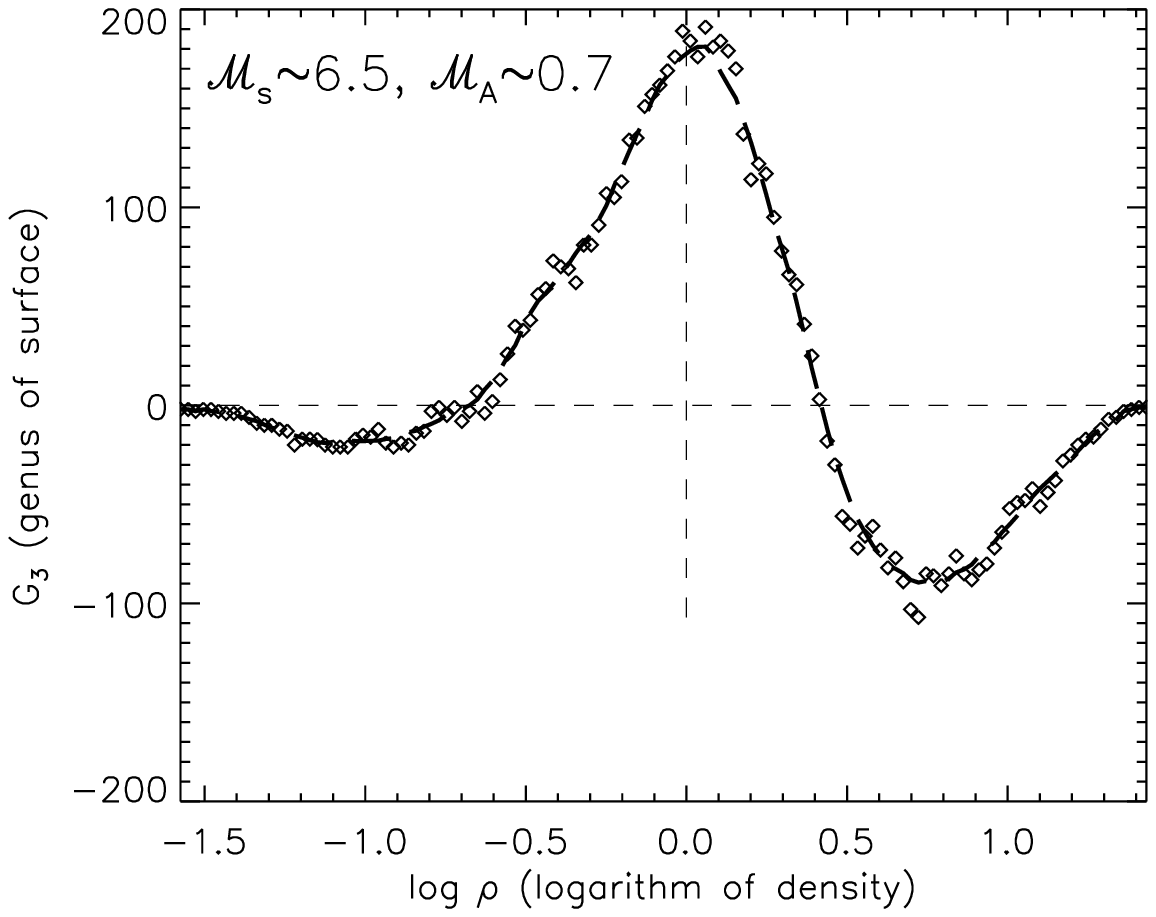}{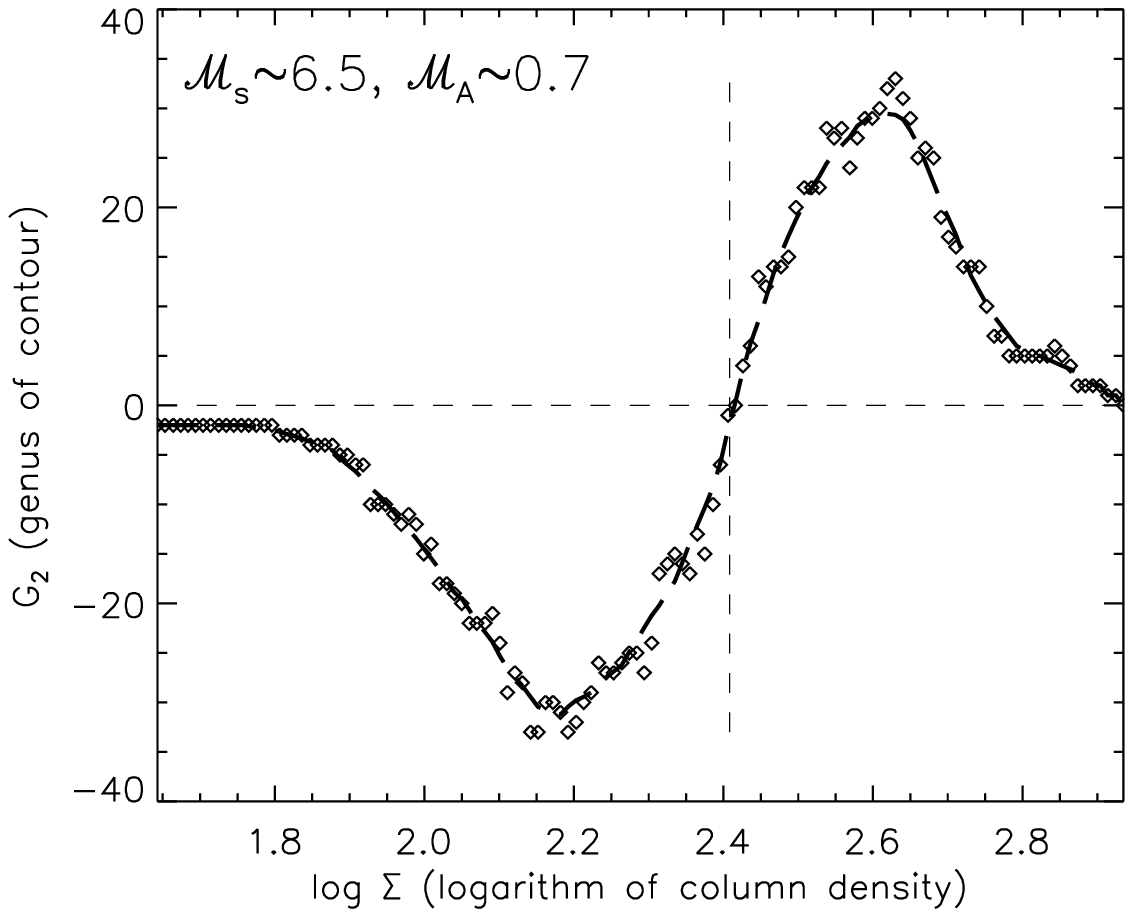}
 \caption{Topology of the logarithm of density ({\em left column}) and the logarithm of column density integrated along the line of sight ({\em right column}). Plots correspond to models from Fig. \ref{fig:genus_statistics}. In all plots, vertical dashed lines designate the position of the mean density value. \label{fig:genus_logarithm}}
\end{figure}

It was previously discussed that clumps with density much higher than the mean value can distort the power spectrum and hide the structure of the density created by motions at small scales (BLC05). Following BLC05 we applied these techniques to test how the genus statistics would be affected. In Figure \ref{fig:genus_logarithm} we present the genus lines for the logarithm of density ({\em left column}) and the corresponding 2D genus lines for the logarithm of the integrated density ({\em right column}). From the 3D genus plots (Fig. \ref{fig:genus_logarithm}, {\em left column}) we see that the lines are almost symmetric about zero. The most non symmetry reveals itself in the supersonic models, where the maximum shifts to the negative values of $\log \rho$. For integrated density plots (fig. \ref{fig:genus_logarithm}, {\em right column}), we do not see any significant differences between the results for different Mach numbers. For high Mach numbers, the logarithm of density has a higher extremum, which is not unexpected.

We have tested the genus statistics for density with the cutoff of high-density peaks. We do not show the plots for them, because the procedure of cutting off the density above some level simply does not change the line in plots of Figure \ref{fig:genus_statistics} below the cutoff value and only removes the high-density tail.

The interesting information would be the sonic Mach number retrieved from the genus statistics of column density maps. The sonic Mach number can be obtained from the length of the high-density tail of the genus line. It is just another way to measure the maximum value of density, which is strongly related to the value of ${\cal M}_s$, which is explained more in \S\ref{sec:pdfs}. However, more important information that the genus can give is the type of structure. For low sonic Mach number models we have both underdense holes and overdense clumps with approximately the same count of each. This is supported by the top right plot in Figure \ref{fig:genus_statistics}, where we observe distribution a roughly symmetric about the mean value. However, with the increase of sonic Mach number, turbulence produces more high-density structures, extending the right tail of the genus line significantly. This means that the structure starts to look rather like a medium with lower density in the most of volume, but with many singular 1D or 2D structures with much higher density. The reader can compare this picture to Figure \ref{fig:synth_maps}, which supports our claims. We can conclude then that the genus statistics can be a valuable tool to study the structure of molecular clouds.

\section{Discussion}
\label{sec:discussion}

\subsection{Density in MHD turbulence}

We feel that it is important for studies of turbulent density to be understood in the context of broad MHD turbulence. There has long been an understanding that MHD turbulence is anisotropic \cite[e.g.][]{shebalin83}. A very important insight into incompressible MHD turbulence by \cite{goldreich95} (GS95) has been followed by progress in the understanding of compressible MHD turbulence \cite[CL02; ][]{cho03,lithwick01,vestuto03}. In particular, simulations in CL02 showed that the Alfv\'{e}nic cascade evolves on its own and the exchange of energy between Alfv\'{e}n, slow, and fast modes is marginal.\footnote{The expression proposed and tested in \cite{cho03} shows that the coupling of Alfv\'{e}nic and compressible modes is appreciable on the injection scale if the injection velocity is comparable to the {\it total} Mach number of the turbulence, i.e. with $(V_A^2+c_s^2)^{1/2}$, where $V_A$ and $c_s$ are the Alfv\'{e}n and sound velocities respectively. However, the coupling becomes marginal at smaller scales, as turbulence cascades and turbulent velocities get smaller.} This understanding differs from some earlier claims of a high degree of coupling of compressible and incompressible motions. Incidentally, the notion of high coupling was used to explain the fast damping of compressible MHD turbulence reported in \cite{stone98} and in \cite{maclow98}. In view of the current understanding of MHD turbulence we claim that the fast decay of MHD turbulence is due to the fast decay of the Alfv\'{e}nic mode \cite[see][]{cho03}, the effect that is present even in incompressible MHD \cite[see][]{cho02b}.

In view of the aforementioned picture of MHD turbulence, density perturbations arise from the slow and fast modes. Alfv\'{e}n modes, however, shear and cascade the slow modes (GS95) and therefore affect the density spectrum. Note that for supersonic turbulence the slow modes get modified and produce shocks [the spectrum $E(k)\sim k^{-2}$]. The large-scale shocks produced compress density and produce a shallow spectrum, as we observed in the simulations above (see also BLC05).

In terms of the study of PDFs and spectra, our present work goes beyond the one in BLC05, as we provide a wider parameter space survey. However, we reach conclusions similar to those in BLC05 \cite[see also][]{padoan04}. Interestingly enough, our results are also similar to those obtained in non-magnetized fluids by \cite{kim05}. Indeed, we find that the spectrum of density gets shallower than the Kolmogorov one as the sonic Mach number increases. The reason for this is the formation of small high-density compressed regions that dominate the spectrum. If the logarithm of density is used, these regions do not dominate and the underlying spectrum influenced by anisotropic Alfv\'{e}n mode shearing emerges. As for the PDFs, they are close to lognormal for sub-Alfv\'{e}nic turbulence even if the turbulence is supersonic. For super-Alfv\'{e}nic turbulence the PDFs can be approximated by a lognormal distribution only for low sonic Mach number.

The remarkable regularity of $\log \rho$ should have a physical explanation. A possible one is related to the multiplicative symmetry with respect to density in the equations for isothermal hydrodynamics \cite[see][]{passot98}. This means that if a stochastic process disturbs the density, it results in perturbations of density being multiplied rather than added together. Consequently, the distribution of density for a Gaussian driving of turbulence tends to be lognormal, which is consistent with our PDF measurements. Magnetic forces should affect the multiplicative symmetry above. However, they do not affect the compressions of gas parallel to the magnetic field. Compressions in magnetically dominated fluid will be of the highest intensity and therefore most important. They are sheared by Alfv\'{e}nic modes, as their own evolution will be slower than that imposed by Alfv\'{e}nic cascade \cite[see theoretical considerations in GS95,][]{cho03}. We note that shearing does not affect the PDFs, but it does affect the spectra and the anisotropy of the turbulence. This explains the close relationship between the properties of velocity in our simulations and those of density.

\subsection{Properties of 3D density}

We confirm the existence for density of a rather mysterious property of turbulence called ``extended self-similarity''. While our spectra have a limited inertial range, the moments of density normalized by the third moment show nice power laws. Interestingly enough, we find that the extended self-similarity is not true if one uses the normalization over the third moment of velocity. While we still may wonder about the origin of this phenomenon, the extended self-similarity is an indispensable tool for studyint turbulence intermittency.

Our study of the intermittency of density and the logarithm of density reveals the dependence of the She-L\'{e}v\^{e}que exponents on both sonic and Alfv\'{e}n Mach numbers. Moreover, we note a rather weak dependence on the intermittency when ${\cal M}_s < 1$. For supersonic turbulence the dependency is much stronger. The intermittency changes with the strength of the external magnetic field may frequently be more important. We observe less intermittent structures when $B_\mathrm{ext}$ is strong (see Figure \ref{fig:dens_expons}).

The question of anisotropy of structures is in a natural way related to the presence of magnetic field. When no magnetic field is present, there is a lack of any force that could organize these structures. On the other hand, the introduction of the external magnetic field to the turbulent system induces the Alfv\'{e}nic mode, which is able to imprint its structure on random densities. In MHD turbulence, one direction is chosen, the direction of $B_\mathrm{ext}$, so the anisotropy should be observed. In our sub-Alfv\'{e}nic simulations the anisotropy of density is difficult to distinguish due to large dispersion of the structure functions. The situation improves in the case of models with a weaker magnetic field. Here the results are in accordance with the GS95 model (i.e. $l_\parallel \sim l_\perp^{2/3}$). The anisotropy of the logarithm of density and velocity structures show similar GS95 behavior in all models with observed dependence on the scale and independence of ${\cal M}_s$. However, the structures of $\log \rho$ and velocity for sub-Alfv\'{e}nic models seem to be less anisotropic at the small scales.

Some basic properties of the density distribution in turbulence have been already studied. However, most papers were focused mainly on the variance of distribution. We analyzed here also two higher order statistical moments: skewness and kurtosis. We observe a strong dependence of these moments on ${\cal M}_s$. For instance, skewness grows with the sonic Mach number for mid- and supersonic turbulence. For highly subsonic turbulence we observe the growth of skewness but with a decreasing ${\cal M}_s$. Kurtosis in this case is zero and starts to grow rapidly when the value of ${\cal M}_s$ exceeds about 0.7. Combining these two measures can allow one to determine the value of ${\cal M}_s$ in turbulence. We do not observe any significant dependency of these moments on ${\cal M}_A$.

Genus statistics reveal the changes of the structure of density with sonic Mach number. For lower values of ${\cal M}_s$ density has a normal distribution, signifying roughly the same number of separate rarefactions and condensations. For higher sonic Mach numbers, overdense structures start to dominate over the underdense ones. This changes the structure of density significantly. For low sonic Mach numbers, the structure of the density is spongelike. In the supersonic turbulence, the shocks produce separated high-density 1D and 2D structures embedded in the medium of much lower density. In this sense, the genus statistics confirm the topology with the intermittency.

\subsection{What can be found from observations?}

Many surveys provide extensive information on column density statistics. Those include H{\footnotesize I} surveys, e.g. the GALFA (\url{http://www.naic.edu/alfa/galfa/}) survey of galactic H{\footnotesize I} distribution. The main goal of another observational project, the Wisconsin H-Alpha Mapper (WHAM), is to provide a survey of H$\alpha$ emission from the ISM over the entire northern sky \citep{haffner03,hill05}. The observations provide the dispersion and emission measures, which can be interpreted as the column density and the integral of the square of the electron density along the entire LOS. The survey supported by the numerical studies helps in better understanding the density and distribution of diffuse plasma in a major component of our Galaxy, the warm ionized medium. Combining data from different surveys and using the techniques in this paper, it is possible to get valuable insight into processes taking place in different ISM phases.

If the observable column density distributions are concerned, our study confirms the earlier conclusion in \cite{ostriker03} that the PDFs are affected by the LOS distribution of density, and therefore, the relation between the column density PDFs and the underlying 3D PDFs is rather involved. On the contrary, the relation between the 3D and 2D density spectra is very straightforward, as was shown earlier by many authors \cite[see][]{stutzki99,lazarian95}. This makes the observable column density spectrum a good measure of the underlying 3D density spectrum if the value of the spectral index is concerned. However, the anisotropies of the density distribution that can be revealed for low sonic Mach number by density itself and for high Mach number by the distribution of $\log\rho$  are affected by the LOS integration that obtaining the column densities involves. The problem here is that the GS95 anisotropies of the magnetic field and velocities that induce anisotropies of the density are defined in the system of reference aligned with the {\it local} magnetic field \cite[see discussion in ][]{clv03}. LOS integration averages the local field directions, and therefore, only the anisotropies have to be calculated with respect to the mean magnetic field. This both decreases the degree of anisotropy and erases the scale dependence of the anisotropy. The residual anisotropy is the anisotropy of the turbulence at the injection scale. Therefore, for super-Alfv\'{e}nic turbulence for which magnetic fields are dynamically unimportant at the injection scale, the anisotropy of the column density statistics vanishes altogether.

\cite{padoan03} studied maps of $^{13}$CO intensity of molecular clouds in Taurus and Perseus in terms of structure functions. The integrated intensity maps can be translated into column densities by simple multiplication by a factor \citep[see][]{ossenkopf02}. \cite{padoan03} show that the scaling exponents of the structure functions, normalized to the third order, follow the velocity scaling of the supersonic turbulence. This suggests that the density scaling can be described by a similar hierarchical model, like the one proposed by \cite{she94}. Our study shows that the intermittencies of densities and velocities are intrinsically different and one should exercise care in relating the relevant She-L\'{e}v\^{e}que exponents. Although the intermittencies can reflect the sonic Mach number, this dependence that we observe is weak for subsonic models and the changes are below the errors (see Figure \ref{fig:col_dens_scal}). The low-resolution models showed the weak dependence of the intermittency of density on the sonic Mach number. However, the intermittency is very sensitive to the resolution, which was stressed in \S\ref{sec:resolution}. The growth of error bars with the order of exponent $p$ is due to the amplification of any irregularities in the structure function by higher order moments. Another difficulty is the change of the intermittency while we switch from density to column density, as seen by the comparison of the scaling exponents in Figures \ref{fig:dens_expons} and \ref{fig:dens_int_expon}. We note, that the intermittency of the column densities in general is lower than the intermittency of density. In such situations we can only speak about some rough predictions taken from the observational data.

In addition, we see that the effects of cloud boundaries change the intermittency itself depending on ${\cal M}_s$. For instance, the highly subsonic and supersonic turbulence reveal more intermittent structure after including the boundary. The effect is especially strong for turbulence with high ${\cal M}_s$ and grows with its value. The strength of the external magnetic field does not influence significantly the effects of the cloud boundary.

We also present a number of different measurements that can be helpful in the determination of the properties of density structures in real cosmic objects, like molecular clouds. We compared the statistical measures applied for the 3D density and corresponding 2D column densities. By such comparison, we were able to regain information about the spatial structure from the synthetic maps of the projected density. For instance, we have shown, that the PDFs of density and column density can be related in particular cases.

We find that the statistical moments skewness and kurtosis, which are easy to obtain, can be very useful. They can be employed to determine the sonic Mach number for supersonic turbulence. The subsonic regime of turbulence should be treated more carefully, because the cloud boundary effects might be very important in this case, as was shown in \S\ref{sec:boundary}. Boundary effects influence the PDFs and the scaling of density structure (see Figures \ref{fig:dens_sph_pdfs} and \ref{fig:dens_sph_expon}, respectively). In the case of PDFs, however, some information can still be regained, because the boundary of the cloud mostly influences the low-density tail of the distribution. Three-dimensional density spectra can be recovered successfully from observations of column densities.

Another promising technique for the topology of density and column density study is genus statistics. Although it is a completely different technique than one- or two-point statistics, we have shown that it can both help us with understanding the topology of density structures in turbulence and relate the results obtained for density and column density (see \S\ref{sec:topology}).

Some measures presented in the article, like statistical moments and spectra, should be easily obtained from the molecular maps. Such observation of quantities can be directly compared with our numerical results, allowing quick determination of the conditions in investigated maps. Other measurements, like scaling exponents and genus statistics require more effort to calculate, but can provide more information related to the density structure. Thus, the techniques of the study of turbulence using 2D maps of observed column densities are very useful.

\section{Summary}
\label{sec:summary}

In this paper we investigated the statistics of density structure in compressible MHD turbulence using numerical experiments. We analyzed PDFs, spectra, structure functions, and the intermittency of density and the logarithm of density for experiments with different sonic ${\cal M}_s$ and Alfv\'{e}n ${\cal M}_A$ Mach numbers. We addressed the issue of whether the numerical results can be compared with observations. We found:

\begin{itemize}
\item The amplitude of density fluctuations strongly depends on ${\cal M}_s$
      both in weakly and strongly magnetized turbulent plasmas.
 \begin{enumerate}
 \item[a)] PDFs of density fluctuations have a lognormal distribution.
 \item[b)] The width of the PDF of $\log \rho$ grows with ${\cal M}_s$.
 \item[c)] The flattening observed in density spectra is due to the contribution of
           the highly dense small-scale structures generated in the supersonic
           turbulence.
 \end{enumerate}
\item Fluctuations of the logarithm of density are much more regular than those
      of density. The logarithm of density exhibits the GS95 scalings and
      anisotropies. The PDFs are close to Gaussian.
\item Structure functions of density and the logarithm of density reveal
      self-similarity. The density field is not self-similar to the velocity
      one.
\item The intermittency of density structure depends on both ${\cal M}_s$
      and ${\cal M}_A$. We report only weak dependence on ${\cal M}_s$ for
      subsonic turbulence. In the case of supersonic models the dependence is
      much stronger.
\item Additional care about resolution is necessary for intermittency studies.
      The She-L\'{e}v\^{e}que exponents results can be affected by the
      resolution.
\item Similar to their 3D counterparts, the 2D measures of column densities,
      i.e. the variance, skewness and kurtosis, depend on ${\cal M}_s$ and can
      be used to determine the value of ${\cal M}_s$ from observations.
\item The cloud boundary effects are important and should be accounted for by the
      study. The most sensitive to the effects of cloud boundaries are the
      PDFs and intermittency studies.
\item The genus statistic is sensitive to ${\cal M}_s$. The 2D genus analysis is a promising tool for studying the topology of the ISM.
\end{itemize}

\acknowledgments
A. L. acknowledges the discussions with Annick Pouquet. The research of G. K. and A. L. is supported by NSF grant AST 03-07869, the Vilas Professorship Award, and the Center for Magnetic Self-Organization in Astrophysical and Laboratory Plasmas; A. B. is supported by an Ice Cube grant. We thank the referee for valuable input.


\clearpage

\clearpage

\clearpage

\clearpage

\clearpage

\clearpage

\clearpage

\clearpage

\clearpage

\clearpage

\clearpage

\clearpage

\clearpage

\clearpage

\clearpage

\clearpage

\clearpage

\clearpage

\clearpage


\begin{deluxetable}{cccccccccc}
\tablewidth{0pt}
\tablecolumns{10}
\tabletypesize{\footnotesize}
\tablecaption{List of All Analyzed Models\label{tab:models}}
\tablehead{
 \colhead{Model} &
 \colhead{$B_\mathrm{ext}$} &
 \colhead{$P_\mathrm{ini}$} &
 \colhead{${\cal M}_{s}$} &
 \colhead{${\cal M}_{A}$} &
 \colhead{Resolution} &
 \colhead{Max. Time} &
 \colhead{Number of Snapshots} &
 \colhead{max $\delta \rho$} &
 \colhead{max $\delta V$} }
\startdata
B1P10   & 1.0 & 10.0 & 0.20$^{\pm 0.02}$ & 0.62$^{\pm 0.04}$ & $128^3$ & 25.0 & 16 & $\sim$0.06 & $\sim$0.6 \\
B1P9    & 1.0 &  9.0 & 0.21$^{\pm 0.02}$ & 0.62$^{\pm 0.04}$ & $128^3$ & 25.0 & 16 & $\sim$0.06 & $\sim$0.6 \\
B1P8    & 1.0 &  8.0 & 0.22$^{\pm 0.02}$ & 0.62$^{\pm 0.04}$ & $128^3$ & 25.0 & 16 & $\sim$0.06 & $\sim$0.6 \\
B1P7    & 1.0 &  7.0 & 0.24$^{\pm 0.02}$ & 0.63$^{\pm 0.04}$ & $128^3$ & 25.0 & 16 & $\sim$0.06 & $\sim$0.6 \\
B1P6    & 1.0 &  6.0 & 0.26$^{\pm 0.02}$ & 0.63$^{\pm 0.04}$ & $128^3$ & 25.0 & 16 & $\sim$0.06 & $\sim$0.6 \\
B1P5    & 1.0 &  5.0 & 0.28$^{\pm 0.02}$ & 0.63$^{\pm 0.04}$ & $128^3$ & 25.0 & 16 & $\sim$0.07 & $\sim$0.6 \\
B1P4    & 1.0 &  4.0 & 0.32$^{\pm 0.03}$ & 0.64$^{\pm 0.04}$ & $128^3$ & 25.0 & 16 & $\sim$0.08 & $\sim$0.6 \\
B1P3    & 1.0 &  3.0 & 0.37$^{\pm 0.04}$ & 0.64$^{\pm 0.05}$ & $128^3$ & 25.0 & 16 & $\sim$0.1  & $\sim$0.6 \\
B1P2    & 1.0 &  2.0 & 0.46$^{\pm 0.05}$ & 0.65$^{\pm 0.05}$ & $128^3$ & 25.0 & 16 & $\sim$0.1  & $\sim$0.6 \\
B1P1    & 1.0 &  1.0 & 0.66$^{\pm 0.05}$ & 0.66$^{\pm 0.05}$ & $128^3$ & 25.0 & 16 & $\sim$0.2  & $\sim$0.6 \\

B1P.9   & 1.0 &  0.9 & 0.70$^{\pm 0.05}$ & 0.66$^{\pm 0.05}$ & $128^3$ & 25.0 & 16 & $\sim$0.2  & $\sim$0.6 \\
B1P.8   & 1.0 &  0.8 & 0.74$^{\pm 0.06}$ & 0.66$^{\pm 0.05}$ & $128^3$ & 25.0 & 16 & $\sim$0.3  & $\sim$0.6 \\
B1P.7   & 1.0 &  0.7 & 0.80$^{\pm 0.06}$ & 0.66$^{\pm 0.05}$ & $128^3$ & 25.0 & 16 & $\sim$0.3  & $\sim$0.6 \\
B1P.6   & 1.0 &  0.6 & 0.86$^{\pm 0.07}$ & 0.66$^{\pm 0.05}$ & $128^3$ & 25.0 & 16 & $\sim$0.3  & $\sim$0.6 \\
B1P.5   & 1.0 &  0.5 & 0.94$^{\pm 0.08}$ & 0.66$^{\pm 0.05}$ & $128^3$ & 25.0 & 16 & $\sim$0.3  & $\sim$0.6 \\
B1P.4   & 1.0 &  0.4 & 1.05$^{\pm 0.08}$ & 0.66$^{\pm 0.05}$ & $128^3$ & 25.0 & 16 & $\sim$0.4  & $\sim$0.7 \\
B1P.3   & 1.0 &  0.3 & 1.21$^{\pm 0.09}$ & 0.66$^{\pm 0.05}$ & $128^3$ & 25.0 & 16 & $\sim$0.4  & $\sim$0.7 \\
B1P.2   & 1.0 &  0.2 & 1.47$^{\pm 0.09}$ & 0.66$^{\pm 0.04}$ & $128^3$ & 25.0 & 16 & $\sim$0.5  & $\sim$0.7 \\
B1P.1   & 1.0 &  0.1 & 2.1$^{\pm 0.2}$   & 0.67$^{\pm 0.06}$ & $128^3$ & 25.0 & 16 & $\sim$0.9  & $\sim$0.7 \\

B.1P10  & 0.1 & 10.0  & 0.23$^{\pm 0.01}$ & 7.1$^{\pm 0.3}$ & $128^3$ & 25.0 & 16 & $\sim$0.06 & $\sim$0.7 \\
B.1P9   & 0.1 &  9.0  & 0.24$^{\pm 0.01}$ & 7.1$^{\pm 0.4}$ & $128^3$ & 25.0 & 16 & $\sim$0.06 & $\sim$0.7 \\
B.1P8   & 0.1 &  8.0  & 0.25$^{\pm 0.02}$ & 7.1$^{\pm 0.4}$ & $128^3$ & 25.0 & 16 & $\sim$0.06 & $\sim$0.7 \\
B.1P7   & 0.1 &  7.0  & 0.27$^{\pm 0.02}$ & 7.1$^{\pm 0.4}$ & $128^3$ & 25.0 & 16 & $\sim$0.06 & $\sim$0.7 \\
B.1P6   & 0.1 &  6.0  & 0.29$^{\pm 0.02}$ & 7.2$^{\pm 0.3}$ & $128^3$ & 25.0 & 16 & $\sim$0.06 & $\sim$0.7 \\
B.1P5   & 0.1 &  5.0  & 0.32$^{\pm 0.02}$ & 7.2$^{\pm 0.3}$ & $128^3$ & 25.0 & 16 & $\sim$0.06 & $\sim$0.7 \\
B.1P4   & 0.1 &  4.0  & 0.36$^{\pm 0.02}$ & 7.2$^{\pm 0.3}$ & $128^3$ & 25.0 & 16 & $\sim$0.07 & $\sim$0.7 \\
B.1P3   & 0.1 &  3.0  & 0.42$^{\pm 0.02}$ & 7.3$^{\pm 0.3}$ & $128^3$ & 25.0 & 16 & $\sim$0.09 & $\sim$0.7 \\
B.1P2   & 0.1 &  2.0  & 0.52$^{\pm 0.03}$ & 7.3$^{\pm 0.4}$ & $128^3$ & 25.0 & 16 & $\sim$0.1  & $\sim$0.6 \\
B.1P1   & 0.1 &  1.0  & 0.72$^{\pm 0.03}$ & 7.2$^{\pm 0.3}$ & $128^3$ & 25.0 & 16 & $\sim$0.2  & $\sim$0.7 \\

B.1P.9  & 0.1 &  0.9 & 0.77$^{\pm 0.03}$ & 7.3$^{\pm 0.3}$ & $128^3$ & 25.0 & 16 & $\sim$0.3  & $\sim$0.7 \\
B.1P.8  & 0.1 &  0.8 & 0.82$^{\pm 0.03}$ & 7.3$^{\pm 0.2}$ & $128^3$ & 25.0 & 16 & $\sim$0.3  & $\sim$0.7 \\
B.1P.7  & 0.1 &  0.7 & 0.88$^{\pm 0.03}$ & 7.4$^{\pm 0.3}$ & $128^3$ & 25.0 & 16 & $\sim$0.3  & $\sim$0.7 \\
B.1P.6  & 0.1 &  0.6 & 0.94$^{\pm 0.03}$ & 7.3$^{\pm 0.3}$ & $128^3$ & 25.0 & 16 & $\sim$0.3  & $\sim$0.7 \\
B.1P.5  & 0.1 &  0.5 & 1.03$^{\pm 0.03}$ & 7.3$^{\pm 0.3}$ & $128^3$ & 25.0 & 16 & $\sim$0.4  & $\sim$0.7 \\
B.1P.4  & 0.1 &  0.4 & 1.15$^{\pm 0.04}$ & 7.3$^{\pm 0.3}$ & $128^3$ & 25.0 & 16 & $\sim$0.4  & $\sim$0.7 \\
B.1P.3  & 0.1 &  0.3 & 1.33$^{\pm 0.04}$ & 7.3$^{\pm 0.3}$ & $128^3$ & 25.0 & 16 & $\sim$0.5  & $\sim$0.7 \\
B.1P.2  & 0.1 &  0.2 & 1.66$^{\pm 0.06}$ & 7.4$^{\pm 0.3}$ & $128^3$ & 25.0 & 16 & $\sim$0.6  & $\sim$0.8 \\
B.1P.1  & 0.1 &  0.1 & 2.3$^{\pm 0.1}$   & 7.4$^{\pm 0.3}$ & $128^3$ & 25.0 & 16 & $\sim$0.8  & $\sim$0.8 \\

\hline
B1P8    & 1.0 & 8.0  & 0.23$^{\pm 0.01}$ & 0.64$^{\pm 0.02}$ & $256^3$ & 10.0 &  8 & $\sim$0.05 & $\sim$0.6 \\
B1P4    & 1.0 & 4.0  & 0.33$^{\pm 0.01}$ & 0.65$^{\pm 0.02}$ & $256^3$ & 15.5 & 14 & $\sim$0.1  & $\sim$0.7 \\
B1P1    & 1.0 & 1.0  & 0.68$^{\pm 0.03}$ & 0.68$^{\pm 0.03}$ & $256^3$ & 10.1 &  6 & $\sim$0.3  & $\sim$0.6 \\
B1P.1   & 1.0 & 0.1  & 2.20$^{\pm 0.09}$ & 0.69$^{\pm 0.03}$ & $256^3$ & 10.1 &  9 & $\sim$0.9  & $\sim$0.7 \\
B1P.01  & 1.0 & 0.01 & 7.0$^{\pm 0.3}$   & 0.70$^{\pm 0.04}$ & $256^3$ & 10.1 & 10 & $\sim$1.9  & $\sim$0.7 \\

B.1P8   & 0.1 & 8.0  & 0.26$^{\pm 0.03}$ & 7.3$^{\pm 0.7}$ & $256^3$ & 14.5 & 14 & $\sim$0.03 & $\sim$0.7 \\
B.1P4   & 0.1 & 4.0  & 0.36$^{\pm 0.04}$ & 7.2$^{\pm 0.7}$ & $256^3$ & 16.7 & 14 & $\sim$0.07 & $\sim$0.7 \\
B.1P1   & 0.1 & 1.0  & 0.74$^{\pm 0.06}$ & 7.4$^{\pm 0.6}$ & $256^3$ & 21.0 & 16 & $\sim$0.2  & $\sim$0.7 \\
B.1P.1  & 0.1 & 0.1  & 2.34$^{\pm 0.08}$ & 7.4$^{\pm 0.3}$ & $256^3$ & 22.3 & 17 & $\sim$0.8  & $\sim$0.7 \\
B.1P.01 & 0.1 & 0.01 & 7.1$^{\pm 0.3}$   & 7.1$^{\pm 0.3}$ & $256^3$ & 20.9 & 16 & $\sim$1.5  & $\sim$0.7 \\
\hline
B1P1    & 1.0 & 1.0   & $\sim$0.6 & $\sim$0.6 & $512^3$ &  7.2 &  1 & $\sim$0.3  & $\sim$0.6 \\
\enddata
\end{deluxetable}
\clearpage

\begin{deluxetable}{ccc|ccc}
\tablewidth{0pt}
\tablecolumns{6}
\tablecaption{Slopes of the Power Spectrum of Density and the Logarithm of Density Fluctuations \label{tab:slopes}}
\tablehead{
 \colhead{} &
 \colhead{${\cal M}_{A}\sim0.7$}  &
 \colhead{} &
 \colhead{} &
 \colhead{${\cal M}_{A}\sim7$} &
 \colhead{}
\\
 \colhead{${\cal M}_{s}$} &
 \colhead{$\alpha_\rho$}  &
 \colhead{$\alpha_{\log\rho}$} &
 \colhead{${\cal M}_{s}$} &
 \colhead{$\alpha_\rho$}  &
 \colhead{$\alpha_{\log\rho}$}
}
\startdata
0.23$^{\pm0.01}$ & -2.3$^{\pm0.3}$ & -2.3$^{\pm0.3}$ & 0.26$^{\pm0.03}$ & -1.7$^{\pm0.3}$ & -1.7$^{\pm0.3}$\\
0.33$^{\pm0.01}$ & -2.2$^{\pm0.3}$ & -2.2$^{\pm0.3}$ & 0.36$^{\pm0.04}$ & -1.7$^{\pm0.3}$ & -1.7$^{\pm0.3}$\\
0.68$^{\pm0.03}$ & -2.0$^{\pm0.3}$ & -2.1$^{\pm0.3}$ & 0.74$^{\pm0.06}$ & -1.6$^{\pm0.2}$ & -1.6$^{\pm0.3}$\\
2.20$^{\pm0.03}$ & -1.3$^{\pm0.2}$ & -2.0$^{\pm0.2}$ & 2.34$^{\pm0.08}$ & -1.2$^{\pm0.2}$ & -1.6$^{\pm0.2}$\\
7.0$^{\pm0.3}$   & -0.5$^{\pm0.1}$ & -1.7$^{\pm0.2}$ & 7.1$^{\pm0.3}$   & -0.6$^{\pm0.2}$ & -1.5$^{\pm0.2}$
\enddata
\tablecomments{The values have been estimated within the inertial range for models with ${\cal M}_A\sim0.7$ ({\em left}) and ${\cal M}_A\sim7$ ({\em right}). Errors of spectral indices combine the errors of estimation at each time snapshot and the standard deviation of variance in time. Errors for sonic Mach numbers are the standard deviation of their variance in time calculated over the period starting from $t \ge 5$ to the last available snapshot.}
\end{deluxetable}

\clearpage


\begin{thebibliography}{}
\bibitem[Armstrong et al. (1995)]{armstrong95}
  Armstrong, J.~W., Rickett, B.~J. \& Spangler, S.~R., 1995, \apj, 443, 209
\bibitem[Benzi et al. (1993)]{benzi93}
  Benzi, R., Ciliberto, S., Tripiccione, R., Baudet, C., Massaioli, F. \& Succi, S., 1993, \pre, 48, 29
\bibitem[Beresnyak et al. (2005)]{beresnyak05}
  Beresnyak, A., Lazarian, A. \& Cho, J., 2005, \apj, 624, L93, BLC05
\bibitem[Biskamp (2003)]{biskamp03}
  Biskamp, D., 2003, {\em Magnetohydrodynamic Turbulence}, Cambridge University Press
\bibitem[Boldyrev et al. (2002)]{boldyrev02}
  Boldyrev, S., Nordlund, \AA. \& Padoan, P., 2002, \apj, 573, 678
\bibitem[Cho \& Vishniac (2000)]{cho00}
  Cho, J. \& Vishniac, E.~T., 2000, \apj, 539, 273
\bibitem[Cho \& Lazarian (2002)]{cho02a}
  Cho, J. \& Lazarian, A., 2002, Phys. Rev. Letters, 88, 245001, CL02
\bibitem[Cho et al. (2002)]{cho02b}
  Cho, J., Lazarian, A. \& Vishniac, E. T., 2002, \apj, 564, 291
\bibitem[Cho \& Lazarian (2003)]{cho03}
  Cho, J. \& Lazarian, A., 2003, \mnras, 345, 325, CL03
\bibitem[Cho et al. (2003)]{clv03}
  Cho, J., Lazarian, A. \& Vishniac, E. T., 2003, Lecture Notes in Physics, 614, 56
\bibitem[Cho \& Lazarian (2004)]{cho04}
  Cho, J. \& Lazarian, A., 2004, Proc. Summer Program 2004, Center for Turbulence Research, 75, http://ctr.stanford.edu/SP04.html
\bibitem[Cho \& Lazarian (2005)]{cho05}
  Cho, J. \& Lazarian, A., 2005, Theor. and Comp. Fl. Dynam., 19, 127
\bibitem[Cox \& Smith (1974)]{cox74}
  Cox, D.~P. \& Smith, B.~W., 1974, \apjl, 189, L105
\bibitem[Dickman (1985)]{dickman85}
  Dickman, R.~L., 1985, {\em Turbulence in Molecular Clouds}, Tucson, University of Arizona Press
\bibitem[Esquivel \& Lazarian (2005)]{esquivel05}
  Esquivel, A. \& Lazarian, A., 2005, \apj, 631, 320
\bibitem[Falgarone \& Perault (1988)]{falgarone88}
  Falgarone, E. \& Perault, M., 1988, \aap, 205, L1
\bibitem[Falgarone et al. (1995)]{falgarone95}
  Falgarone, E., Pineau des Forets, G. \& Roueff, E, 1995, \aap, 300, 870
\bibitem[Falgarone et al. (2005)]{falgarone05}
  Falgarone, E., Hily-Blant, P., Pety, J. \& Pineau Des For{\^e}ts, G., 2005, AIP Conf. Proc. 784: {\em Magnetic Fields in the Universe: From Laboratory and Stars to Primordial Structures.}, 299
\bibitem[Goldreich \& Sidhar (1995)]{goldreich95}
  Goldreich, P., Sidhar, S., 1995, \apj, 438, 763, GS95
\bibitem[Gott et al. (1986)]{gott86}
  Gott, J. R., Melott, A. L. \& Dickinson, M., 1986, \apj, 306, 341
\bibitem[Haffner et al. (2003)]{haffner03}
  Haffner, L.~M., Reynolds, R.~J., Tufte, S.~L., Madsen, G.~J., Jaehnig, K.~P., Percival, J.~W., 2003, \apjs, 149, 405
\bibitem[Hill et al. (2005)]{hill05}
  Hill, A.~S., Benjamin, R.~A., Reynolds, R.~J. \& Haffner, L.~M., 2005, Bulletin of the American Astronomical Society, 1302
\bibitem[Kallenberg (1997)]{kallenberg97}
  Kallenberg, O., 1997, {\em Foundations of Modern Probability}, New York: Springer-Verlag
\bibitem[Kaplan \& Pikelner (1970)]{kaplan70}
  Kaplan, S.~A. \& Pikelner, S.~B., 1970, {\em The Interstellar Medium}, Cambridge: Harvard University Press
\bibitem[Kim \& Ryu (2005)]{kim05}
  Kim, J. and Ryu, D., 2005, \apjl, 630, L45
\bibitem[Lazarian (1995)]{lazarian95}
  Lazarian, A., 1995, \aap, 293, 507
\bibitem[Lazarian \& Pogosyan (2000)]{lazarian00}
  Lazarian, A. \& Pogosyan, D., 2000, \apj, 537, 720
\bibitem[Lazarian et al. (2002)]{lazarian02}
  Lazarian, A., Pogosyan, D. \& Esquivel, A., 2002, ASP Conf. Ser. 276: {\em Seeing Through the Dust: The Detection of HI and the Exploration of the ISM in Galaxies}
\bibitem[Lazarian (2004)]{lazarian04}
  Lazarian, A., 2004, JKAS, 37, 563
\bibitem[Lazarian \& Beresnyak (2005)]{lazarian05}
  Lazarian, A. \& Beresnyak, A., Proceedings of the conference {\em The Magnetized Plasma in Galaxy Evolution}, Krak\'{o}w, Jagiellonian University Press, 2005
\bibitem[Levy, Puppo \& Russo (1999)]{levy99}
  Levy, D., Puppo, G. \& Russo, G., 1999, Mathematical Modelling and Numerical Analysis, 33, 547
\bibitem[Lithwick \& Goldreich (2001)]{lithwick01}
  Lithwick, Y. \& Goldreich, P., 2001, \apj, 562, 279
\bibitem[Liu \& Osher (1998)]{liu98}
  Liu, X.-D. \& Osher, S., 1998, Journal of Computational Physics, 141, 1
\bibitem[Mac Low et al. (1998)]{maclow98}
  Mac Low, M.-M., Klessen, R.~S., Burkert, A. \& Smith, M.~D., 1998, Physical Review Letters, 80, 2754
\bibitem[Mac Low \& Klessen (2004)]{maclow04}
  Mac Low, M.-M. \& Klessen, R.~S., 2004, Rev. of Mod. Phys., 76, 125
\bibitem[McKee \& Ostriker (1977)]{mckee77}
  McKee, C.~F. \& Ostriker, J.~P., 1977, \apj, 218, 148
\bibitem[McKee \& Tan (2002)]{mckee02}
  McKee, C.~F. \& Tan, J.~C., 2002, \nat, 416, 59
\bibitem[Melott et al. (1989)]{melott89}
  Melott, A. L., Cohen, A. P., Hamilton, A. J. S., Gott, J. R., Weinberg, D. H., 1989, \apj, 345, 618
\bibitem[Melott (1990)]{melott90}
  Melott, A. L., 1990, \physrep, 193, 1, 1
\bibitem[Montgomery et al. (1987)]{montgomery87}
  Montgomery, D., Brown, M.~R., Matthaeus, W.~H., 1987, \jgr, 92, 282
\bibitem[M\"uller \& Biskamp (2000)]{mueller00}
  M\"uller, W. C. \& Biskamp, D., 2000, \prl, 84, 3
\bibitem[Nordlund \& Padoan (1999)]{nordlund99}
  Nordlund, \AA.~K. \& Padoan, P., 1999, {\em Interstellar Turbulence}, proceedings of the 2nd Guillermo Haro Conference, Cambridge University Press
\bibitem[Novikov (1994)]{novikov94}
  Novikov, E.~A., 1994, \pre, 50, 3303
\bibitem[Ossenkopf (2002)]{ossenkopf02}
  Ossenkopf, V., 2002, \aap, 391, 295
\bibitem[Ostriker et al. (2001)]{ostriker01}
  Ostriker, E.~C., Stone, J.~M. \& Gammie, C.~F., 2001, \apj, 546, 980
\bibitem[Ostriker (2003)]{ostriker03}
  Ostriker, E.~C., 2003, LNP, 614, 252
\bibitem[Padoan et al. (2003)]{padoan03}
  Padoan, P., Boldyrev, S., Langer, W. \& Nordlund, \AA.,  2003, \apj, 583, 308
\bibitem[Padoan et al. (2004)]{padoan04}
  Padoan, P., Jimenez, R., Nordlund, \AA. \& Boldyrev, S., 2004, \prl, 92, 1102
\bibitem[Passot \& V{\'a}zquez-Semadeni (1998)]{passot98}
  Passot, T. and V{\'a}zquez-Semadeni, E., 1998, \pre, 58, 4501
\bibitem[Passot \& V{\'a}zquez-Semadeni (2003)]{passot01}
  Passot, T. and V{\'a}zquez-Semadeni, E., 2003, \AA, 398, 845
\bibitem[Porter et al. (2002)]{porter02}
  Porter, D., Pouquet, A. \& Woodward, P., 2002, \pre, 66, 026301
\bibitem[Scalo et al. (1998)]{scalo98}
  Scalo, J., Vazquez-Semadeni, E., Chappell, D. \& Passot, T., 1998, \apj, 504, 835
\bibitem[She \& L\'{e}v\^{e}que (1994)]{she94}
  She, Z. \& L\'{e}v\^{e}que, E., 1994, \prl, 72, 3
\bibitem[Shebalin et al. (1983)]{shebalin83}
  Shebalin, J.~V., Matthaeus, W.~H. \& Montgomery, D., 1983, JPlPh, 29, 525
\bibitem[Stone et al. (1998)]{stone98}
  Stone, J.~M., Ostriker, E.~C. \& Gammie, C.~F., 1998, \apjl, 508, L99
\bibitem[Stutzki (1999)]{stutzki99}
  Stutzki, J., 1999, {\em Plasma Turbulence and Energetic Particles in Astrophysics}, Proceedings of the International Conference, Cracow (Poland), 5-10 September 1999
\bibitem[V{\'a}zquez-Semadeni \& Garc{\'{\i}}a (2001)]{vazquez01}
  V{\'a}zquez-Semadeni, E. \& Garc{\'{\i}}a, N., 2001, \apj, 557, 727
\bibitem[Vestuto et al. (2003)]{vestuto03}
  Vestuto, J.~G., Ostriker, E.~C.\& Stone, J.~M., 2003, \apj, 590, 858
\bibitem[Weinberg (1988)]{weinberg88}
  Weinberg, D. H., 1988, \pasp, 100, 1373
\end{thebibliography}
\end{document}